\documentclass[12pt]{article}
\usepackage{a4wide, amsmath, latexsym, epsfig, graphicx, rotating, fancyhdr, tabularx, afterpage}
\usepackage[T2A]{fontenc}
\usepackage[utf8]{inputenc}
\usepackage[english]{babel}
\usepackage{axodraw}
\usepackage{subfigure}
\begin{document}
\begin{titlepage}
\begin{center}
{\Large \bf Impact of QED radiative corrections on Parton Distribution Functions}
\end{center}

\vspace*{0.5cm}
\begin{center}
Renat R. Sadykov

\vspace*{0.5cm}
{\em Joint Institute for Nuclear Research, Joliot-Curie str. 6, Dubna, 141980, Russia}
\end{center}

\vspace*{0.5 cm}
\begin{center}
{\bf Abstract}
\end{center}

The level of precision achieved by the experimental measurements at the LHC requires the inclusion
of higher order electroweak effects to the processes of $ pp $ scattering. In particular the
photon-induced process $ \gamma\gamma \to \ell^+\ell^- $ make a significant contribution
($ \sim 10 \%$) to the dilepton invariant mass distribution. To evaluate the cross-section of this
process one need to know the parton distribution function (PDF) of the photon in the proton
$ \gamma (x,\mu^2) $. The aim of the current study is to investigate the impact of QED
corrections on PDFs and describe the implementation of QED-modified evolution equations
into beta release of new version of {\tt QCDNUM} program. The {\tt APPLGRID} interface to
{\tt SANC} Monte Carlo generator for fast evaluation of photon-induced cross-section is also
outlined. The results were cross-checked with {\tt partonevolution} program, {\tt MRST2004QED} PDF
set and {\tt APFEL} program. The described developments are planned to include into
{\tt HERAFitter} package and can be used to determine the photon PDF using new data from the LHC
experiments.
\end{titlepage}
\section{Introduction}
With an abundance of high precision measurements available from the LHC experiments, the theoretical
precision in matching data accuracy have become crucial. Recent studies on high mass Drell-Yan
production in ATLAS \cite{Aad:2013iua} have shown that the size of the photon-induced contribution is
as large as the uncertainties arising from the different choice of PDF set. Therefore it become
important to address the impact of QED corrections to PDFs.

Currently two PDF sets are available which incorporate the photon contribution of the proton. These
are old {\tt MRST2004QED }\cite{Martin:2004dh} and most recent {\tt NNPDF2.3QED}
\cite{Ball:2013hta}.

In this paper an alternative approach is presented. The QED-modified evolution is implemented into
beta release of new version of {\tt QCDNUM} \cite{Botje:2010ay} program (v17-01-0b) which allows to
solve arbitrary number of coupled DGLAP evolution equations. This implementation together with the
{\tt APPLGRID} \cite{Sutton:2010zz} interface to {\tt SANC} \cite{Andonov:2008ga} Monte Carlo
generator for fast evaluation of photon-induced cross-section are planned to include into QCD
data fit package {\tt HERAFitter} \cite{Abramowicz:1900rp}. The advantage of this approach is the
possibility of using different data sets for QCD fits within a single framework.

The paper is organized as follows. Section 2 describes the modification of QCD evolution
required to accommodation for QED effects, section 3 refers to implementation of fast-evaluation of
photon-induced process cross-section, in section 4 the comparison of {\tt QCDNUM+QED} with
{\tt partonevolution} \cite{Weinzierl:2002mv} program (v1.1.3), {\tt MRST2004QED} PDF set and
{\tt APFEL} \cite{Bertone:2013vaa} program (v2.0.0) are presented. Conclusions are given in
section 5.
\section{QED evolution of parton densities}
If the PDFs are known at the initial scale $ \mu_0^2 $, then the evaluation of PDFs at the scale
$ \mu^2 $ in pQCD is performed using DGLAP evolution equations. QED-modified DGLAP evolution
equations for PDF of quarks $ q_i(x,\mu_F^2) $,
anti-quarks $ \bar{q}_i(x,\mu_F^2) $, gluon $ g(x,\mu_F^2) $ and photon $ \gamma (x,\mu_F^2) $ can
be written as:
\begin{equation}
\begin{split}
&\frac{\partial q_i}{\partial \ln\mu^2} = \sum_{j=1}^{n_f} P_{q_i q_j} \otimes q_j
+ \sum_{j=1}^{n_f} P_{q_i \bar{q}_j} \otimes \bar{q}_j
+ P_{q_i g} \otimes g + P_{q_i \gamma} \otimes \gamma,\\
&\frac{\partial \bar{q}_i}{\partial \ln\mu^2} = \sum_{j=1}^{n_f} P_{\bar{q}_i q_j} \otimes q_j
+ \sum_{j=1}^{n_f} P_{\bar{q}_i \bar{q}_j} \otimes \bar{q}_j
+ P_{\bar{q}_i g} \otimes g + P_{\bar{q}_i \gamma} \otimes \gamma,\\
&\frac{\partial g}{\partial \ln\mu^2} = \sum_{j=1}^{n_f} P_{g q_j} \otimes q_j
+ \sum_{j=1}^{n_f} P_{g \bar{q}_j} \otimes \bar{q}_j
+ P_{g g} \otimes g,\\
&\frac{\partial \gamma}{\partial \ln\mu^2} = \sum_{j=1}^{n_f} P_{\gamma q_j} \otimes q_j
+ \sum_{j=1}^{n_f} P_{\gamma \bar{q}_j} \otimes \bar{q}_j
+ P_{\gamma \gamma} \otimes \gamma,\\
\end{split}
\label{eq1}
\end{equation}
where $ \otimes $-operation denotes the Mellin convolution defined as
\begin{equation}
\left[f \otimes g\right] (x) \equiv \int_x^1 \frac{dz}{z} f\left(\frac{x}{z}\right) g(z)
= \int_x^1 \frac{dz}{z} f(z) g\left(\frac{x}{z}\right).
\label{eq2}
\end{equation}

The splitting functions at NLO QCD and LO QED are expressed as
\begin{equation}
\begin{split}
&P_{q_i q_j} = P_{\bar{q}_i \bar{q}_j} = a_s\delta_{ij}P_{q q}^{(0)}
+ a_s^2\left(\delta_{ij}\frac{P_{+}^{(1)}+P_{-}^{(1)}}{2}
+ \frac{P_{q q}^{(1)}-P_{+}^{(1)}}{2n_f} \right) + a\delta_{ij}e_ie_j\tilde{P}_{q q}^{(0)},\\
&P_{q_i \bar{q}_j} = P_{\bar{q}_i q_j} = a_s^2\left(\delta_{ij}\frac{P_{+}^{(1)}-P_{-}^{(1)}}{2}
+ \frac{P_{q q}^{(1)}-P_{+}^{(1)}}{2n_f} \right),\\
&P_{q_i g} = P_{\bar{q}_i g} = a_s \frac{P_{q g}^{(0)}}{2n_f} + a_s^2 \frac{P_{q g}^{(1)}}{2n_f},\\
&P_{q_i \gamma} = P_{\bar{q}_i \gamma} = ae_i^2 \frac{P_{q \gamma}^{(0)}}{2n_f},\\
&P_{g q_i} = P_{g \bar{q}_i} = a_s P_{g q}^{(0)} + a_s^2 P_{g q}^{(1)},\\
&P_{g g} = a_s P_{g g}^{(0)} + a_s^2 P_{g g}^{(1)},\\
&P_{\gamma q_i} = P_{\gamma \bar{q}_i} = ae_i^2 P_{\gamma q}^{(0)},\\
&P_{\gamma \gamma} = a P_{\gamma \gamma}^{(0)},
\end{split}
\label{eq3}
\end{equation}
where LO splitting kernels are given by
\begin{equation}
\begin{split}
&P_{qq}^{(0)} = \frac{4}{3}\left(\frac{1+x^2}{(1-x)_{+}}
+ \frac{3}{2}\delta(1-x)\right), \quad \tilde{P}_{qq}^{(0)} = \frac{3}{4} P_{qq}^{(0)},\\
&P_{qg}^{(0)} = n_f\left(x^2 + (1-x)^2 \right), \quad
P_{q\gamma}^{(0)} = 2 P_{qg}^{(0)},\\
&P_{gq}^{(0)} = \frac{4}{3}\left(\frac{1 + (1-x)^2}{x}\right), \quad
P_{\gamma q}^{(0)} = \frac{3}{4} P_{gq}^{(0)},\\
&P_{gg}^{(0)} = 6\left(\frac{x}{(1-x)_{+}} + \frac{1-x}{x} + x(1-x)
+\left(\frac{11}{12} - \frac{n_f}{18}\right)\delta(1-x)\right),\\
&\tilde{P}_{\gamma\gamma}^{(0)} = -\frac{2}{3}\sum_{i=1}^{n_f}e_i^2\delta(1-x),
\end{split}
\label{eq4}
\end{equation}
and the expressions for NLO singlet splitting kernels $ P_{qq}^{(1)}$, $ P_{qg}^{(1)}$,
$ P_{gq}^{(1)}$, $ P_{gg}^{(1)}$ can be found in \cite{Furmanski:1980cm} (see footnote on p.9 of
\cite{Botje:2010ay} concerning the known misprints in these expressions) and those for NLO
non-singlet splitting kernels $ P_{+}^{(1)}$, $P_{-}^{(1)} $ can be found in \cite{Curci:1980uw}.

For pure QCD evolution the equations (\ref{eq1}) can be simplified using singlet and
non-singlet combinations of quark densities. Then the singlet quark density obeys the coupled
evolution equations with the gluon density and non-singlet combinations evolve independently. This
decomposition is not suitable for QED-modified evolution since up- and down- quarks have different
electric charges. Instead for QED evolution we use the following basis of distribution functions:
\begin{equation}
\begin{split}
&f_1 = \Delta = u+\bar{u}+c+\bar{c}-d-\bar{d}-s-\bar{s}-b-\bar{b},\\
&f_2 = \Sigma = u+\bar{u}+c+\bar{c}+d+\bar{d}+s+\bar{s}+b+\bar{b},\\
&f_3 = g,\\
&f_4 = \gamma,\\
&f_5 = d_v = d-\bar{d},\\
&f_6 = u_v = u-\bar{u},\\
&f_7 = \Delta_{ds} = d+\bar{d}-s-\bar{s},\\
&f_8 = \Delta_{uc} = u+\bar{u}-c-\bar{c},\\
&f_9 = \Delta_{sb} = s+\bar{s}-b-\bar{b}.
\end{split}
\label{eq5}
\end{equation}

In this basis we have $ 4 $ coupled and $ 5 $ uncoupled evolution equations:
\begin{equation}
\begin{split}
&\frac{\partial}{\partial \ln\mu^2}
\left(
\begin{array}{c}
f_1\\
f_2\\
f_3\\
f_4
\end{array}
\right)
=
\left(
\begin{array}{cccc}
P_{11} & P_{12} & P_{13} & P_{14}\\
P_{21} & P_{22} & P_{23} & P_{24}\\
P_{31} & P_{32} & P_{33} & P_{34}\\
P_{41} & P_{42} & P_{43} & P_{44}
\end{array}
\right)
\otimes
\left(
\begin{array}{c}
f_1\\
f_2\\
f_3\\
f_4
\end{array}
\right),\\
&\frac{\partial f_i}{\partial \ln\mu^2} = P_{ii} \otimes f_i, \quad i = 5,...,9.
\end{split}
\label{eq6}
\end{equation}

We find after simple calculations that the expressions for splitting kernels $ P_{ii} $ in this
basis are given by
\begin{equation}
\begin{array}{ll}
P_{11} = a_s P_{qq}^{(0)} + a_s^2 P_{+}^{(1)} + { \frac{e_u^2+e_d^2}{2}a\tilde{P}_{qq}^{(0)}}, &
P_{33} = a_s P_{gg}^{(0)} + a_s^2 P_{gg}^{(1)},\\
P_{12} = \frac{n_u-n_d}{n_f}a_s^2(P_{qq}^{(1)}-P_{+}^{(1)})
+ { \frac{e_u^2-e_d^2}{2}a\tilde{P}_{qq}^{(0)}}, &
P_{34} = 0,\\
P_{13} = \frac{n_u-n_d}{n_f}(a_sP_{qg}^{(0)} + a_s^2P_{qg}^{(1)}), &
P_{41} = { \frac{e_u^2-e_d^2}{2}a P_{\gamma q}^{(0)}},\\
P_{14} = { \frac{n_ue_u^2-n_de_d^2}{n_f}aP_{q\gamma}^{(0)}}, &
P_{42} = { \frac{e_u^2+e_d^2}{2}a P_{\gamma q}^{(0)}},\\
P_{21} = { \frac{e_u^2-e_d^2}{2}a\tilde{P}_{qq}^{(0)}}, &
P_{43} = 0,\\
P_{22} = a_s P_{qq}^{(0)} + a_s^2 P_{qq}^{(1)}
+ { \frac{e_u^2+e_d^2}{2}a\tilde{P}_{qq}^{(0)}}, &
P_{44} = { a P_{\gamma\gamma}^{(0)}},\\
P_{23} = a_sP_{qg}^{(0)} + a_s^2P_{qg}^{(1)}, &
P_{55} = a_s P_{qq}^{(0)} + a_s^2 P_{-}^{(1)} + { a e_d^2 \tilde{P}_{qq}^{(0)}},\\
P_{24} = { \frac{n_ue_u^2+n_de_d^2}{n_f}aP_{q\gamma}^{(0)}}, &
P_{66} = a_s P_{qq}^{(0)} + a_s^2 P_{-}^{(1)} + { a e_u^2 \tilde{P}_{qq}^{(0)}},\\
P_{31} = 0, &
P_{77} = P_{99} = a_s P_{qq}^{(0)} + a_s^2 P_{+}^{(1)} + { a e_d^2 \tilde{P}_{qq}^{(0)}},\\
P_{32} = a_s P_{gq}^{(0)} +a_s^2 P_{gq}^{(1)}, &
P_{88} = a_s P_{qq}^{(0)} + a_s^2 P_{+}^{(1)} + {a e_u^2 \tilde{P}_{qq}^{(0)}}.
\end{array}
\label{eq7}
\end{equation}

The new beta version of {\tt QCDNUM} program has routines to solve arbitrary number of coupled
evolution equations and is therefore suitable for solving of equations (\ref{eq6}). The flavour
decomposition (\ref{eq5}) and the splitting kernels (\ref{eq7}) were implemented into user module
of this version of {\tt QCDNUM}.
\section{Photon-induced process}
Corrections to the neutral current Drell-Yan cross-section due to photon-induced process
$ \gamma \gamma \to \ell^+\ell^- $ can reach up to $ 10-20 \% $ for high invariant mass
$ M_{\ell^+\ell^-} $ with the appropriate choice of kinematic cuts \cite{Dittmaier:2009cr}.
Therefore it is important to use this contribution for data fits when extracting photon PDF
from LHC data.

3-differential cross-section of the process $ p[\gamma]p[\gamma] \to \ell^+\ell^- $ at LO reads
\begin{equation}
\begin{split}
&\frac{d\sigma(p[\gamma]p[\gamma] \to \ell^+\ell^-)}{dx \, dy \, dz} =\\
&= \frac{4\pi\alpha^2}{s_0}
f_\gamma\left(\frac{M_{min}}{\sqrt{s_0}}e^{x+y},\mu_F^2\right)
f_\gamma\left(\frac{M_{min}}{\sqrt{s_0}}e^{x-y},\mu_F^2\right)\left(1+\tanh^2{z}\right),\\
&x = \ln{\frac{M_{\ell^+\ell^-}}{M_{min}}}, \quad y = Y_{\ell^+\ell^-}, \quad
z = -\ln{\tan{\frac{\theta}{2}}}.
\end{split}
\label{eq8}
\end{equation}

For fast evaluation of this cross-section the {\tt APPLGRID} interface to the {\tt SANC} Monte Carlo
generator was created. The interface is organized as follows:
\begin{itemize}
\item[-] {\tt SANC} generator produces unweighted events which are saved to data file. Each event
$ i $ is assigned a weight
\[
w_i = \frac{\sigma_{tot}}{N} \cdot \frac{1}{f_\gamma(x_{1_i},\mu^2_i)f_\gamma(x_{2_i},\mu^2_i)},
\]
where $ \sigma_{tot} = \int d\sigma $ is the total cross-section and $ N $ is total number of
generated events. {\tt MRST2004QED} PDF set is used to evaluate the photon PDF.

\item[-] These events are used to fill grids for different observables with help of {\tt APPLGRID}
interface.

\item[-] These grids then can be used for fast convolution with arbitrary photon PDF to get the
cross-section.
\end{itemize}
The standard {\tt APPLGRID} package was modified so one can handle one extra PDF (photon) in
addition to 13 standard PDFs (gluon and quarks).
\section{Validation and performance}
The described implementation of QED-modified DGLAP evolution equations (which we denote as
{\tt QCDNUM+QED}) was cross-checked  with {\tt partonevolution} program in fixed flavour number
scheme (FFNS) and with {\tt MRST2004QED} PDF set and {\tt APFEL} program in variable flavour
number scheme (VFNS). It takes $ \sim 5 $ s of CPU time (Intel Core i7-3630QM, 2.4 GHz) to fill
weight tables of splitting functions and $ \sim 0.5 $ s to evolve the distributions (\ref{eq5}) on
$ 100 \times 100 $ grid in $ x $ and $ \mu^2 $.

\subsection{Comparison of {\tt QCDNUM+QED} and {\tt partonevolution}}
For the numerical comparison of {\tt QCDNUM+QED} and {\tt partonevolution} ({\tt PE}) codes
the toy model with $ n_f  = 4 $ is adopted. PDFs at initial scale $ \mu_0 = 2 $ GeV are given by:
\begin{equation}
\begin{split}
& xu_v = \frac{35}{16}x^{0.5} (1-x)^3,\\
& xd_v = \frac{315}{256} x^{0.5} (1-x)^4,\\
& x\bar{d} = x\bar{u} = x\bar{s} = xs = \frac{0.673345}{6} x^{-0.2} (1-x)^{7},\\
& x\bar{c} = xc = 0,\\
& xg = 1.90836x^{-0.2}(1-x)^{5},\\
& x\gamma = 0.
\end{split}
\label{eq9}
\end{equation}

The running strong $ a_s = \alpha_s/2\pi $ and electromagnetic $ a = \alpha/2\pi $ couplings
determined by the following expressions:
\begin{equation}
\begin{split}
& a_s(\mu^2) = \frac{1}{\beta_0L}\left(1-\frac{\beta_1}{\beta_0^2}\frac{\ln{L}}{L}\right),\\
\\
& a(\mu^2) = \frac{a(m_\tau^2)}{1-\frac{38}{9}a(m_\tau^2)\ln{\frac{\mu^2}{m_\tau^2}}},
\end{split}
\label{eq10}
\end{equation}
where
\begin{equation}
\begin{split}
&L = \ln{\frac{\mu^2}{\Lambda_{QCD}^2}},
\quad \Lambda_{QCD} = 0.25 \text{ GeV},
\quad \beta_0 = \frac{25}{6},
\quad \beta_1 = \frac{77}{6},\\
&a(m_\tau^2) = \frac{1}{2\pi}\frac{1}{133.4},
\quad m_\tau = 1.777 \text{ GeV}.
\end{split}
\label{eq11}
\end{equation}

\begin{figure}
\begin{center}
\includegraphics[width = 0.45\textwidth]{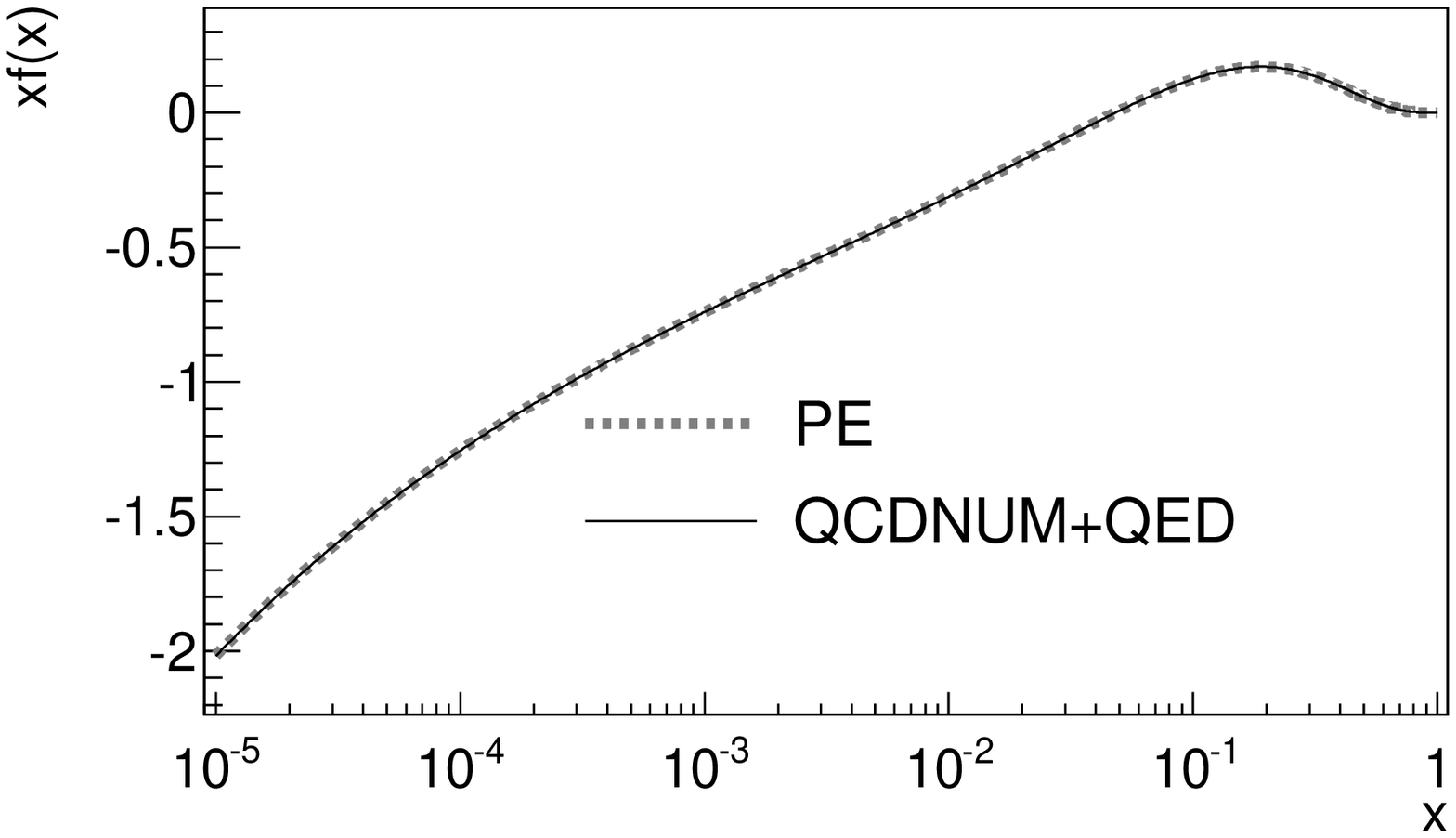}
\includegraphics[width = 0.45\textwidth]{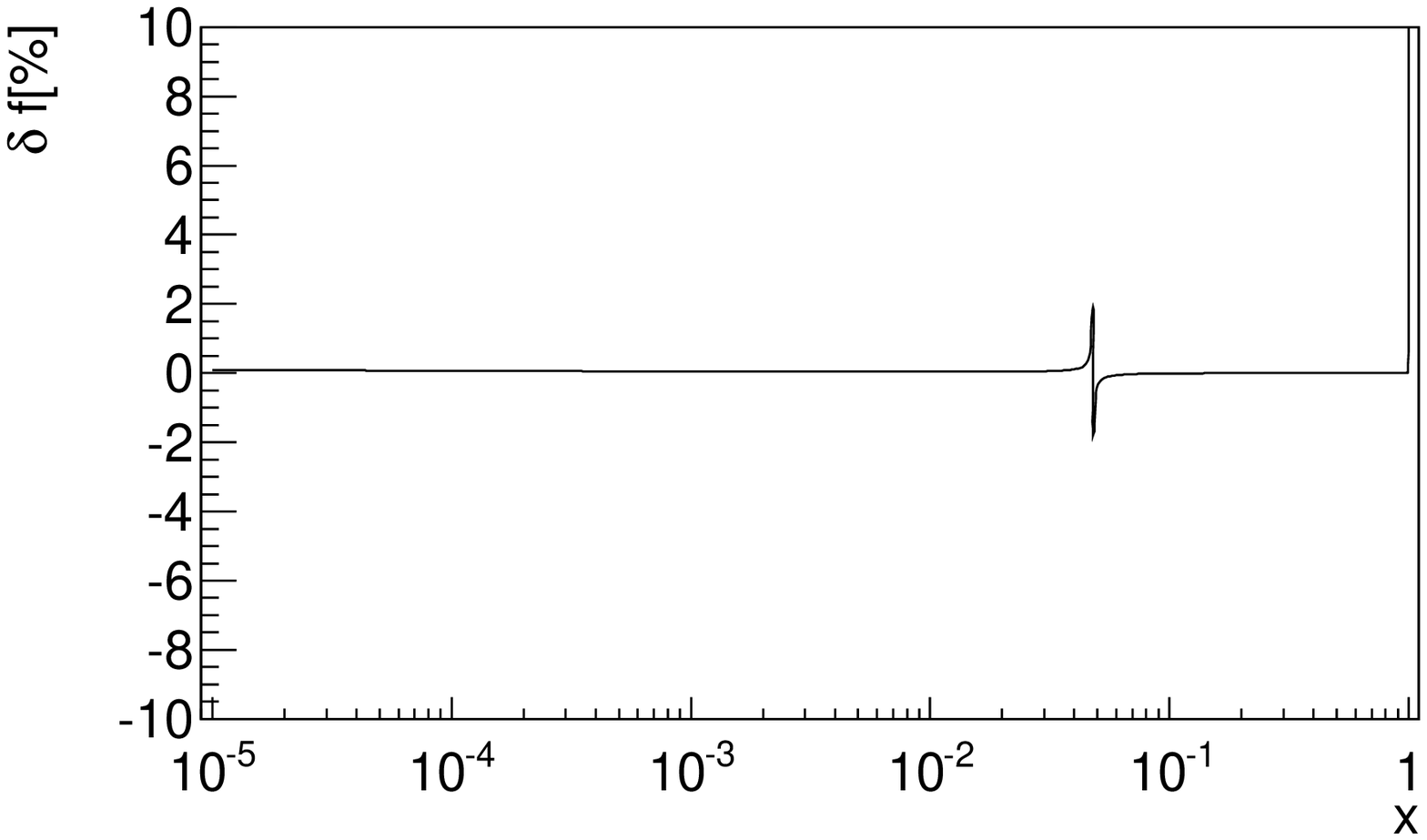}
\end{center}
\caption {Tuned comparison between {\tt QCDNUM+QED} and {\tt partonevolution}.
Left plot shows the momentum distribution of $ \Delta $ at $ \mu^2 = 10^4 \text{ GeV}^2 $.
The corresponding $ \delta f $ is shown on the right plot.}
\label{fig1}
\end{figure}

\begin{figure}
\begin{center}
\includegraphics[width = 0.45\textwidth]{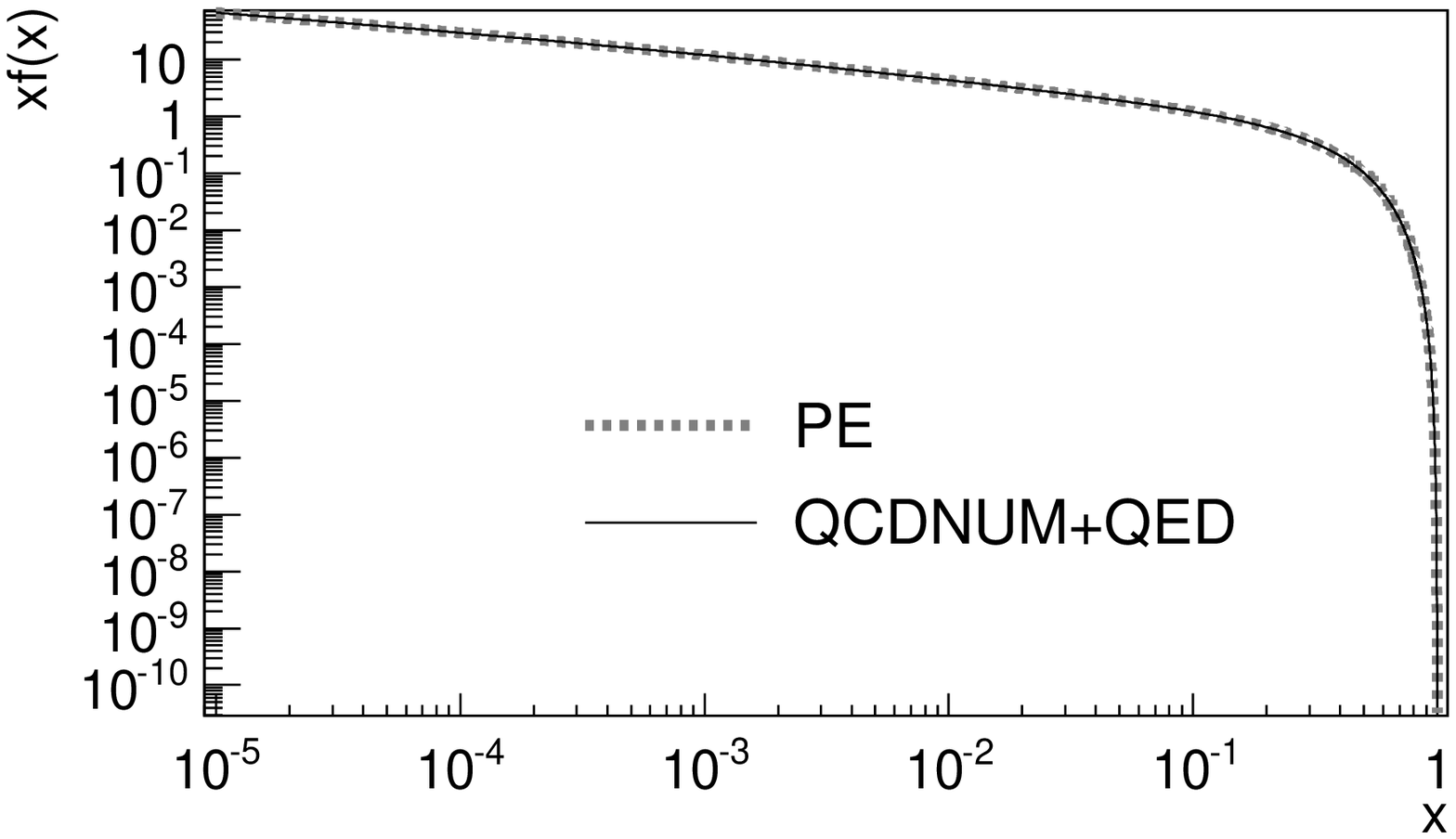}
\includegraphics[width = 0.45\textwidth]{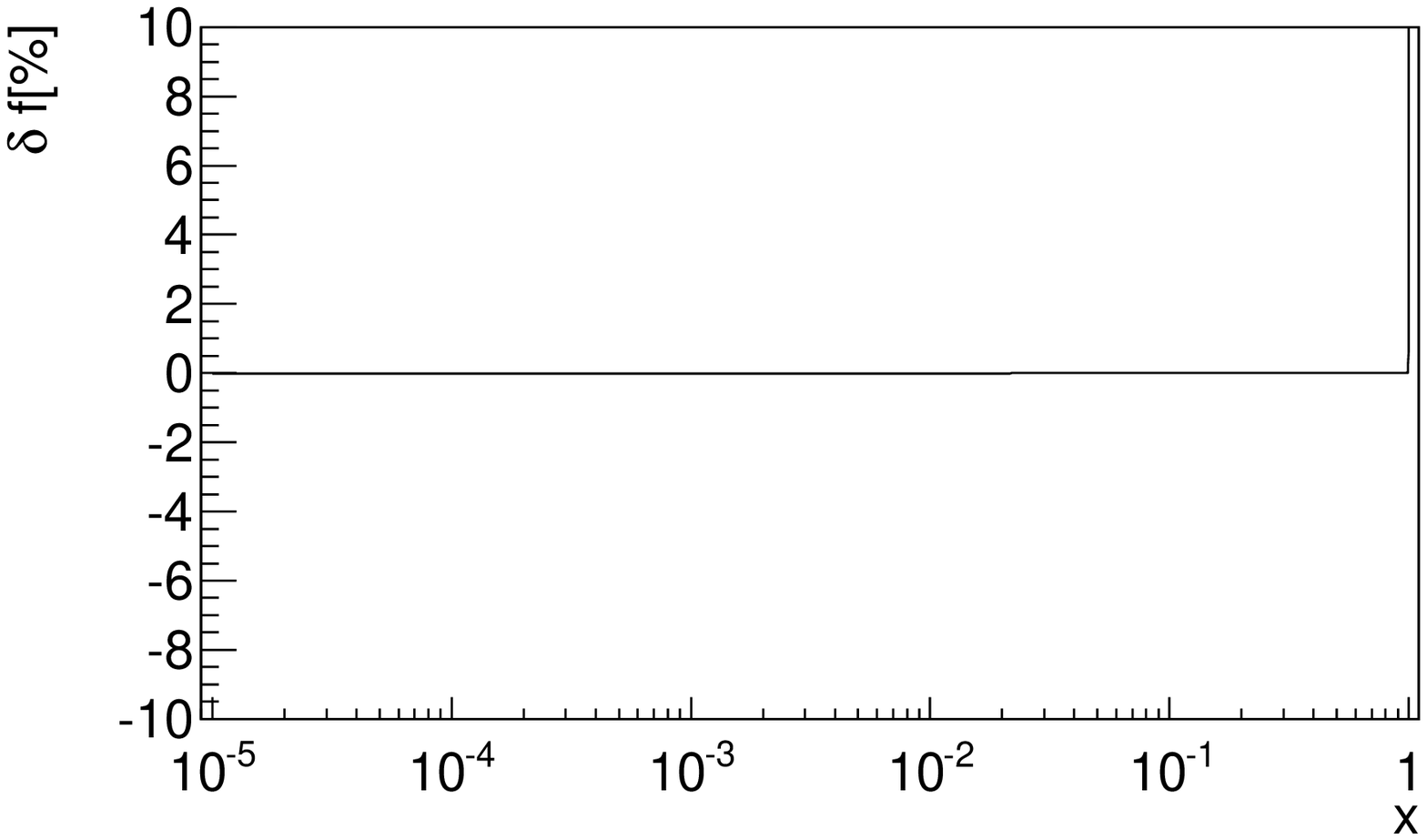}
\end{center}
\caption {Tuned comparison between {\tt QCDNUM+QED} and {\tt partonevolution}.
Left plot shows the momentum distribution of $ \Sigma $ at $ \mu^2 = 10^4 \text{ GeV}^2 $.
The corresponding $ \delta f $ is shown on the right plot.}
\label{fig2}
\end{figure}

\begin{figure}
\begin{center}
\includegraphics[width = 0.45\textwidth]{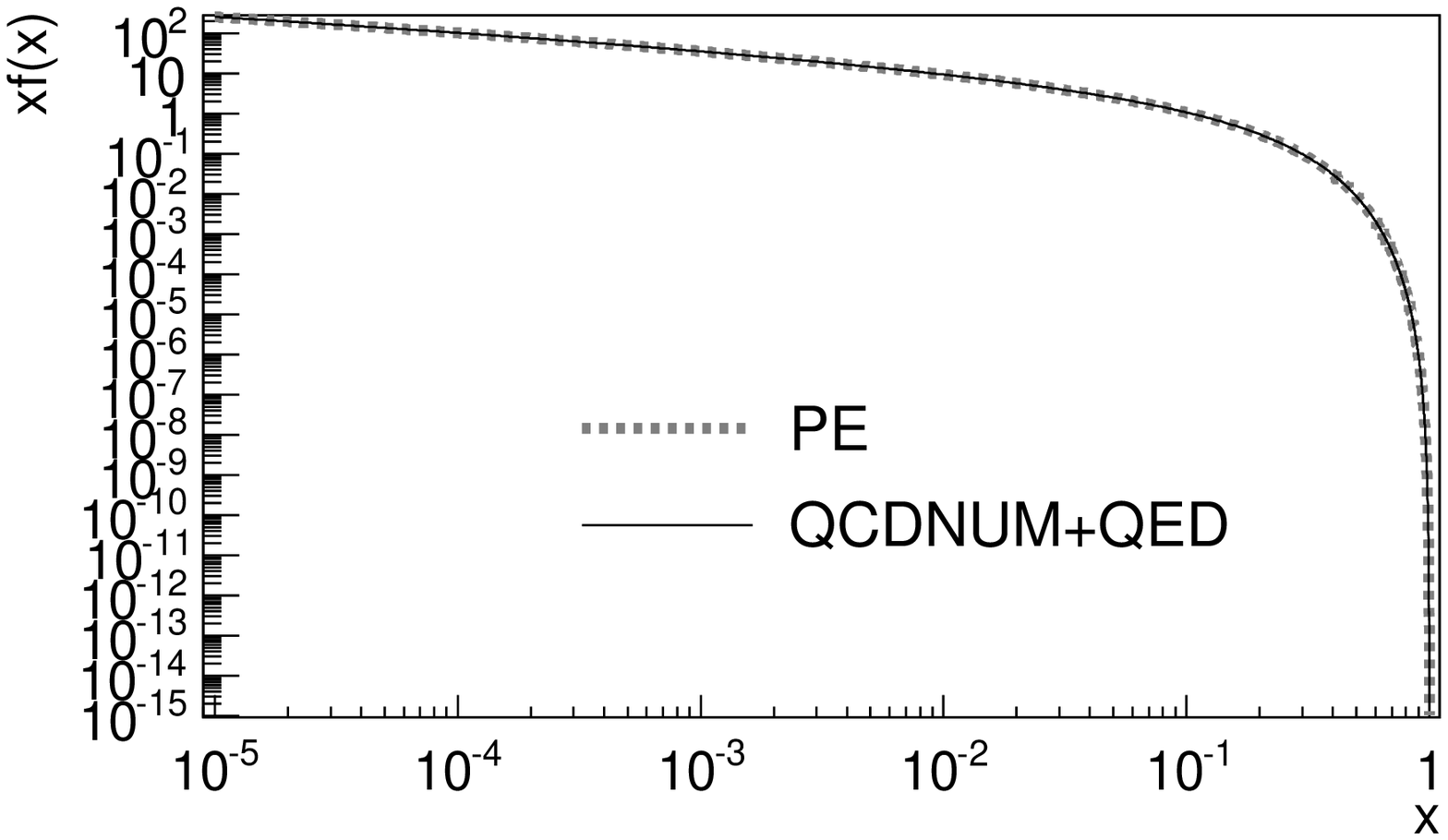}
\includegraphics[width = 0.45\textwidth]{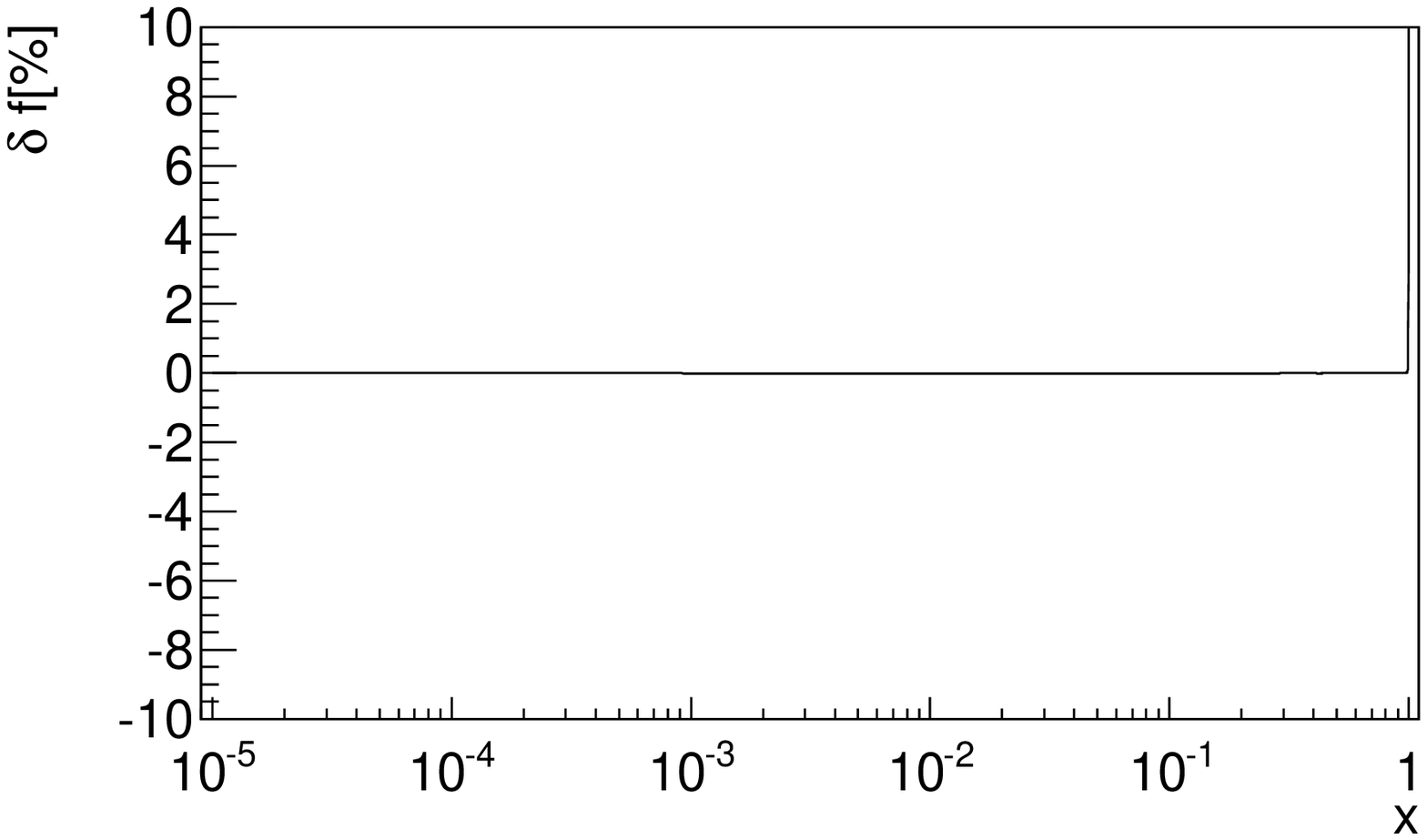}
\end{center}
\caption {Tuned comparison between {\tt QCDNUM+QED} and {\tt partonevolution}.
Left plot shows the momentum distribution of $ g $ at $ \mu^2 = 10^4 \text{ GeV}^2 $.
The corresponding $ \delta f $ is shown on the right plot.}
\label{fig3}
\end{figure}

\begin{figure}
\begin{center}
\includegraphics[width = 0.45\textwidth]{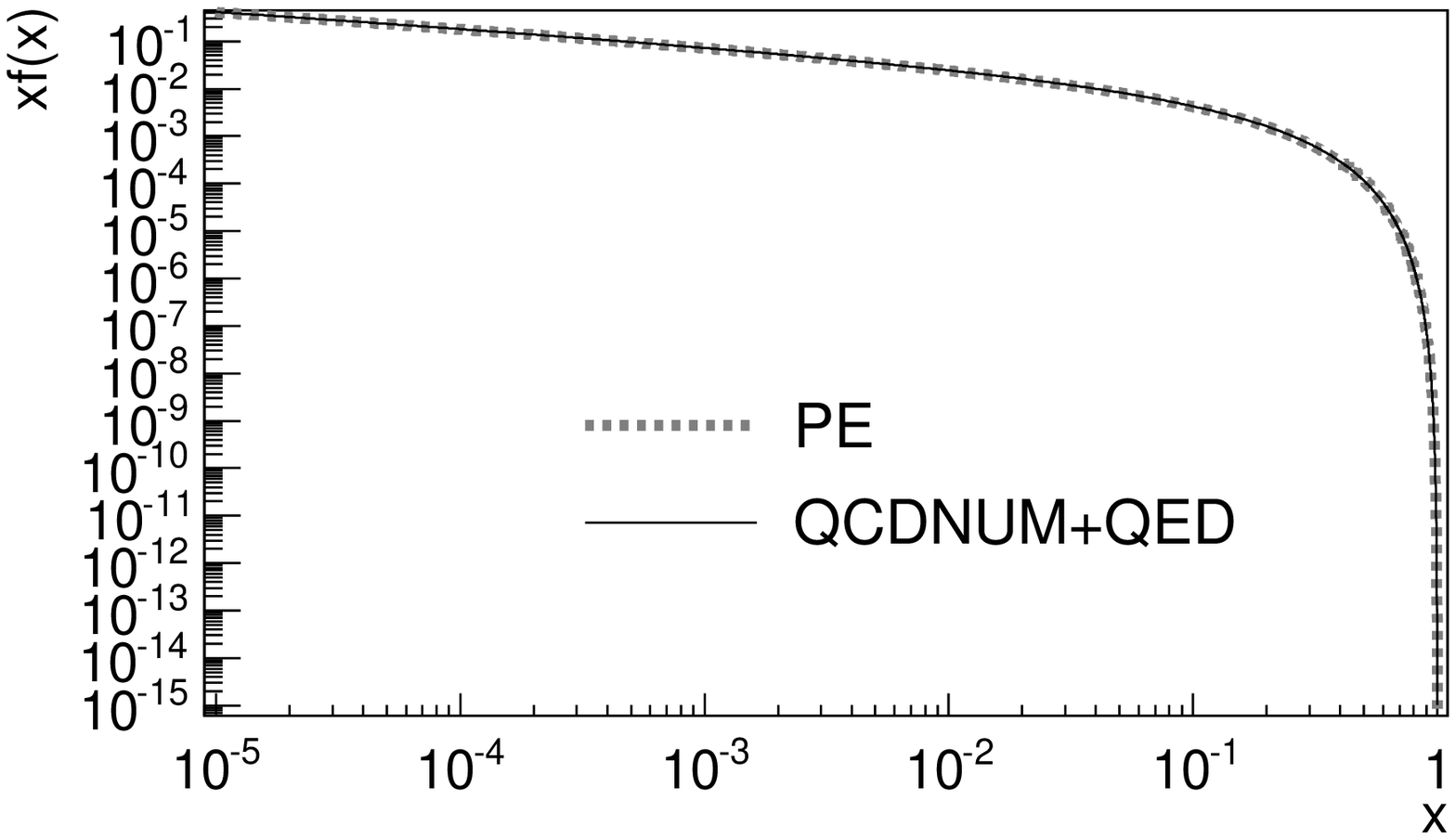}
\includegraphics[width = 0.45\textwidth]{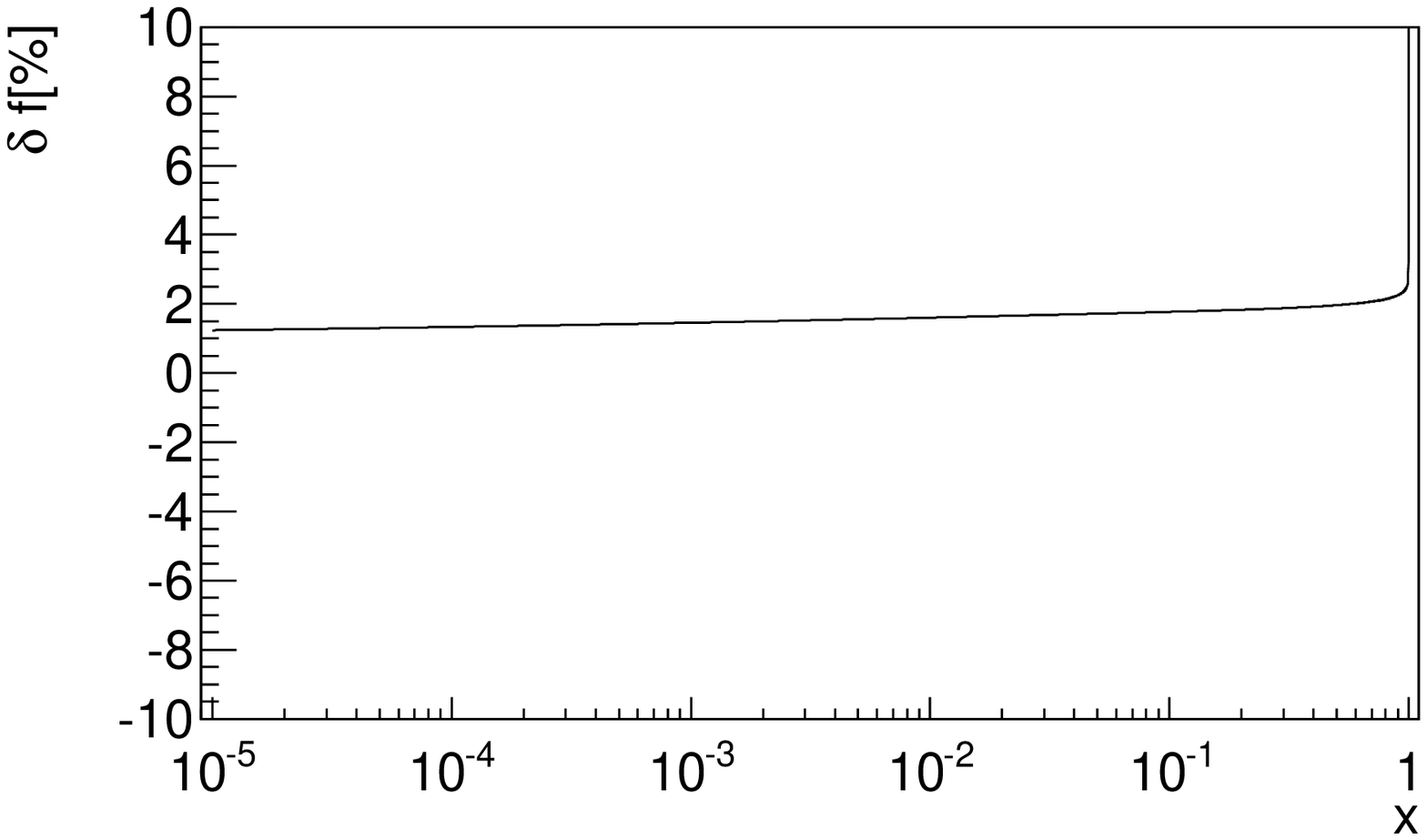}
\end{center}
\caption {Tuned comparison between {\tt QCDNUM+QED} and {\tt partonevolution}.
Left plot shows the momentum distribution of $ \gamma $ at $ \mu^2 = 10^4 \text{ GeV}^2 $.
The corresponding $ \delta f $ is shown on the right plot.}
\label{fig4}
\end{figure}

\begin{figure}
\begin{center}
\includegraphics[width = 0.45\textwidth]{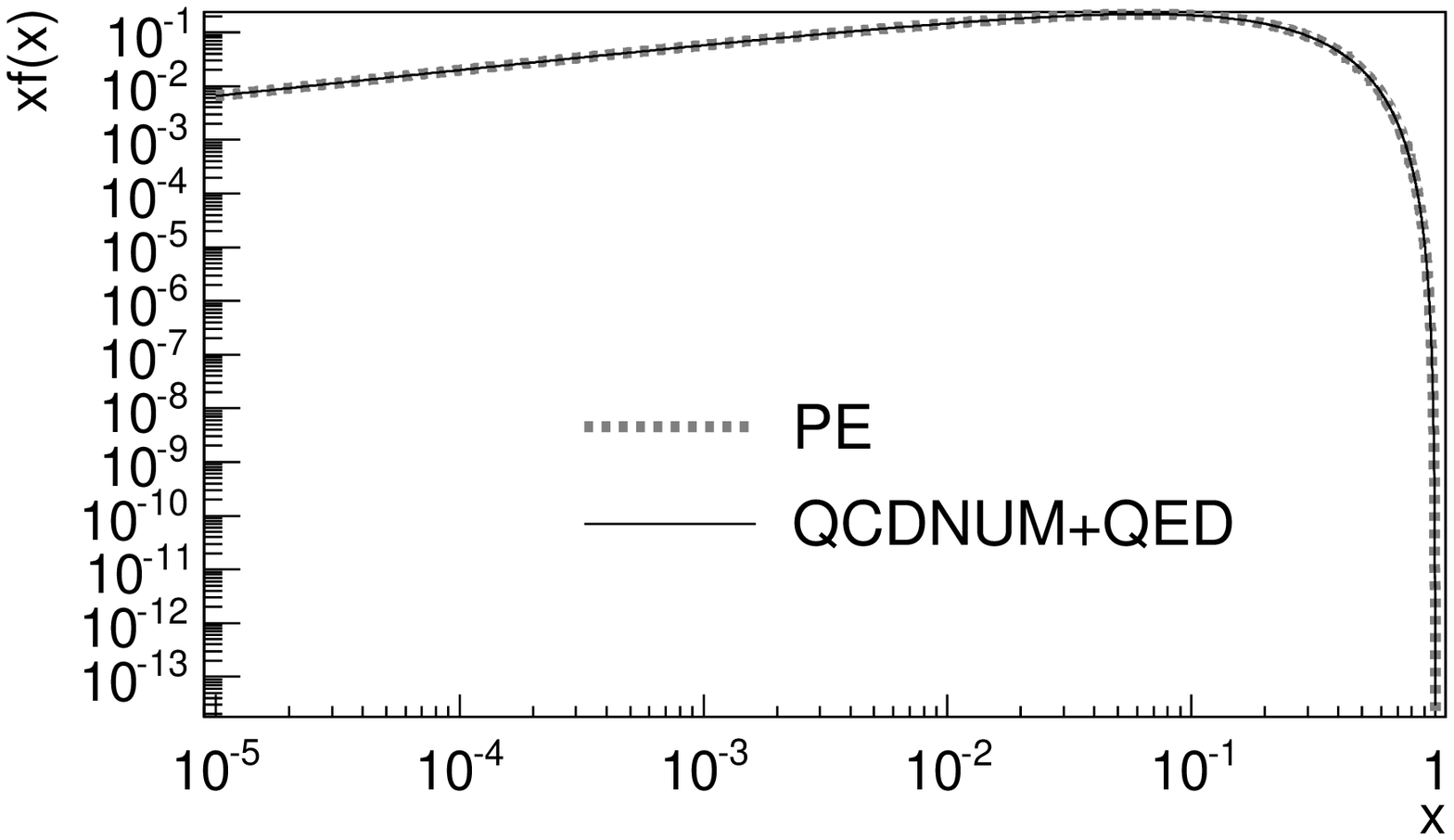}
\includegraphics[width = 0.45\textwidth]{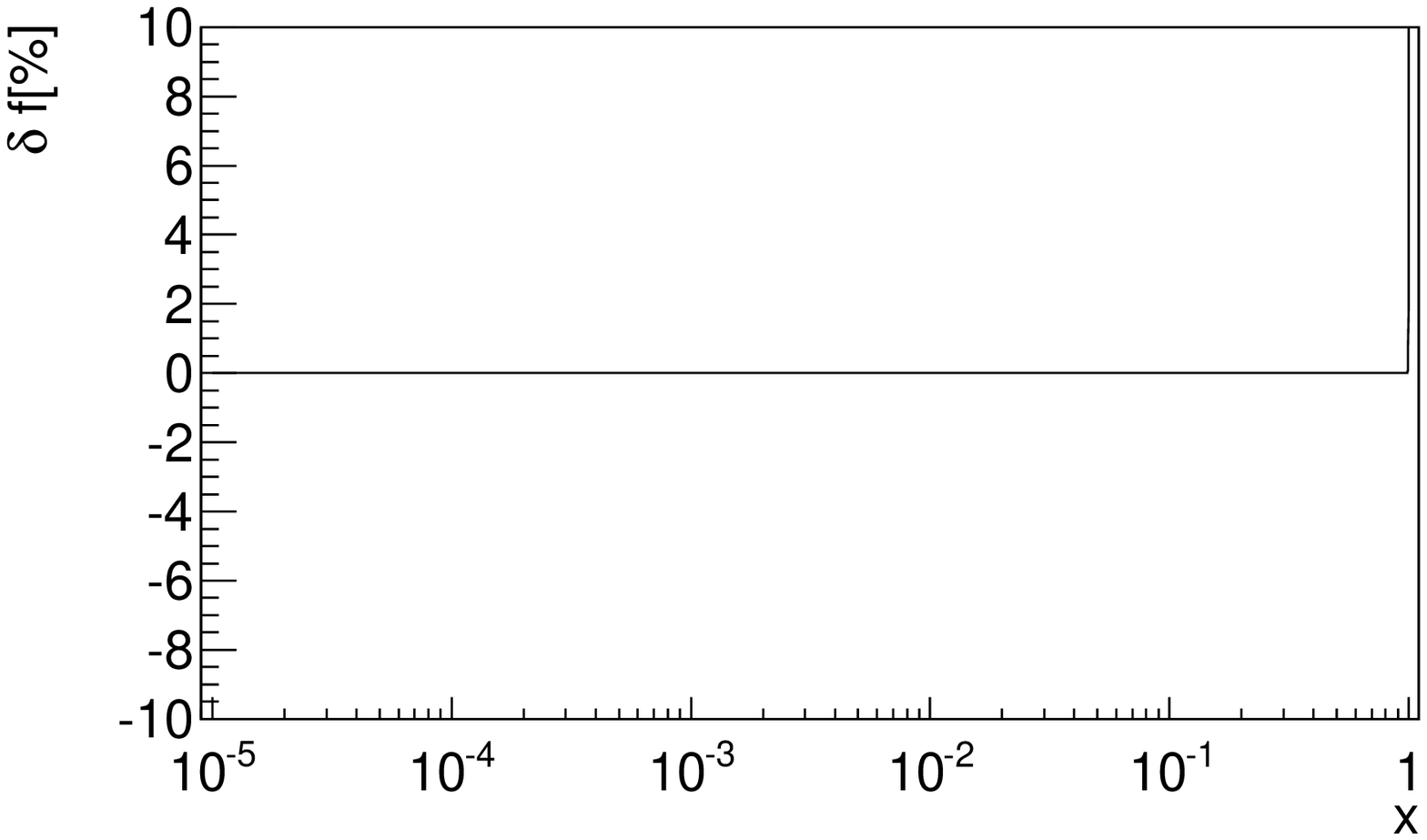}
\end{center}
\caption {Tuned comparison between {\tt QCDNUM+QED} and {\tt partonevolution}.
Left plot shows the momentum distribution of $ d_v $ at $ \mu^2 = 10^4 \text{ GeV}^2 $.
The corresponding $ \delta f $ is shown on the right plot.}
\label{fig5}
\end{figure}

\begin{figure}
\begin{center}
\includegraphics[width = 0.45\textwidth]{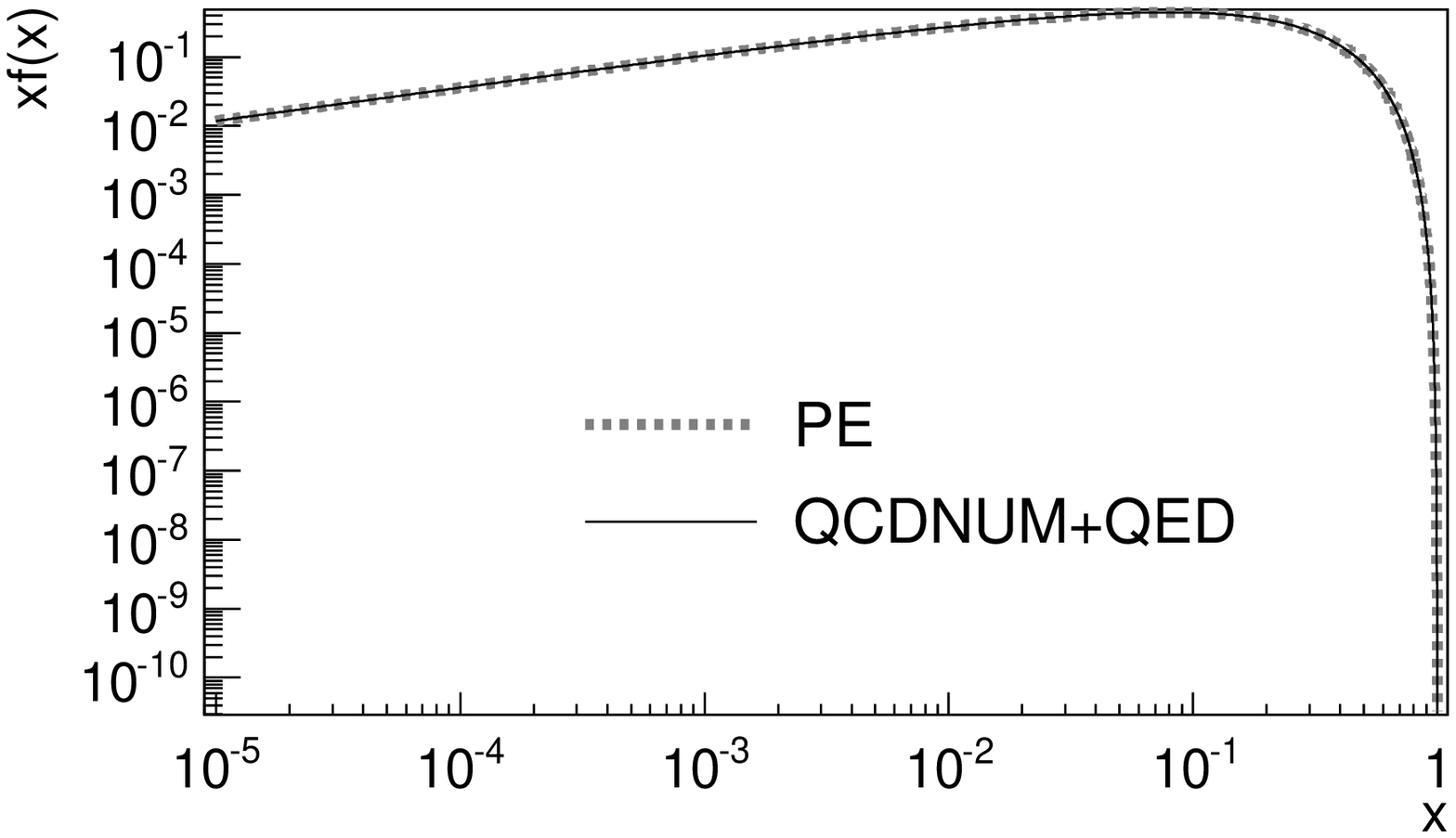}
\includegraphics[width = 0.45\textwidth]{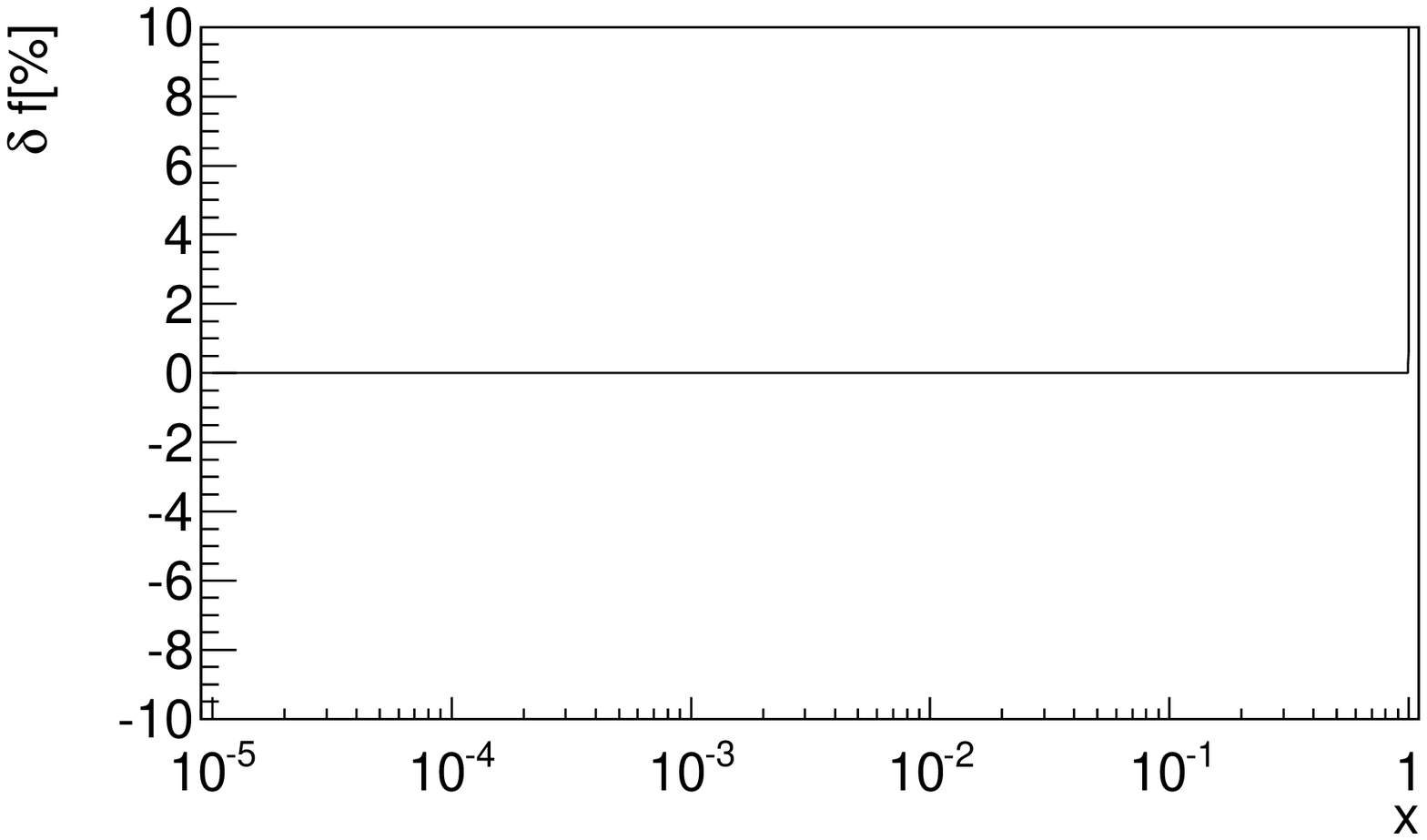}
\end{center}
\caption {Tuned comparison between {\tt QCDNUM+QED} and {\tt partonevolution}.
Left plot shows the momentum distribution of $ u_v $ at $ \mu^2 = 10^4 \text{ GeV}^2 $.
The corresponding $ \delta f $ is shown on the right plot.}
\label{fig6}
\end{figure}

\begin{figure}
\begin{center}
\includegraphics[width = 0.45\textwidth]{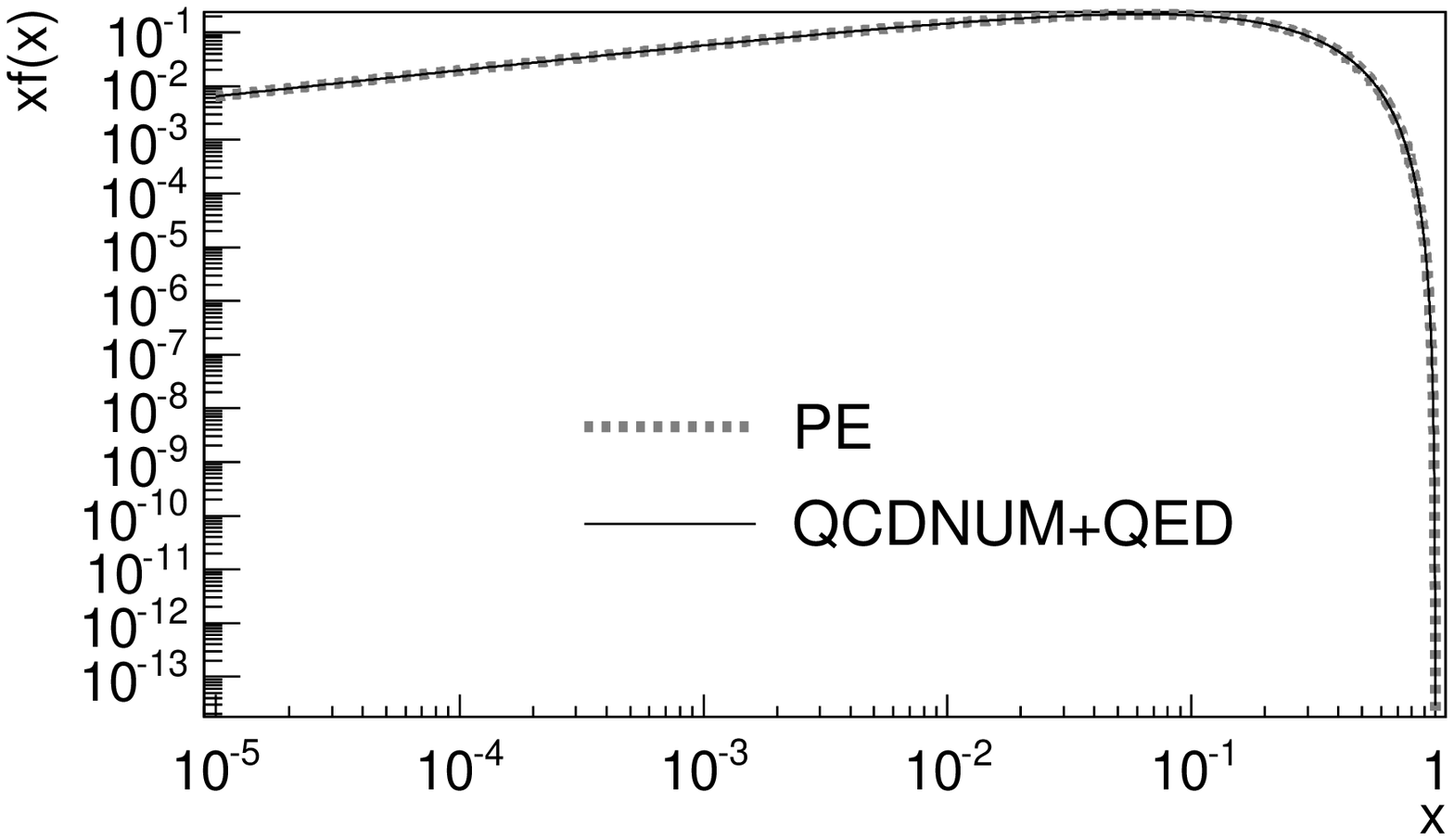}
\includegraphics[width = 0.45\textwidth]{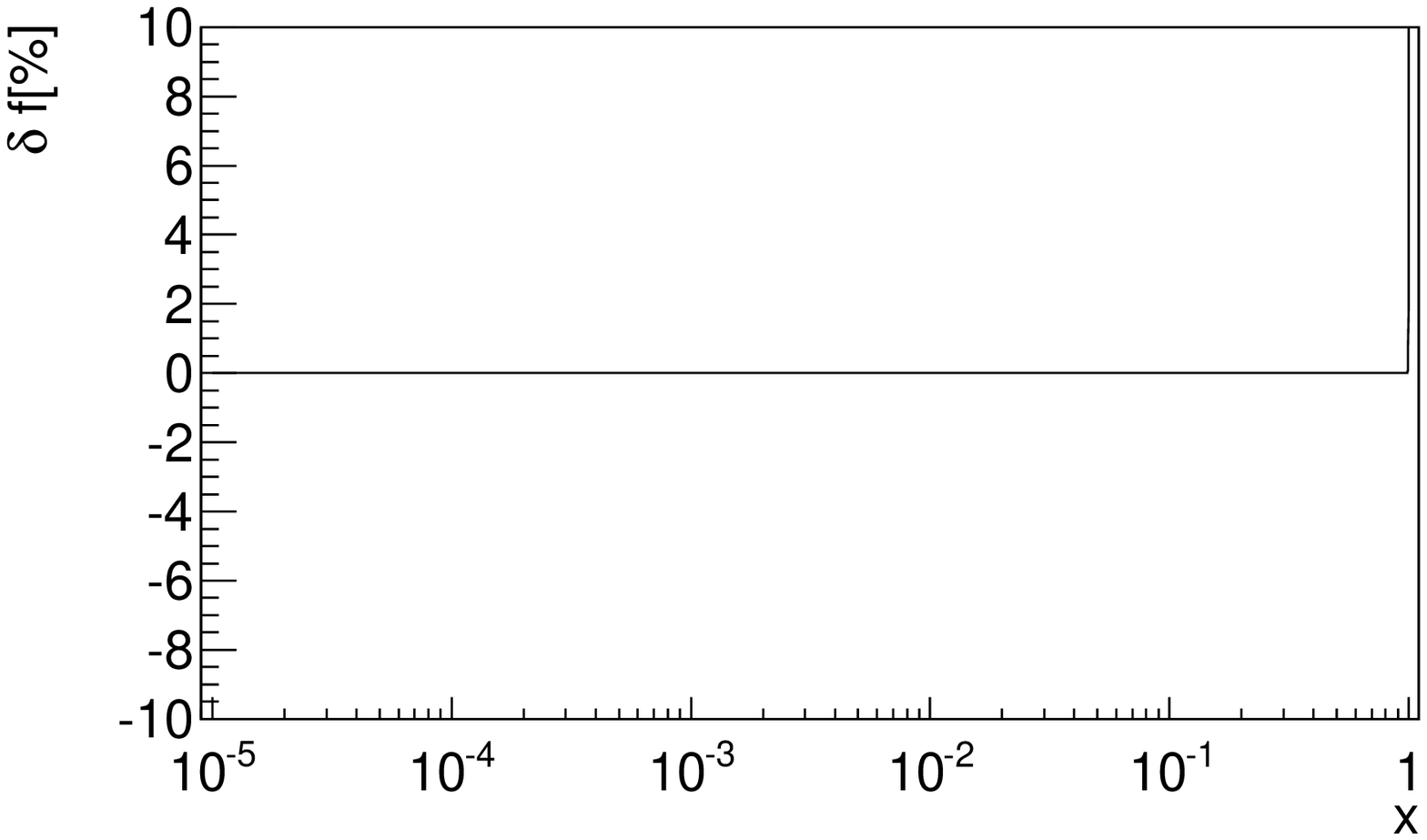}
\end{center}
\caption {Tuned comparison between {\tt QCDNUM+QED} and {\tt partonevolution}.
Left plot shows the momentum distribution of $ \Delta_{ds} $ at $ \mu^2 = 10^4 \text{ GeV}^2 $.
The corresponding $ \delta f $ is shown on the right plot.}
\label{fig7}
\end{figure}

\begin{figure}
\begin{center}
\includegraphics[width = 0.45\textwidth]{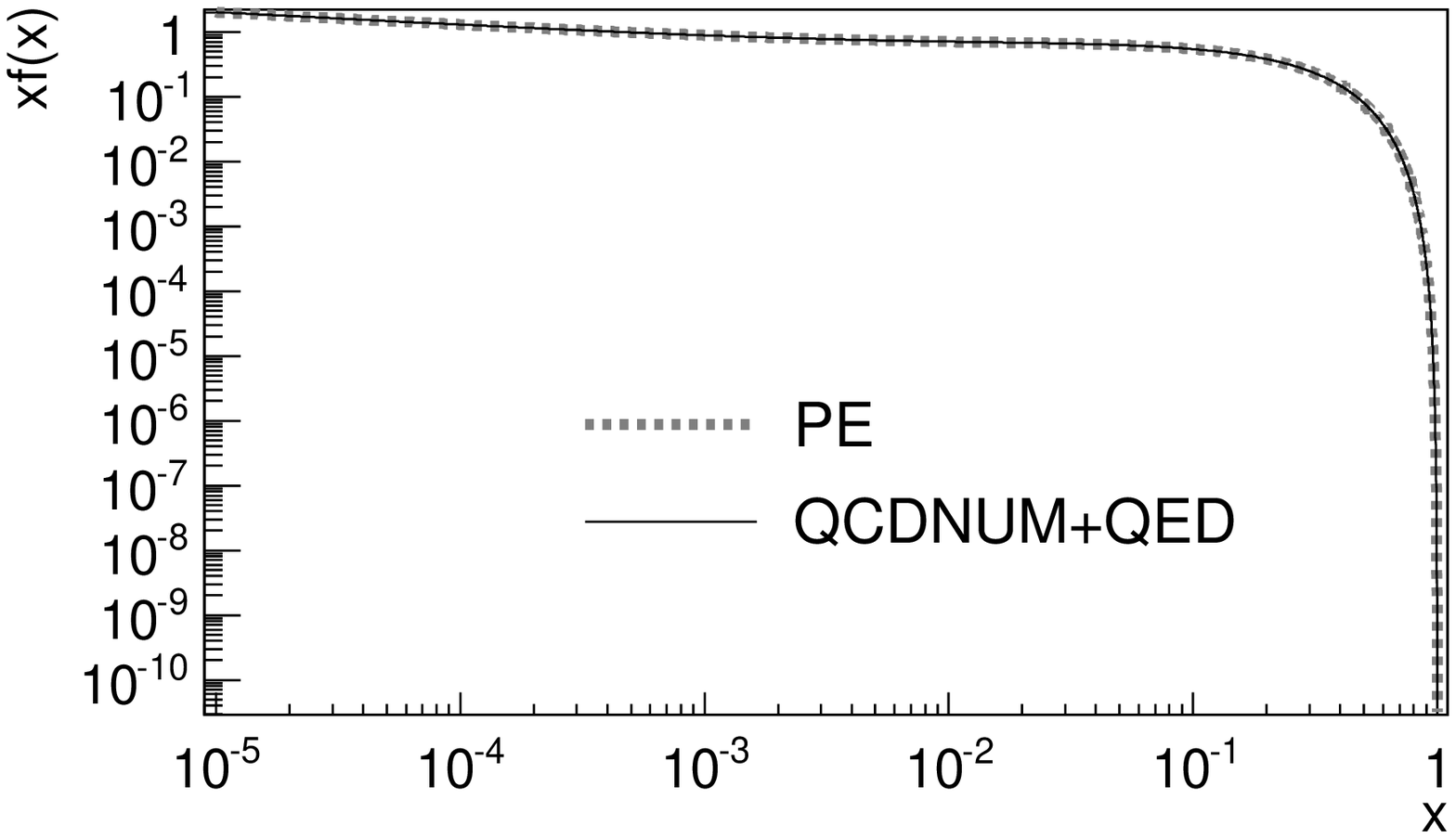}
\includegraphics[width = 0.45\textwidth]{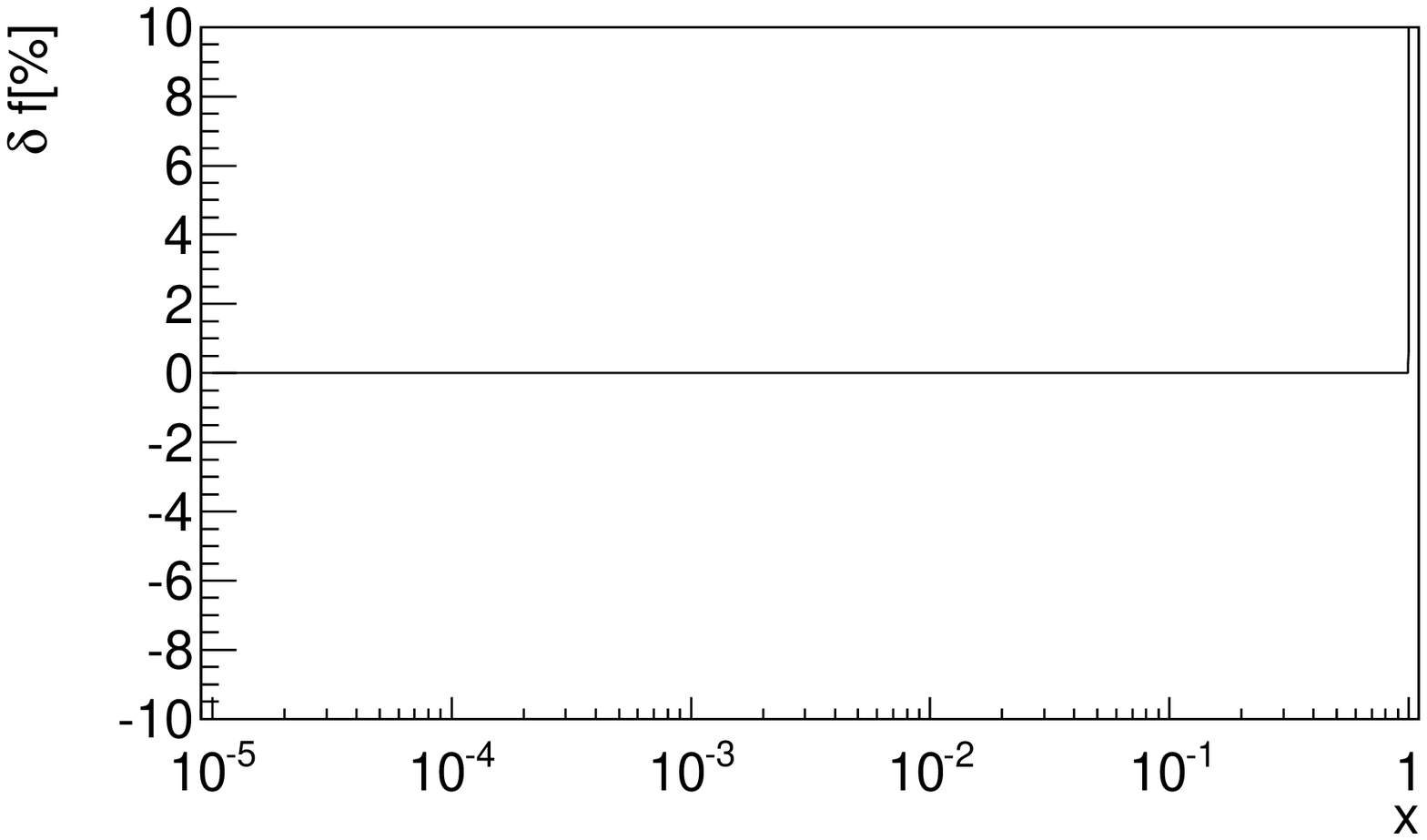}
\end{center}
\caption {Tuned comparison between {\tt QCDNUM+QED} and {\tt partonevolution}.
Left plot shows the momentum distribution of $ \Delta_{uc} $ at $ \mu^2 = 10^4 \text{ GeV}^2 $.
The corresponding $ \delta f $ is shown on the right plot.}
\label{fig8}
\end{figure}

The results of comparison in FFNS for basis distributions (\ref{eq5}) are presented in
Figs. \ref{fig1}-\ref{fig8} for momentum distributions $ xf $ and the corresponding relative
differences which are defined by
\begin{equation}
\delta f =
\frac{xf(\text{\tt QCDNUM+QED}) - xf(\text{\tt PE})}{xf(\text{\tt PE})}.
\label{eq12}
\end{equation}
A very good agreement is observed between two codes. For photon distribution there is about
$ 1-2 \% $ difference which can be explained by the effect of contribution from lepton PDFs that
are taken into account in {\tt PE} code. The discontinuity around $ x = 0.033 $ in the $ \delta f $
plot for $ \Delta $ distribution is due to the $ \Delta $ distribution changes sign from - to + near
this point.
\subsection{Comparison of {\tt QCDNUM+QED} and {\tt MRST2004QED}}
\begin{figure}
\begin{center}
\includegraphics[width = 0.45\textwidth]{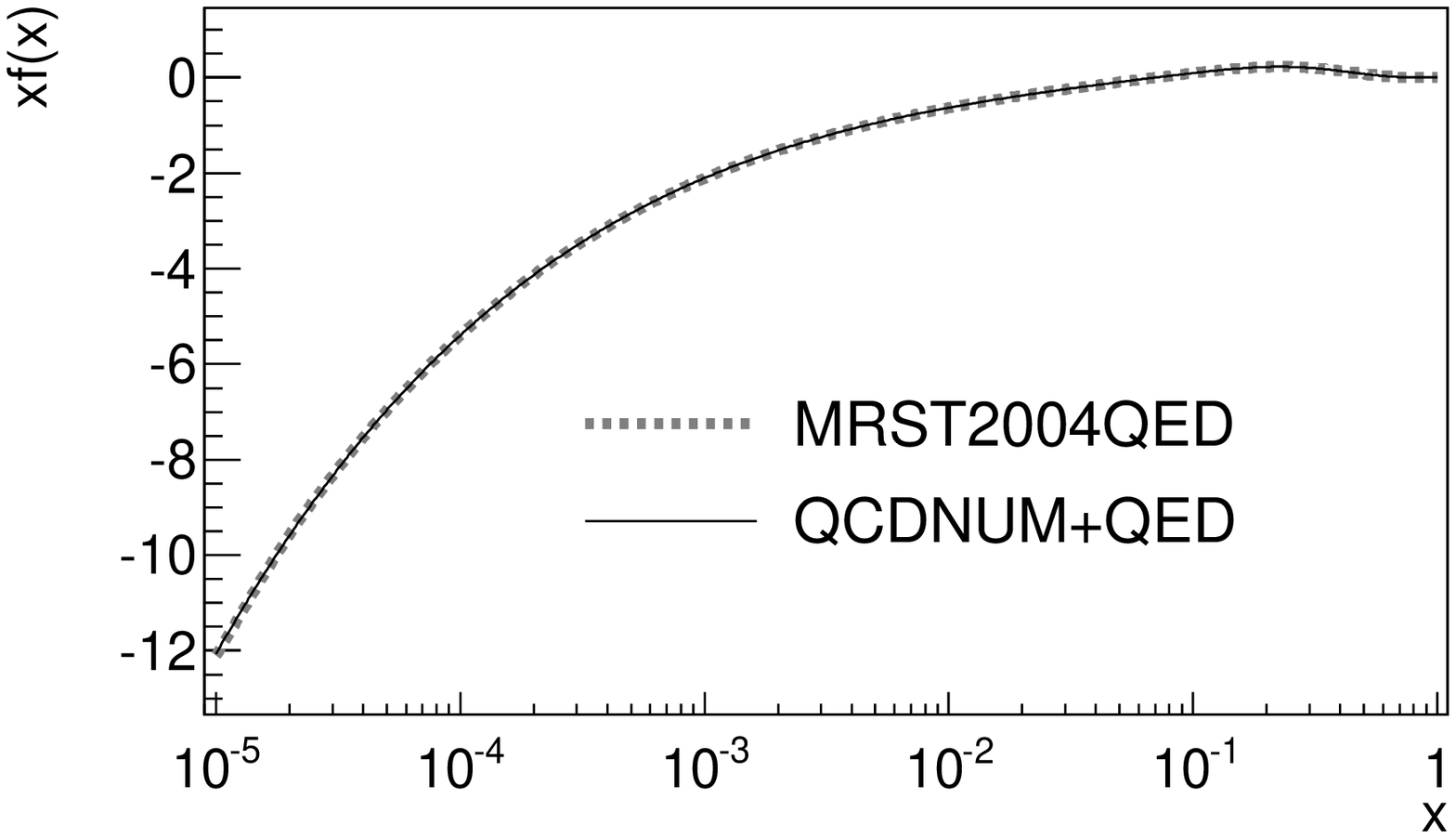}
\includegraphics[width = 0.45\textwidth]{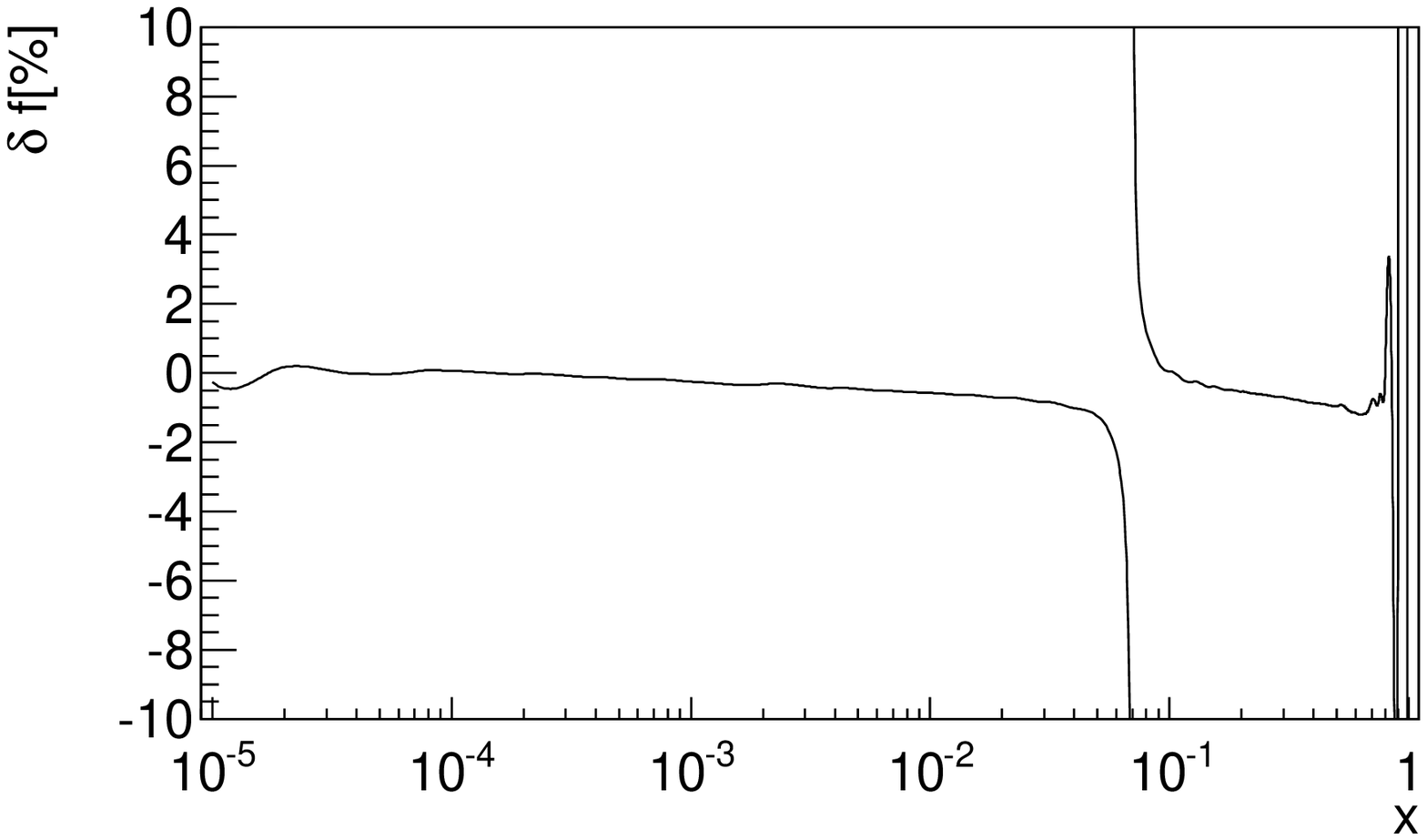}
\end{center}
\caption {Tuned comparison between {\tt QCDNUM+QED} and {\tt MRST2004QED}.
Left plot shows the momentum distribution of $ \Delta $  at $ \mu^2 = 10^4 \text{ GeV}^2 $.
The corresponding $ \delta f $ is shown on the right plot.}
\label{fig9}
\end{figure}

\begin{figure}
\begin{center}
\includegraphics[width = 0.45\textwidth]{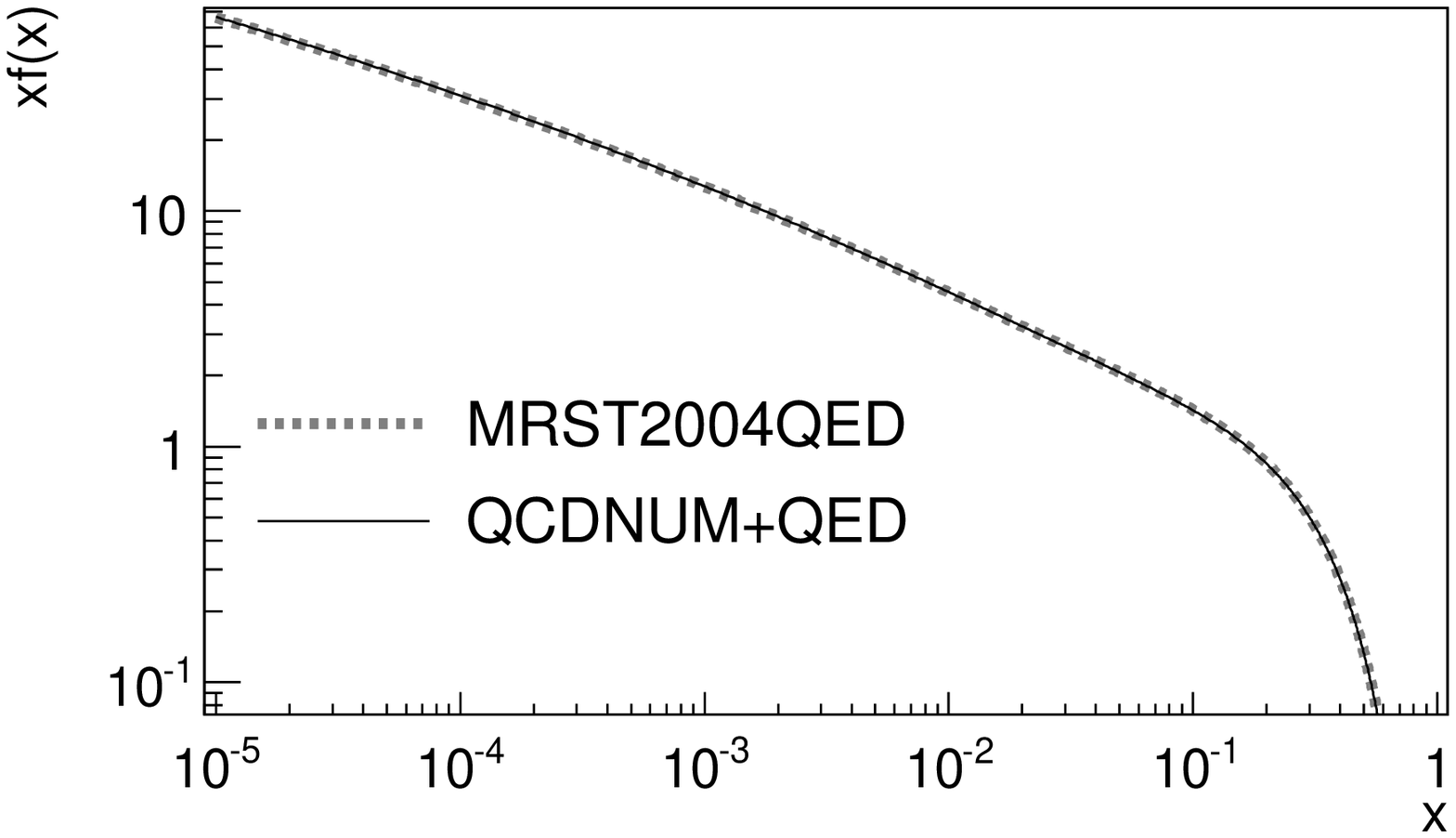}
\includegraphics[width = 0.45\textwidth]{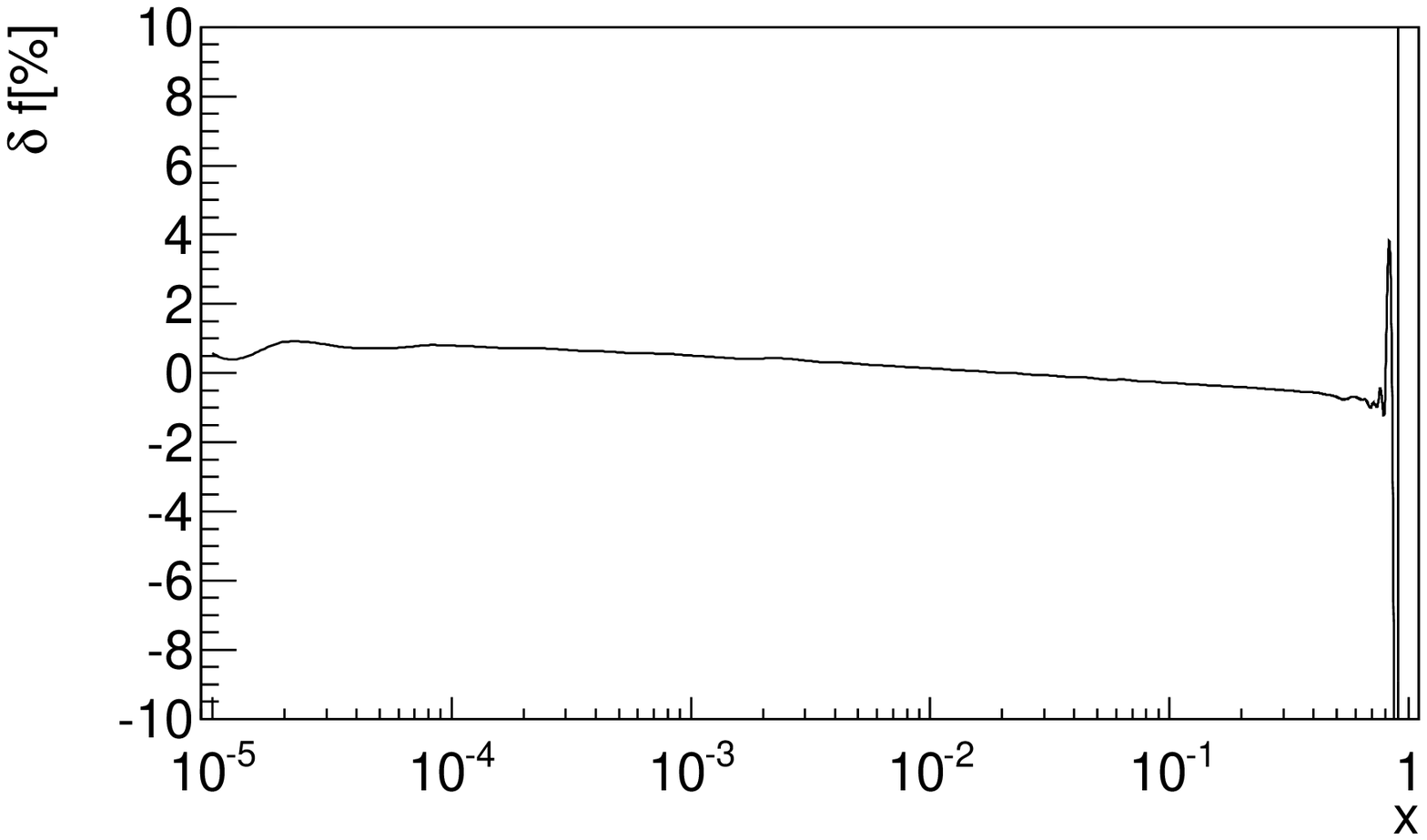}
\end{center}
\caption {Tuned comparison between {\tt QCDNUM+QED} and {\tt MRST2004QED}.
Left plot shows the momentum distribution of $ \Sigma $ at $ \mu^2 = 10^4 \text{ GeV}^2 $.
The corresponding $ \delta f $ is shown on the right plot.}
\label{fig10}
\end{figure}

\begin{figure}
\begin{center}
\includegraphics[width = 0.45\textwidth]{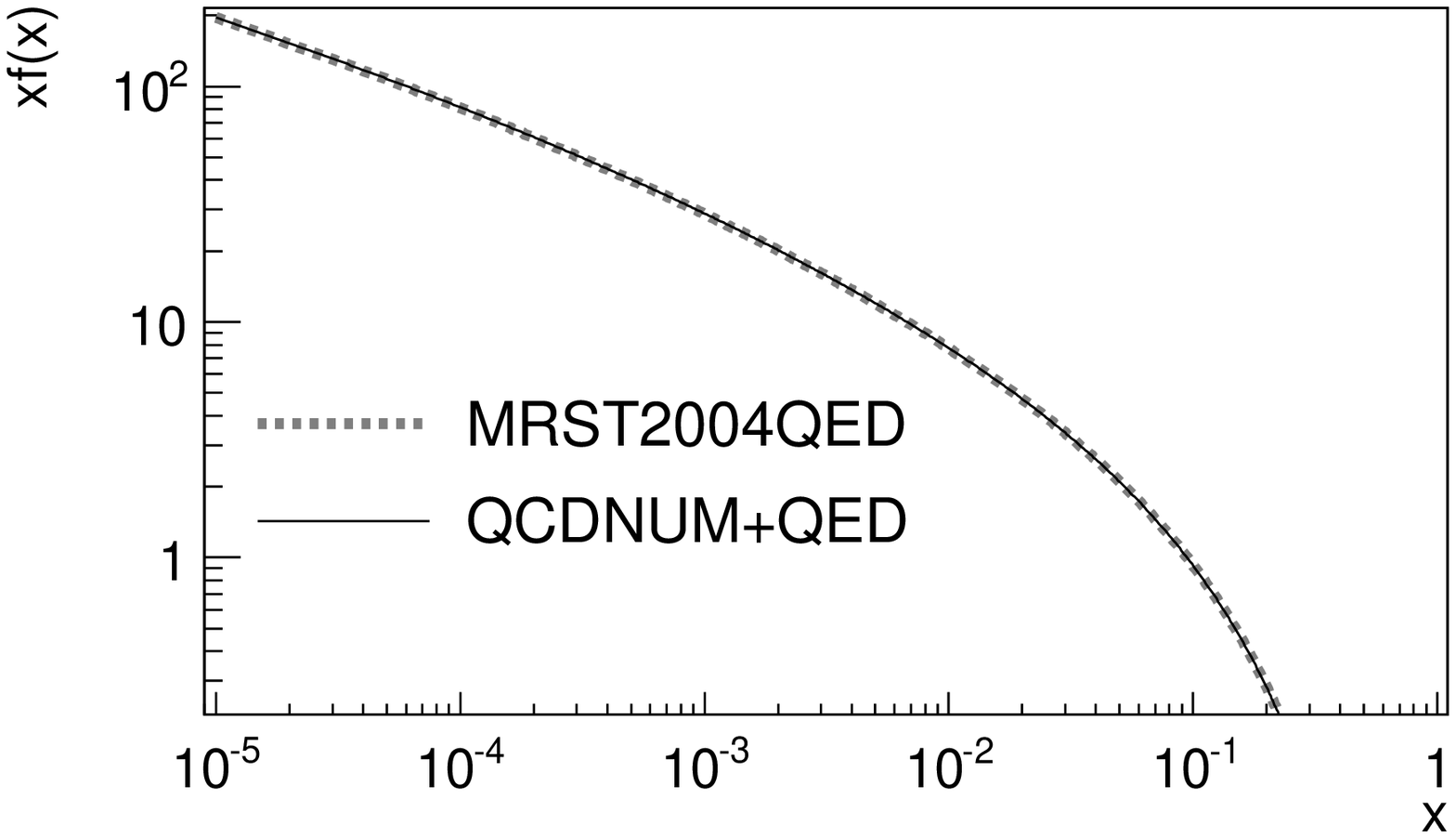}
\includegraphics[width = 0.45\textwidth]{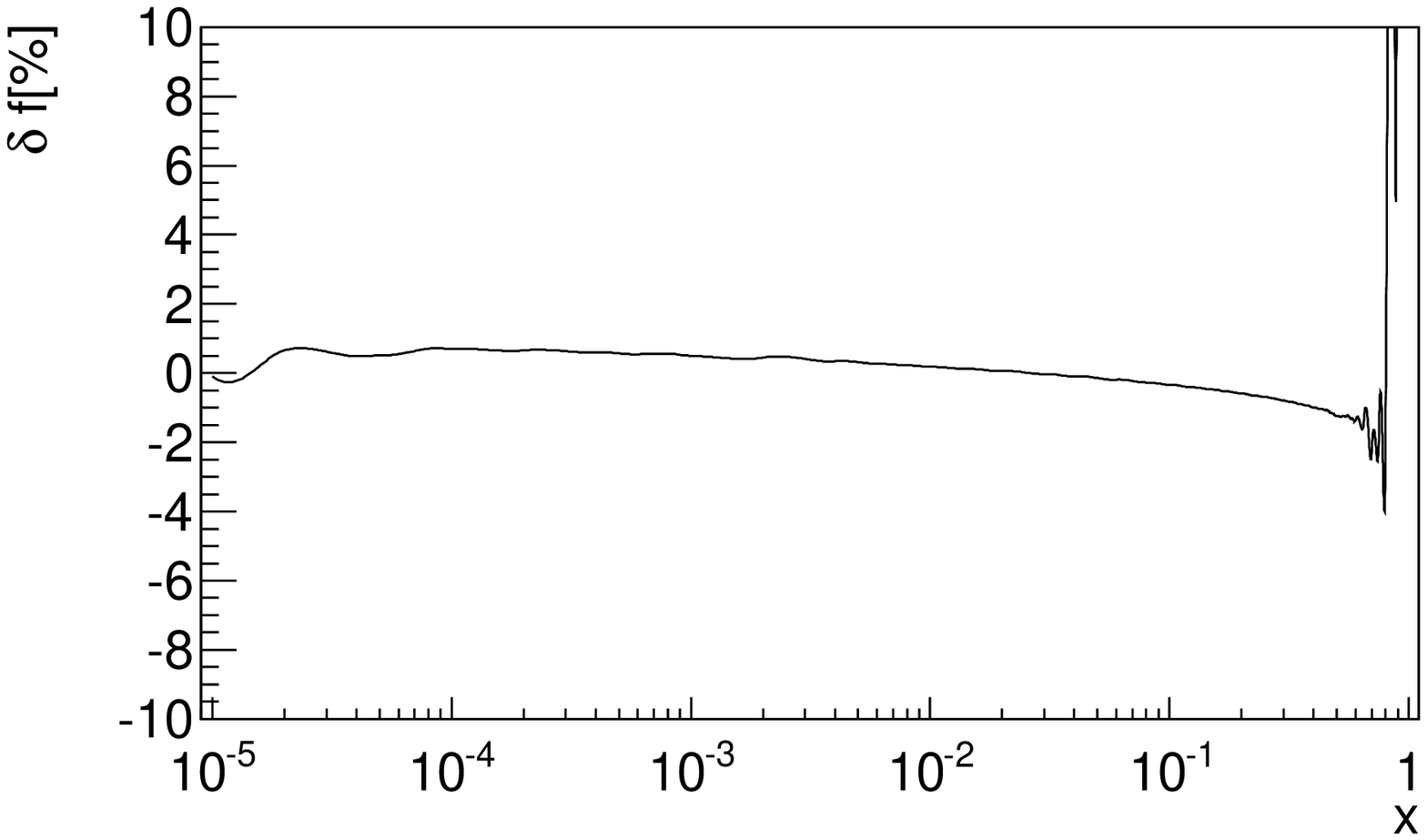}
\end{center}
\caption {Tuned comparison between {\tt QCDNUM+QED} and {\tt MRST2004QED}.
Left plot shows the momentum distribution of $ g $ at $ \mu^2 = 10^4 \text{ GeV}^2 $.
The corresponding $ \delta f $ is shown on the right plot.}
\label{fig11}
\end{figure}

\begin{figure}
\begin{center}
\includegraphics[width = 0.45\textwidth]{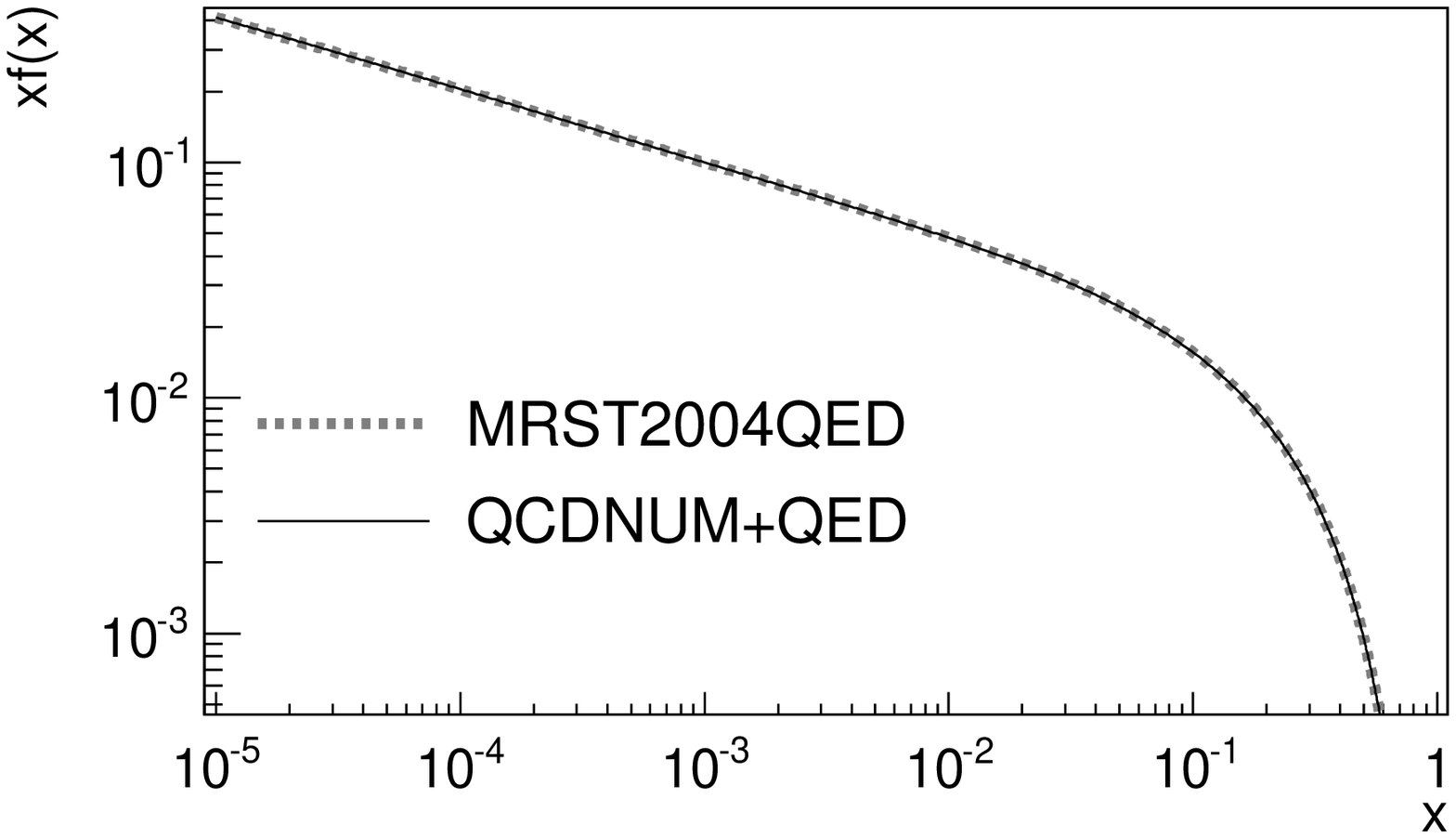}
\includegraphics[width = 0.45\textwidth]{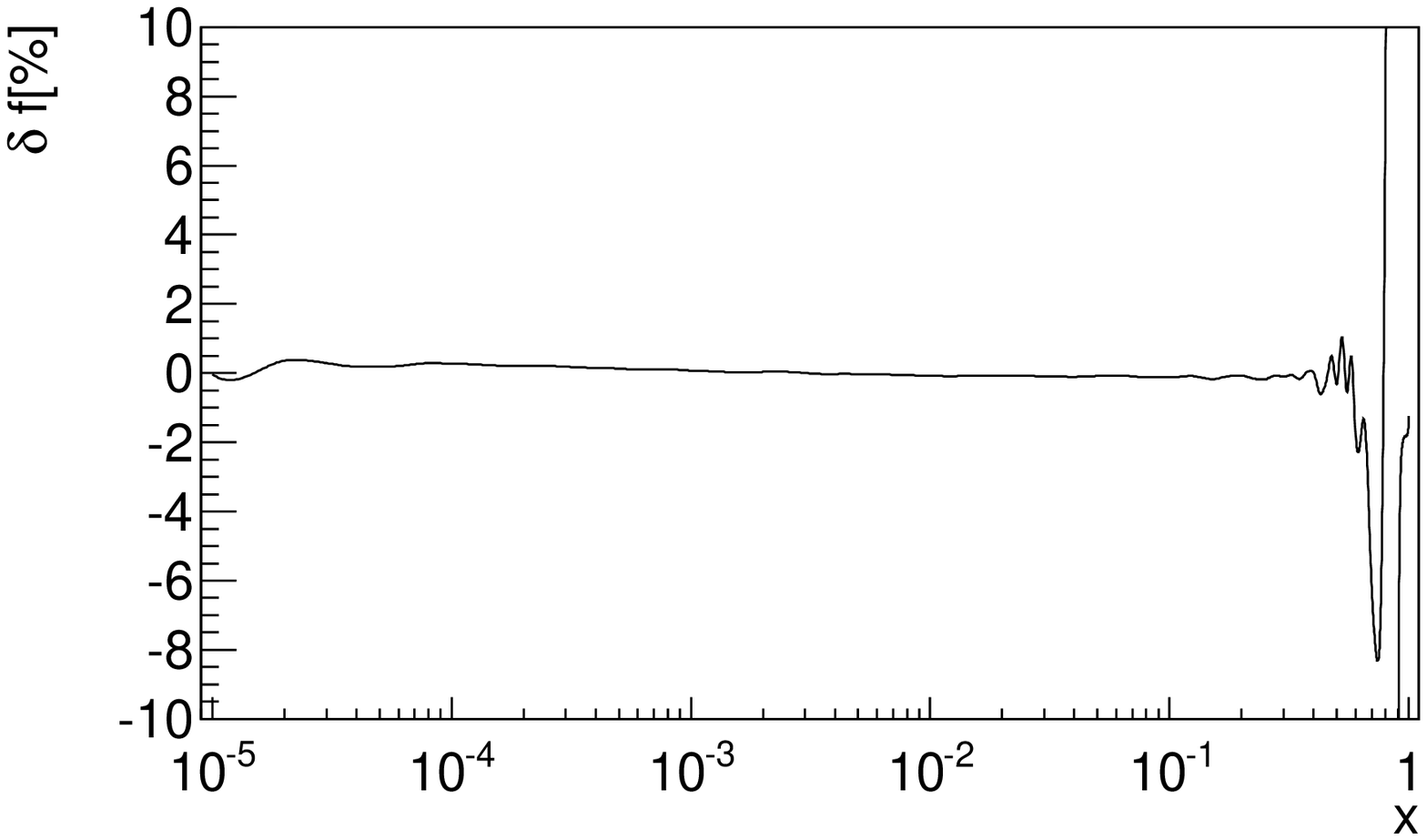}
\end{center}
\caption {Tuned comparison between {\tt QCDNUM+QED} and {\tt MRST2004QED}.
Left plot shows the momentum distribution of $ \gamma $ at $ \mu^2 = 10^4 \text{ GeV}^2 $.
The corresponding $ \delta f $ is shown on the right plot.}
\label{fig12}
\end{figure}

\begin{figure}
\begin{center}
\includegraphics[width = 0.45\textwidth]{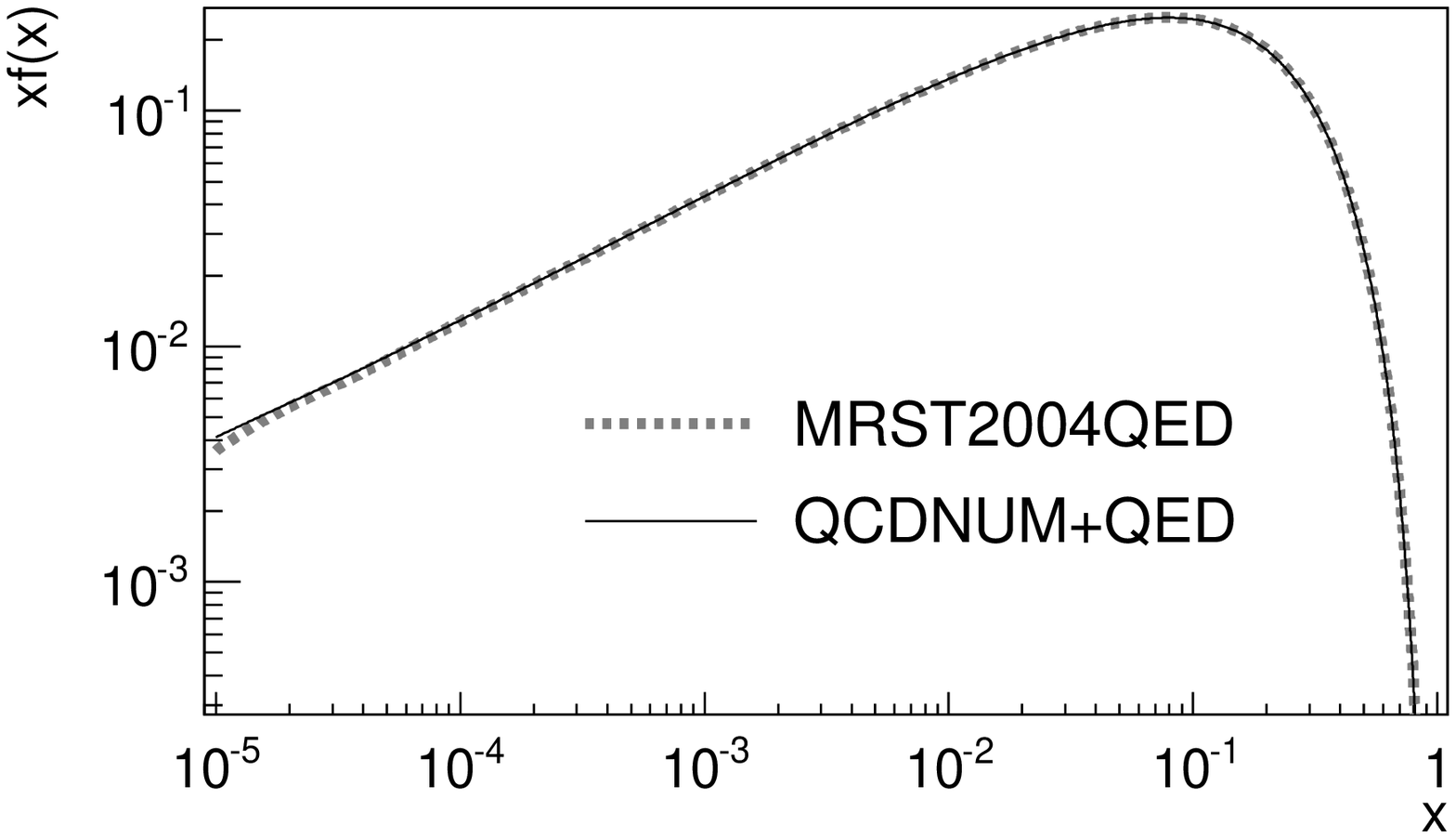}
\includegraphics[width = 0.45\textwidth]{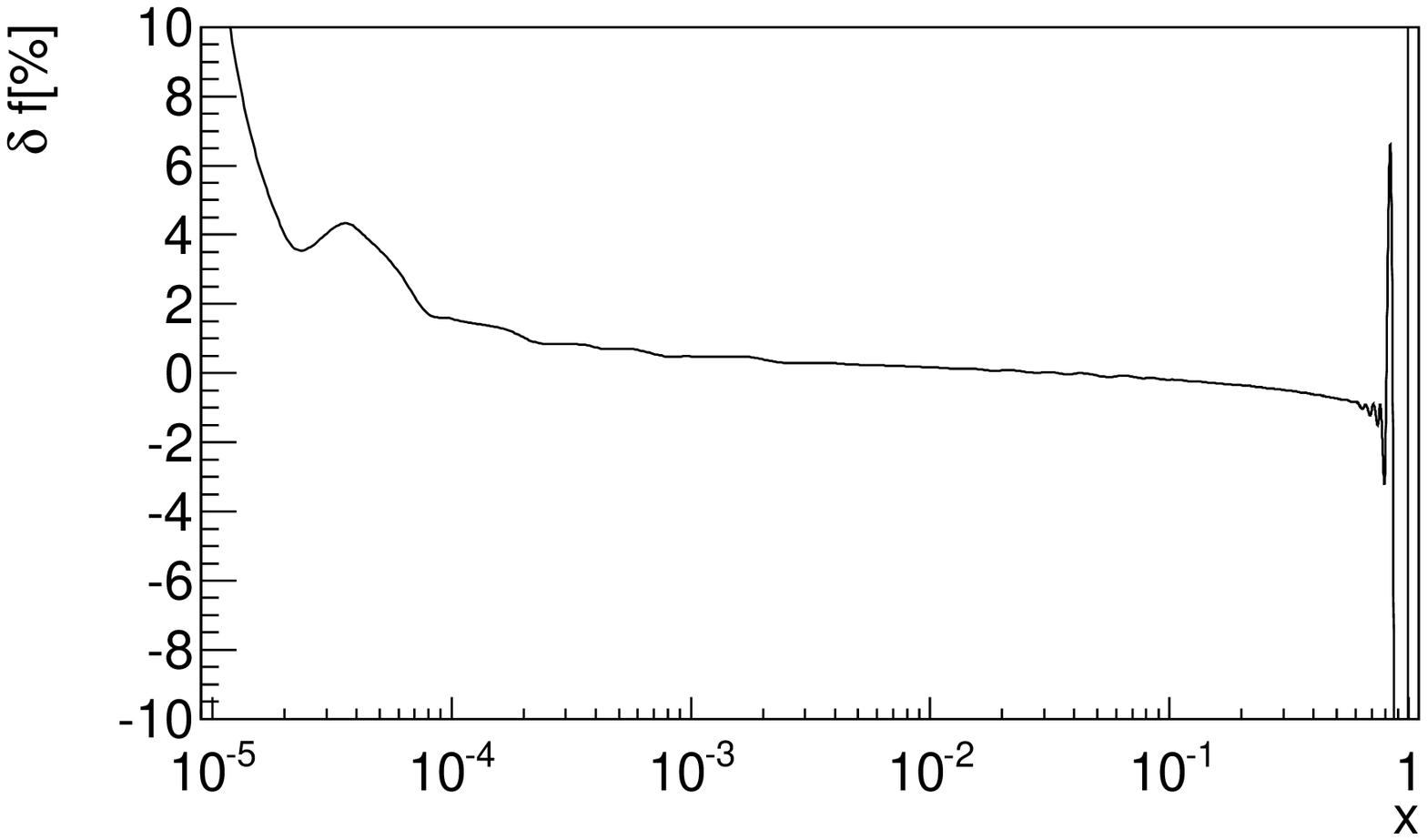}
\end{center}
\caption {Tuned comparison between {\tt QCDNUM+QED} and {\tt MRST2004QED}.
Left plot shows the momentum distribution of $ d_v $ at $ \mu^2 = 10^4 \text{ GeV}^2 $.
The corresponding $ \delta f $ is shown on the right plot.}
\label{fig13}
\end{figure}

\begin{figure}
\begin{center}
\includegraphics[width = 0.45\textwidth]{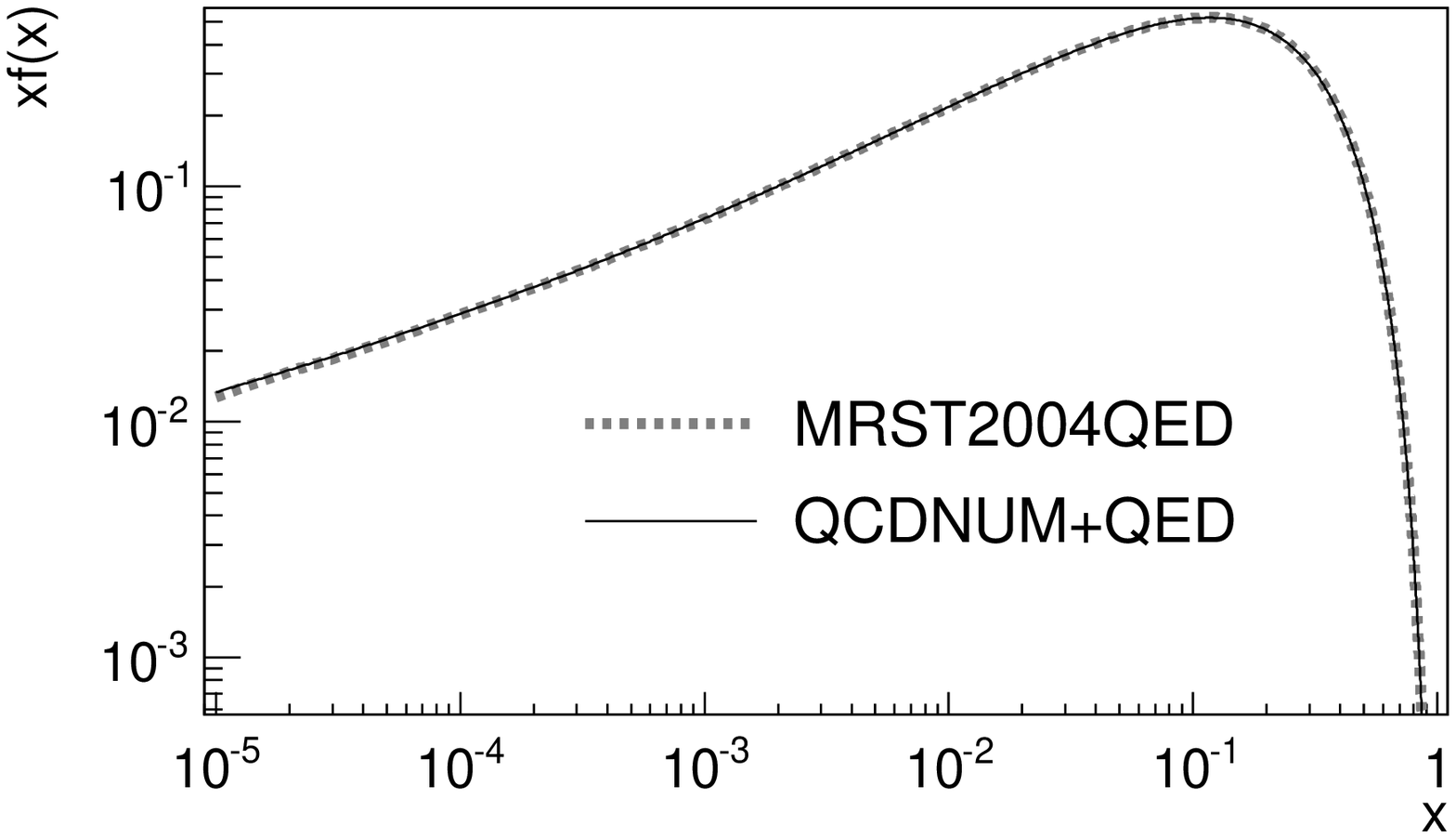}
\includegraphics[width = 0.45\textwidth]{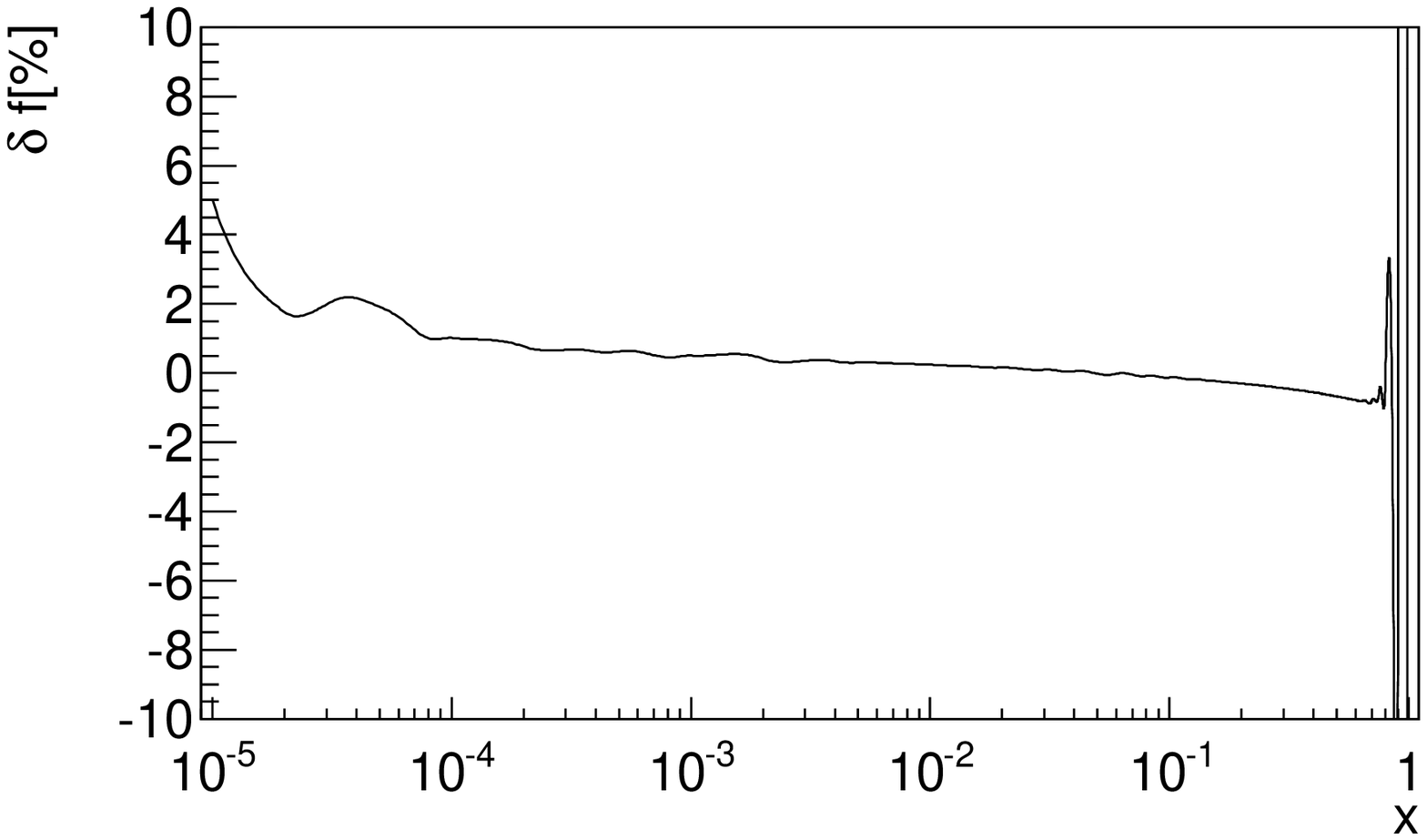}
\end{center}
\caption {Tuned comparison between {\tt QCDNUM+QED} and {\tt MRST2004QED}.
Left plot shows the momentum distribution of $ u_v $ at $ \mu^2 = 10^4 \text{ GeV}^2 $.
The corresponding $ \delta f $ is shown on the right plot.}
\label{fig14}
\end{figure}

\begin{figure}
\begin{center}
\includegraphics[width = 0.45\textwidth]{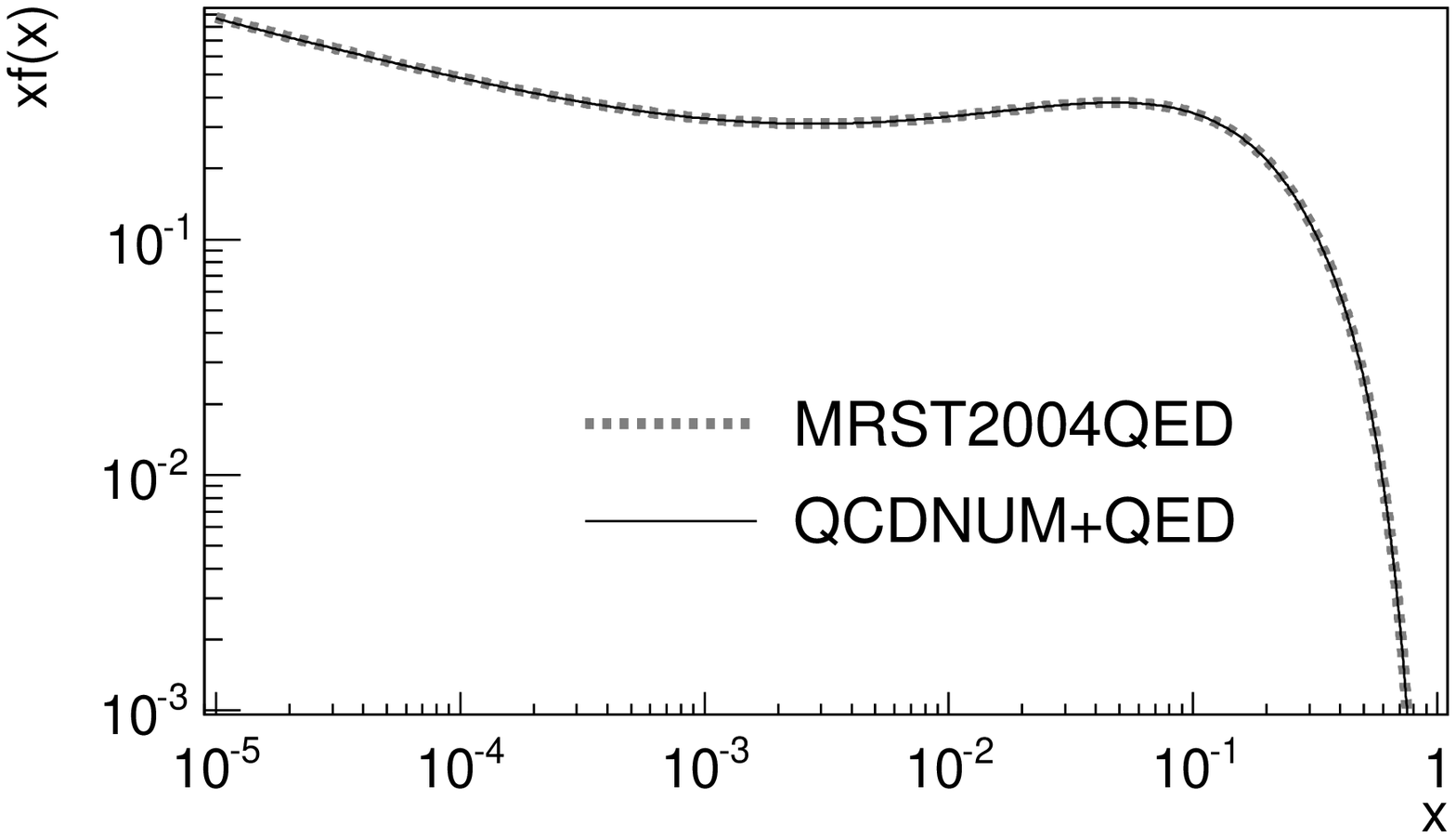}
\includegraphics[width = 0.45\textwidth]{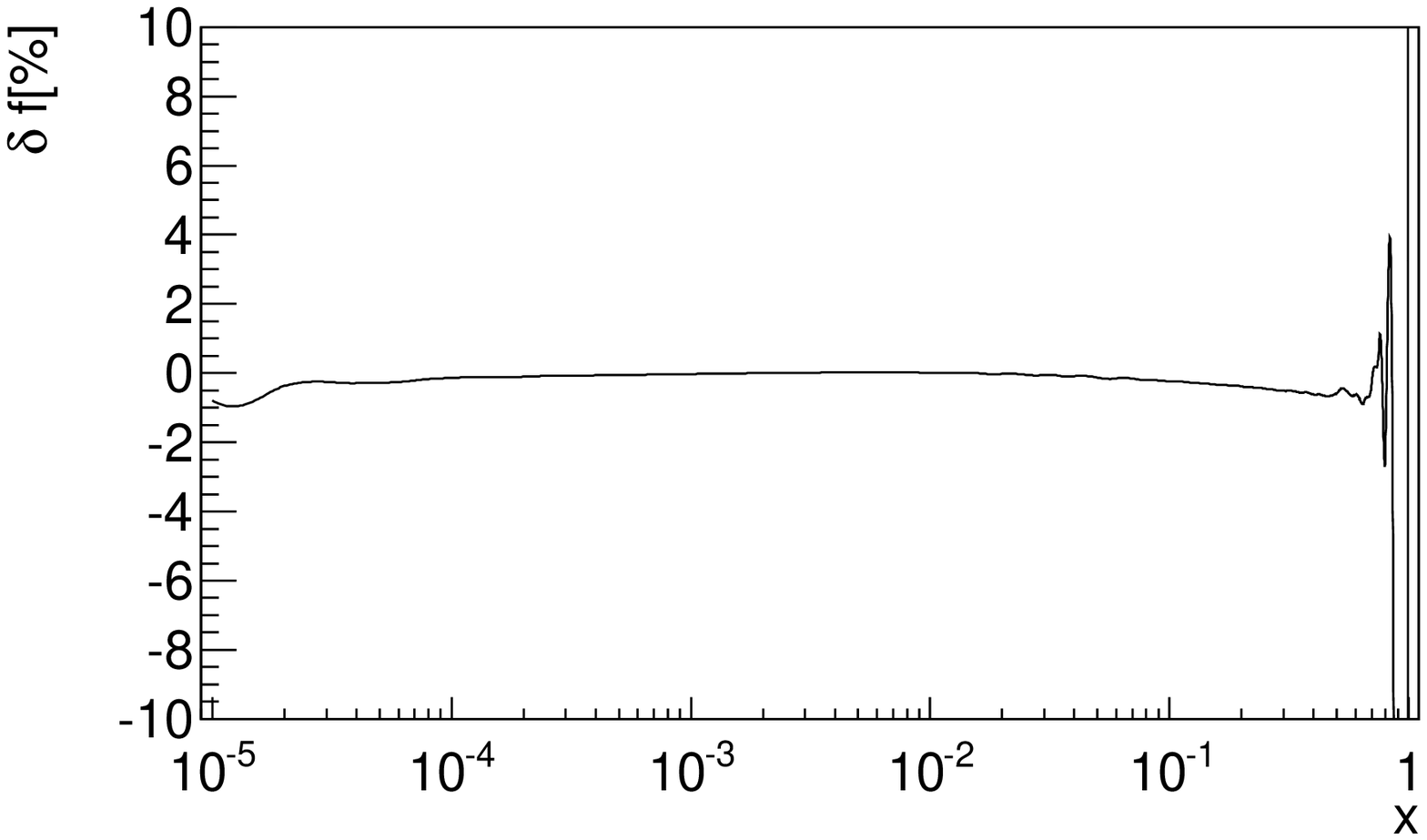}
\end{center}
\caption {Tuned comparison between {\tt QCDNUM+QED} and {\tt MRST2004QED}.
Left plot shows the momentum distribution of $ \Delta_{ds} $ at $ \mu^2 = 10^4 \text{ GeV}^2 $.
The corresponding $ \delta f $ is shown on the right plot.}
\label{fig15}
\end{figure}

\begin{figure}
\begin{center}
\includegraphics[width = 0.45\textwidth]{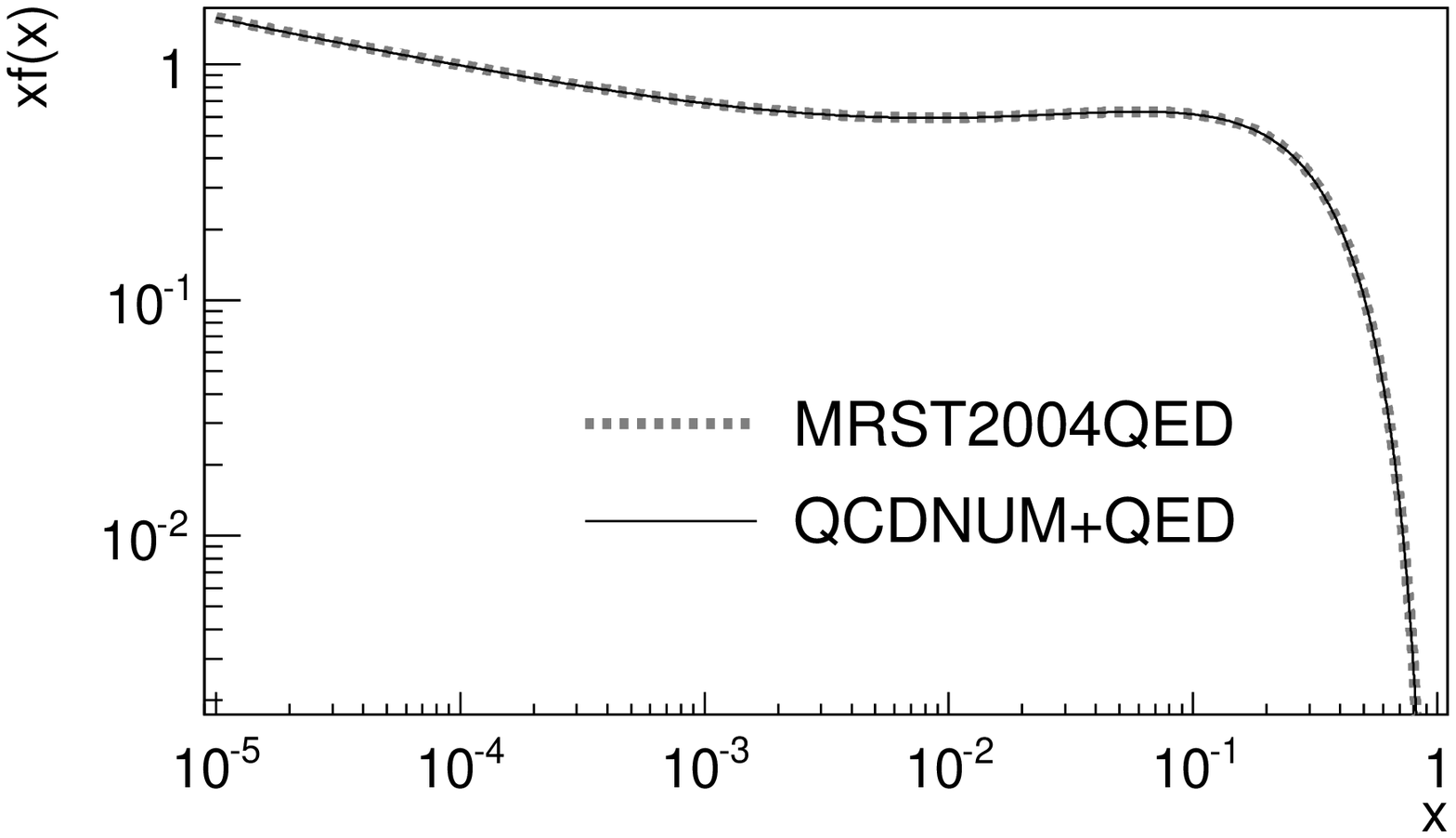}
\includegraphics[width = 0.45\textwidth]{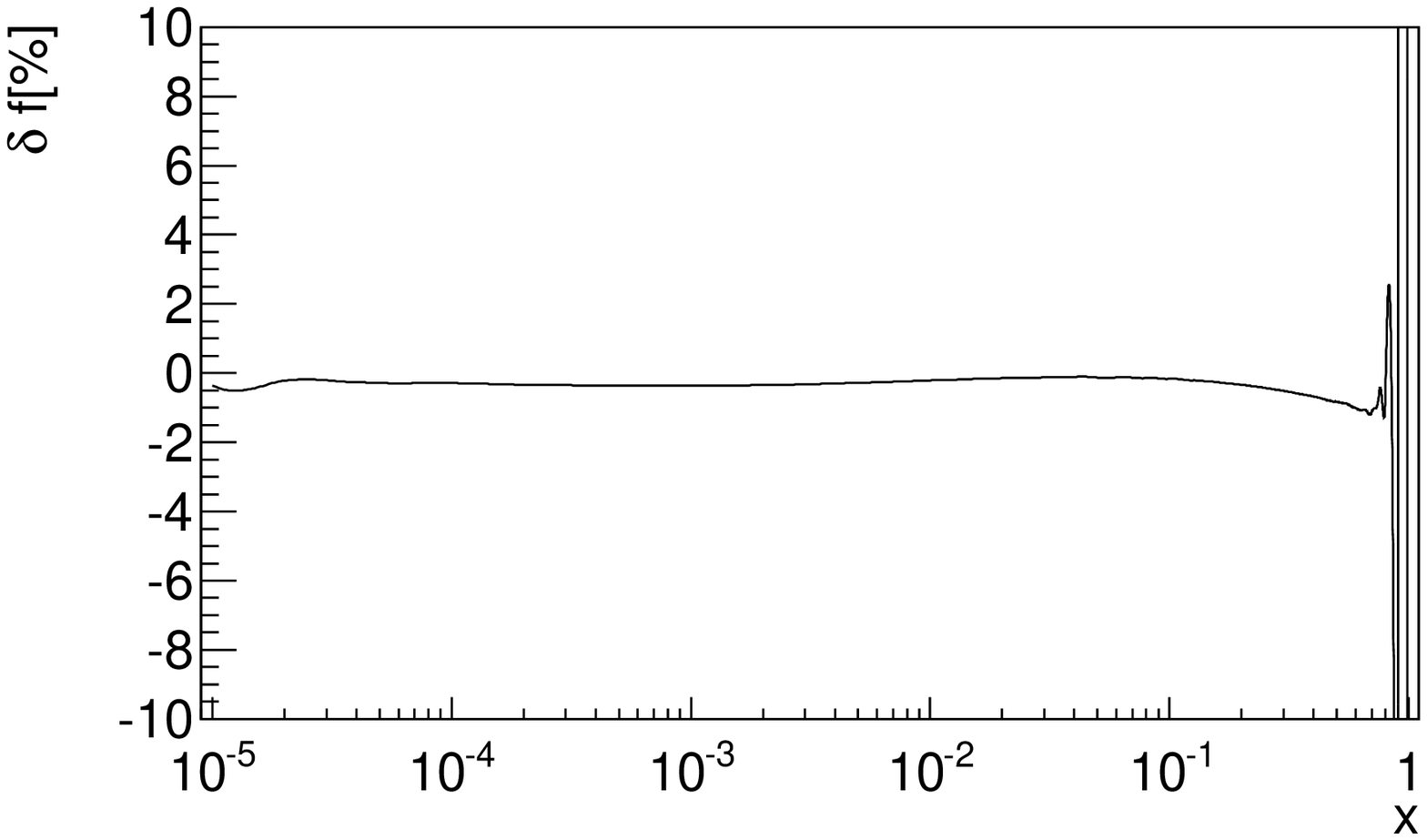}
\end{center}
\caption {Tuned comparison between {\tt QCDNUM+QED} and {\tt MRST2004QED}.
Left plot shows the momentum distribution of $ \Delta_{uc} $ at $ \mu^2 = 10^4 \text{ GeV}^2 $.
The corresponding $ \delta f $ is shown on the right plot.}
\label{fig16}
\end{figure}

\begin{figure}
\begin{center}
\includegraphics[width = 0.45\textwidth]{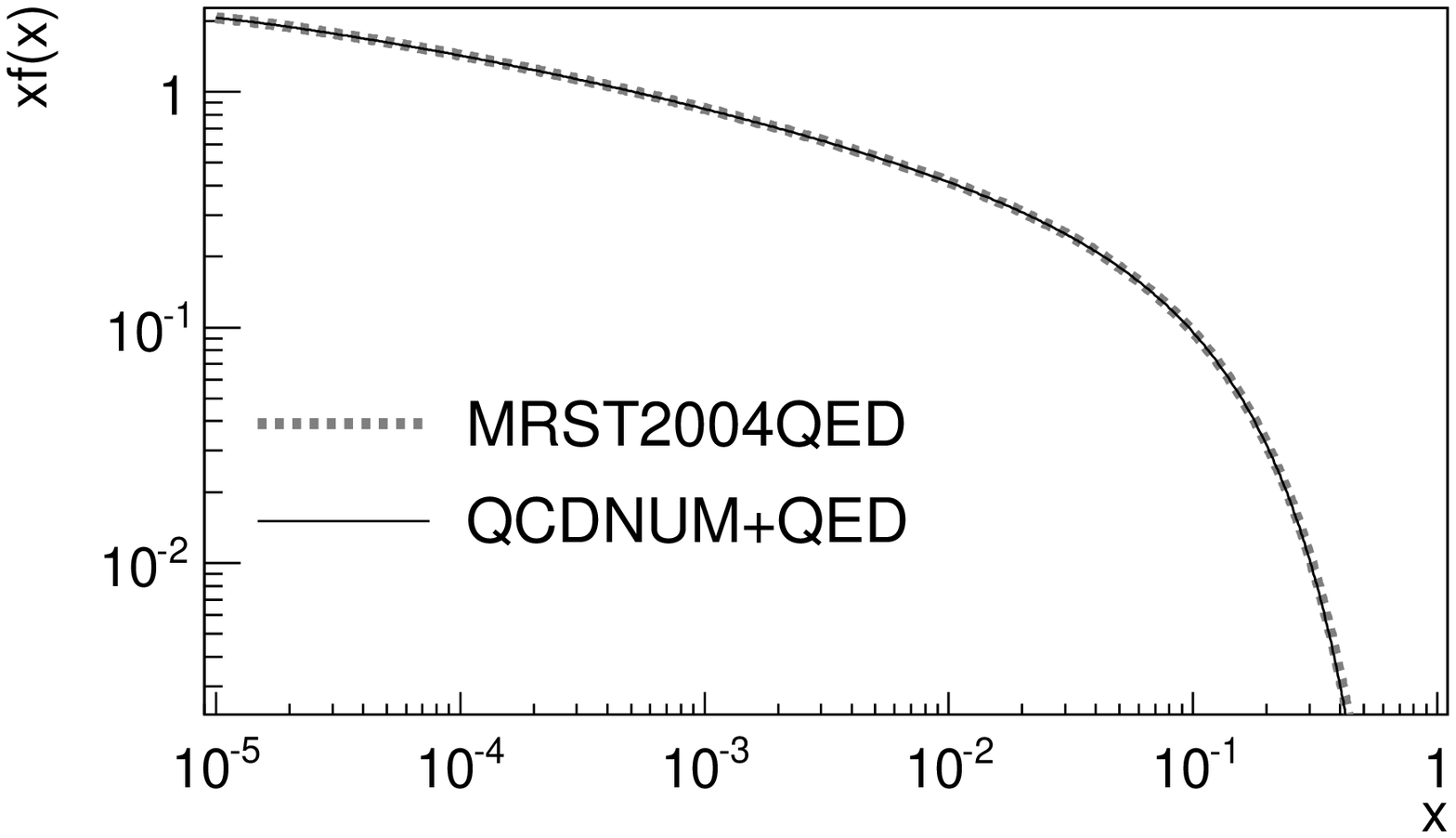}
\includegraphics[width = 0.45\textwidth]{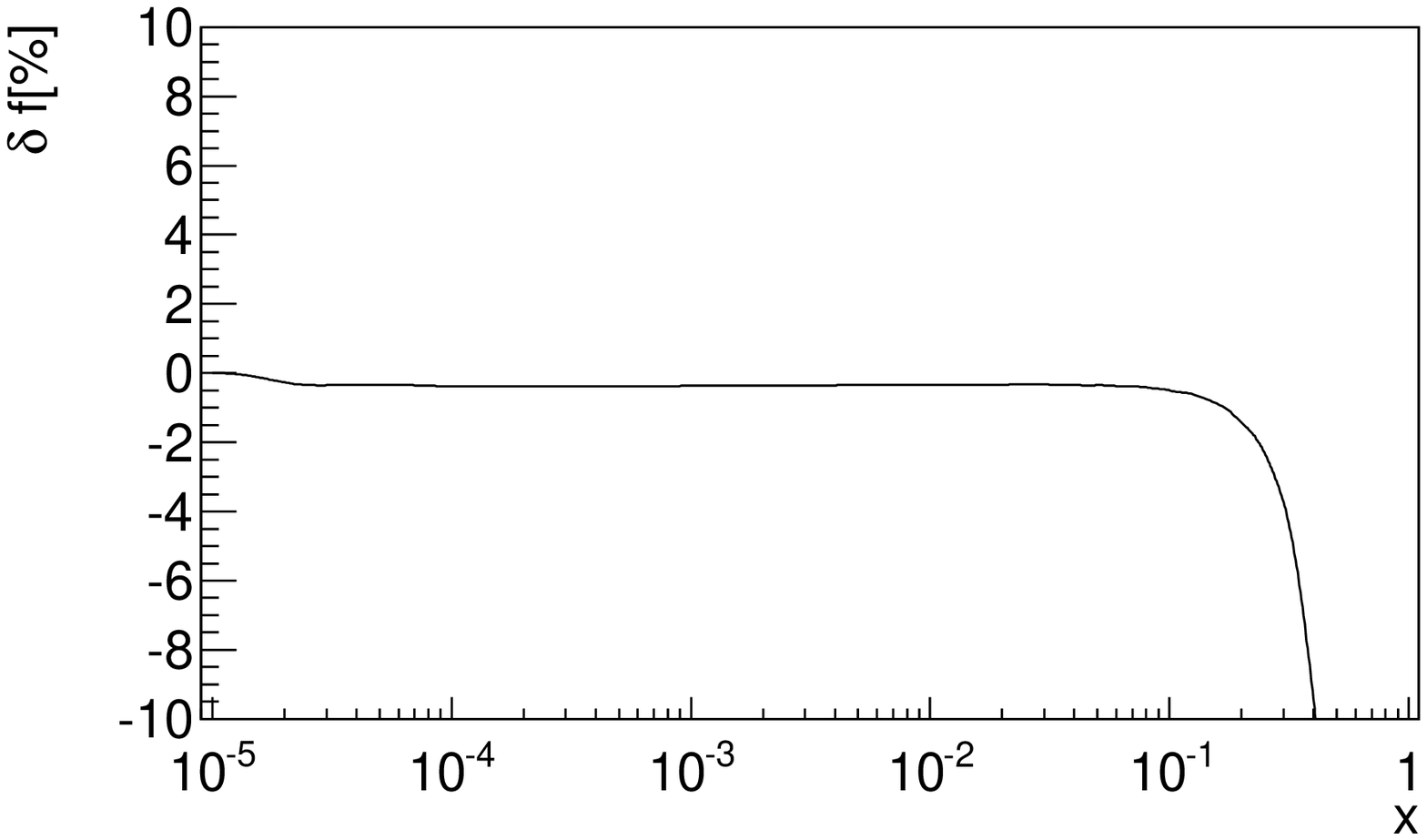}
\end{center}
\caption {Tuned comparison between {\tt QCDNUM+QED} and {\tt MRST2004QED}.
Left plot shows the momentum distribution of $ \Delta_{sb} $ at $ \mu^2 = 10^4 \text{ GeV}^2 $.
The corresponding $ \delta f $ is shown on the right plot.}
\label{fig17}
\end{figure}

To compare {\tt QCDNUM+QED} with {\tt MRST2004QED} PDF the staring distributions were taken
at initial scale $ \mu_0 = 2 $ GeV using {\tt LHAPDF} interface. Then this distributions were
evolved to the scale $ \mu = 100 $ GeV on $ 1000 \times 1000 $ grid in $ x $ and $ \mu^2 $.
The results of comparison in VFNS for momentum distributions $ xf $ and the corresponding relative
differences
\begin{equation}
\delta f =
\frac{xf(\text{\tt QCDNUM+QED}) - xf(\text{\tt MRST2004QED})}{xf(\text{\tt MRST2004QED})}
\label{eq13}
\end{equation}
are presented in Figs. \ref{fig9}-\ref{fig17} where an agreement within $ \sim 1 \% $ is observed
excluding a region of x where PDFs are very small.
\subsection{Comparison of {\tt QCDNUM+QED} and {\tt APFEL}}
\begin{figure}
\begin{center}
\includegraphics[width = 0.45\textwidth]{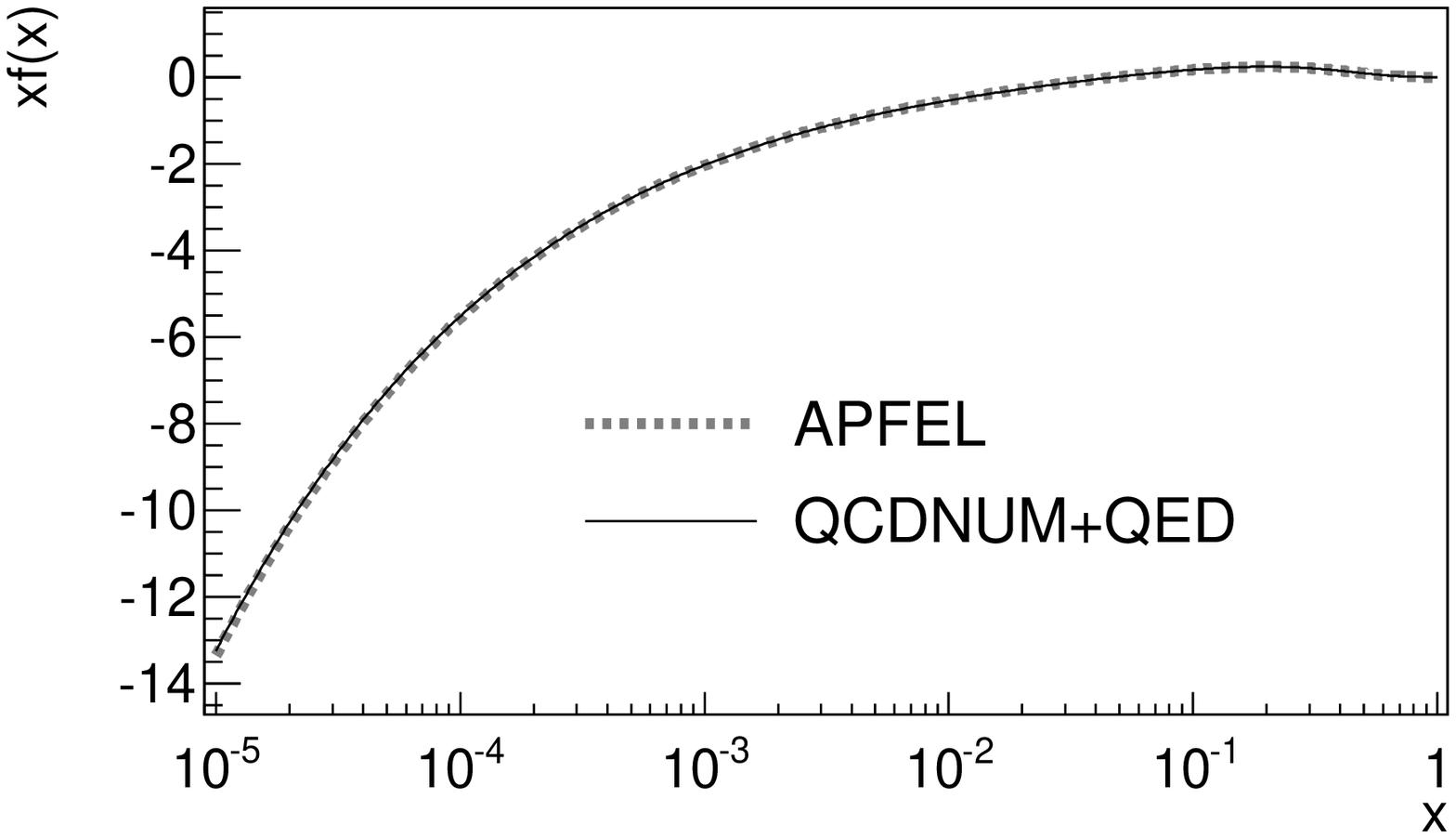}
\includegraphics[width = 0.45\textwidth]{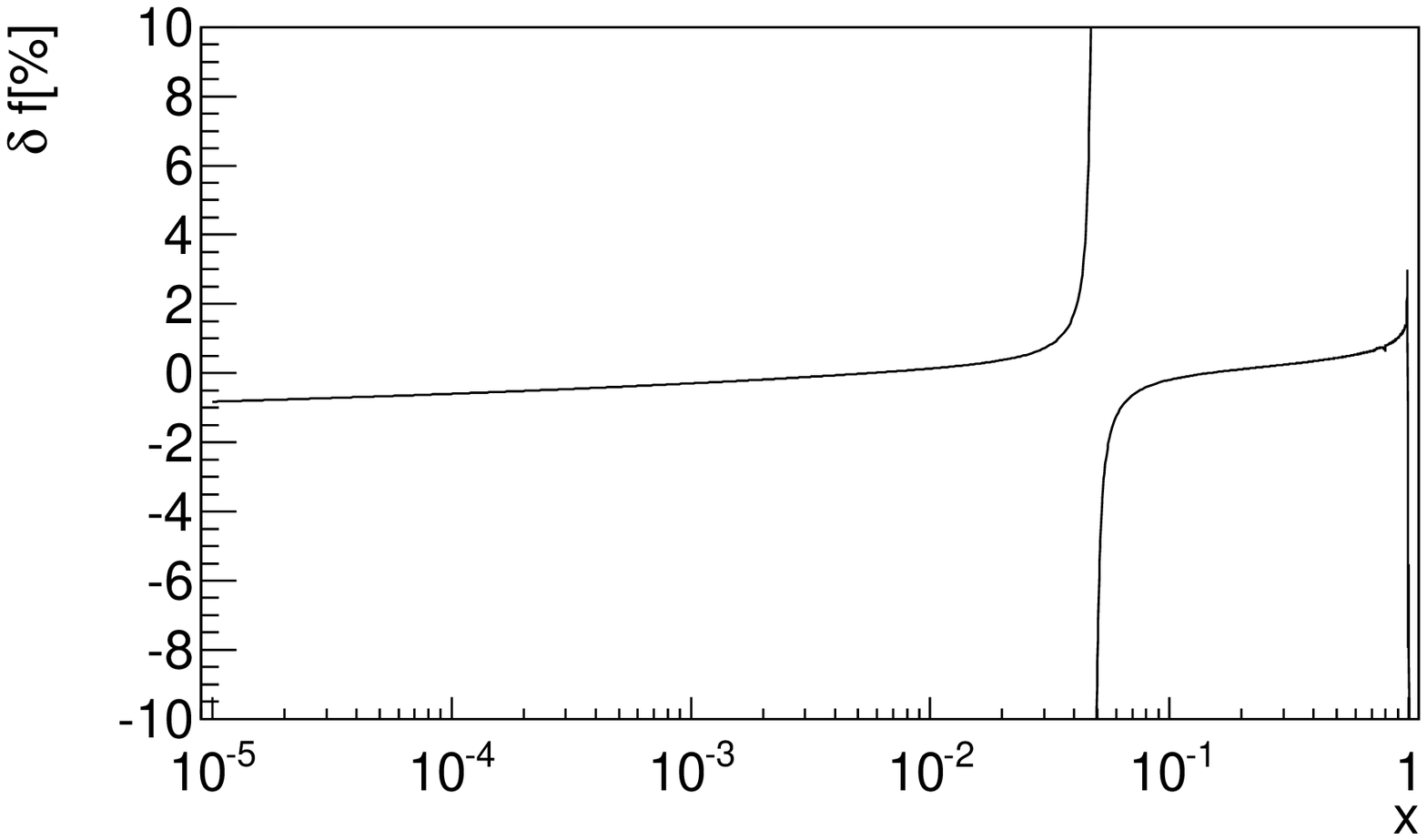}
\end{center}
\caption {Tuned comparison between {\tt QCDNUM+QED} and {\tt APFEL}.
Left plot shows the momentum distribution of $ \Delta $ at $ \mu^2 = 10^4 \text{ GeV}^2 $.
The corresponding $ \delta f $ is shown on the right plot.}
\label{fig18}
\end{figure}

\begin{figure}
\begin{center}
\includegraphics[width = 0.45\textwidth]{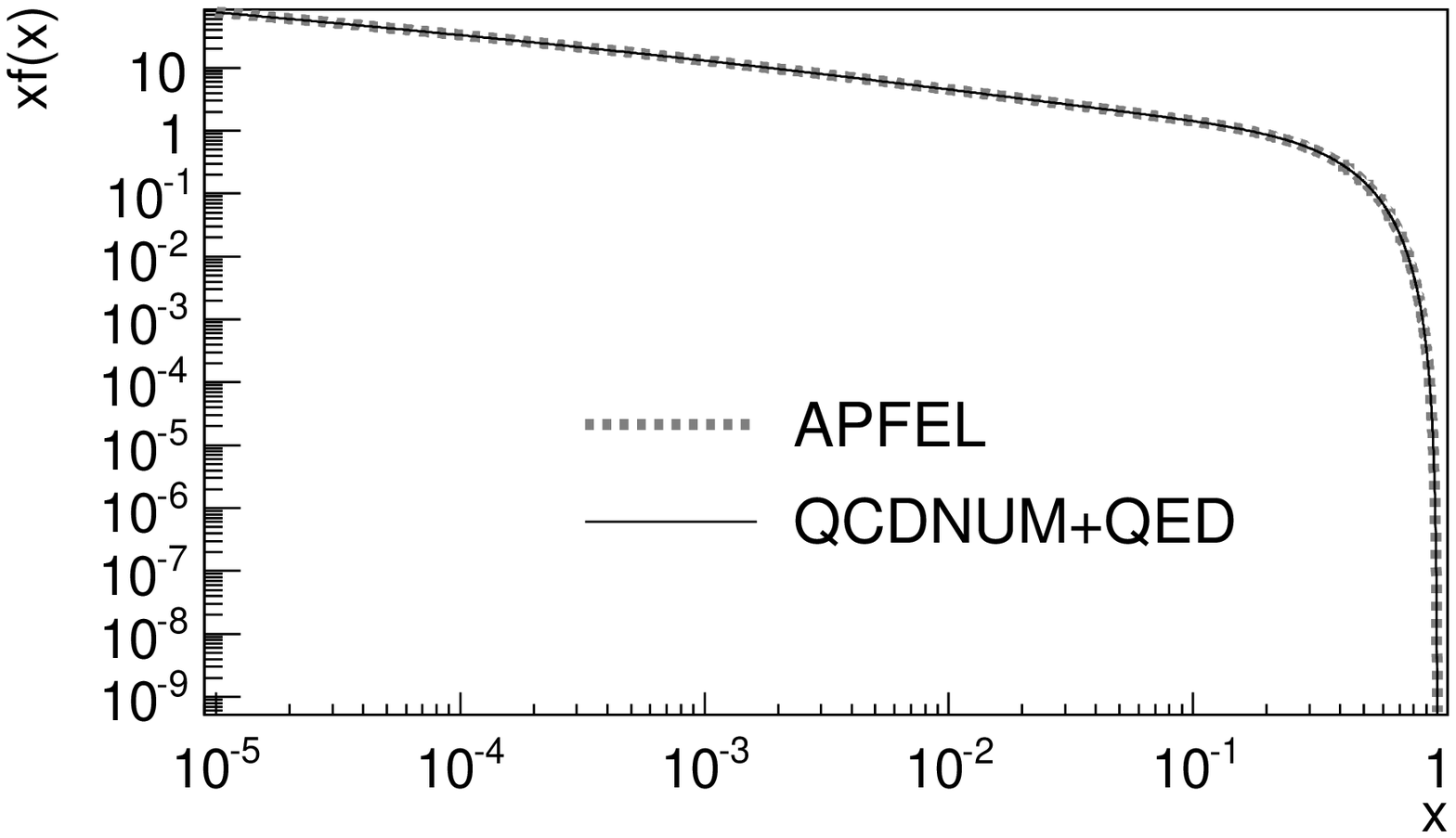}
\includegraphics[width = 0.45\textwidth]{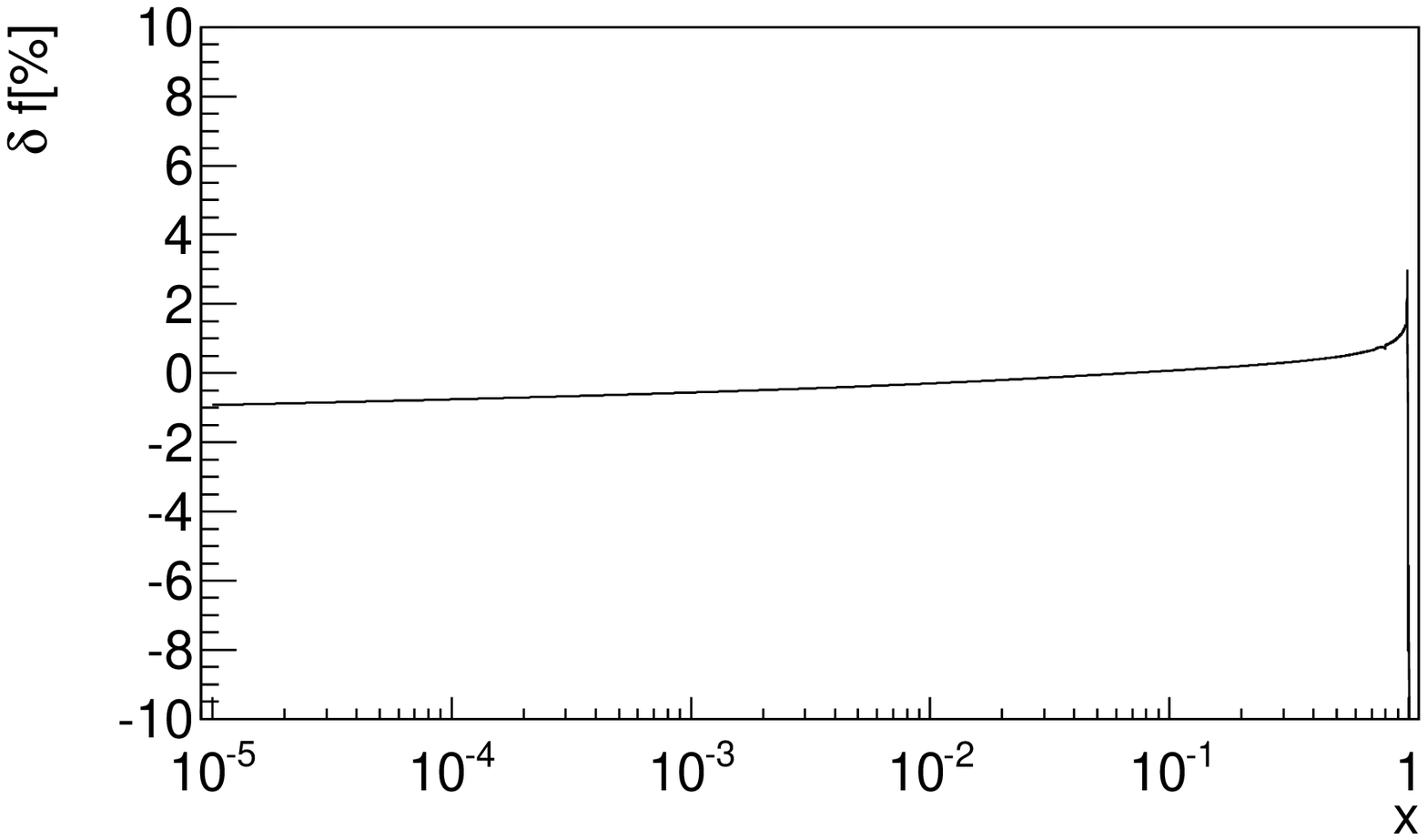}
\end{center}
\caption {Tuned comparison between {\tt QCDNUM+QED} and {\tt APFEL}.
Left plot shows the momentum distribution of $ \Sigma $ at $ \mu^2 = 10^4 \text{ GeV}^2 $.
The corresponding $ \delta f $ is shown on the right plot.}
\label{fig19}
\end{figure}

\begin{figure}
\begin{center}
\includegraphics[width = 0.45\textwidth]{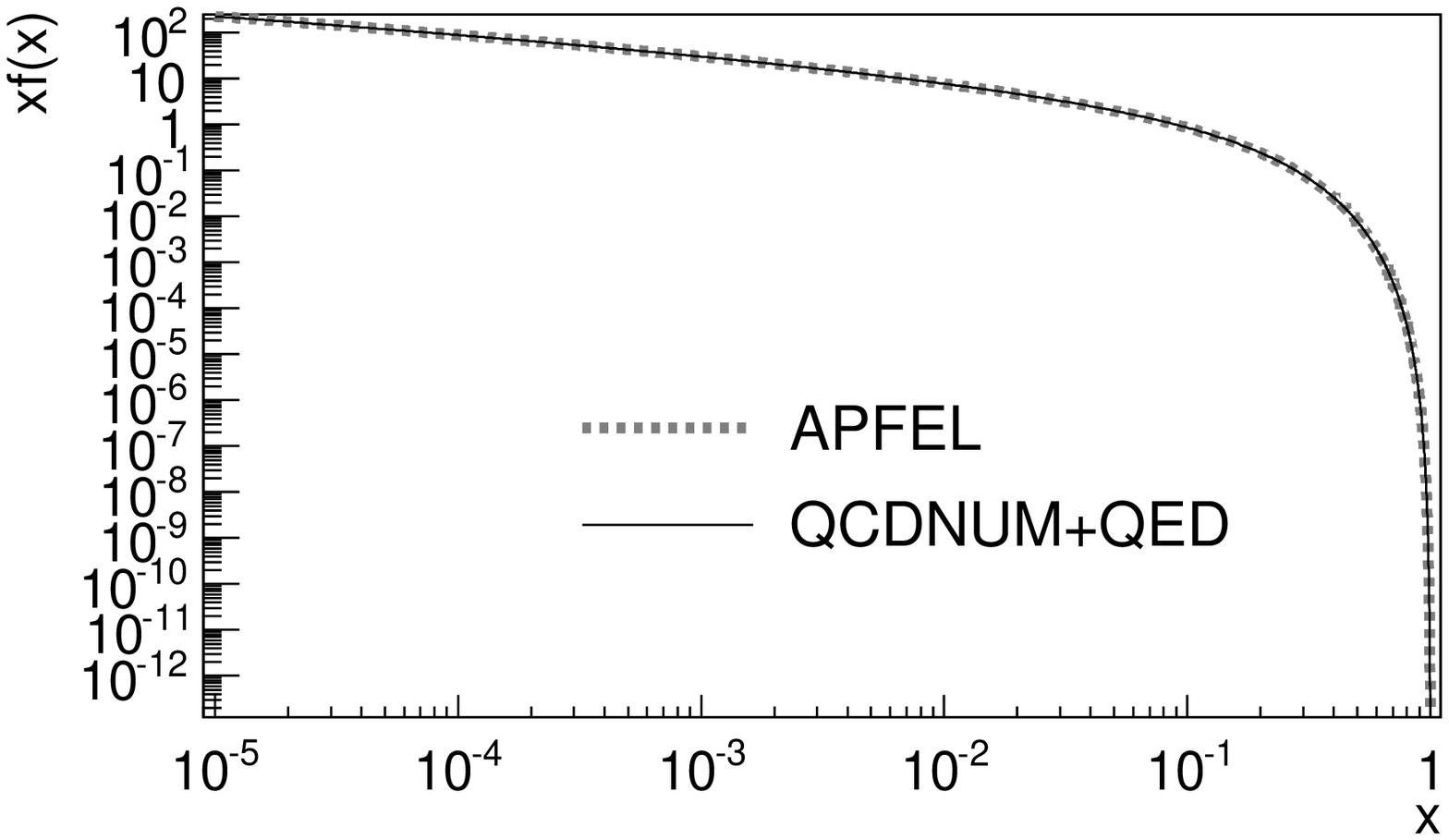}
\includegraphics[width = 0.45\textwidth]{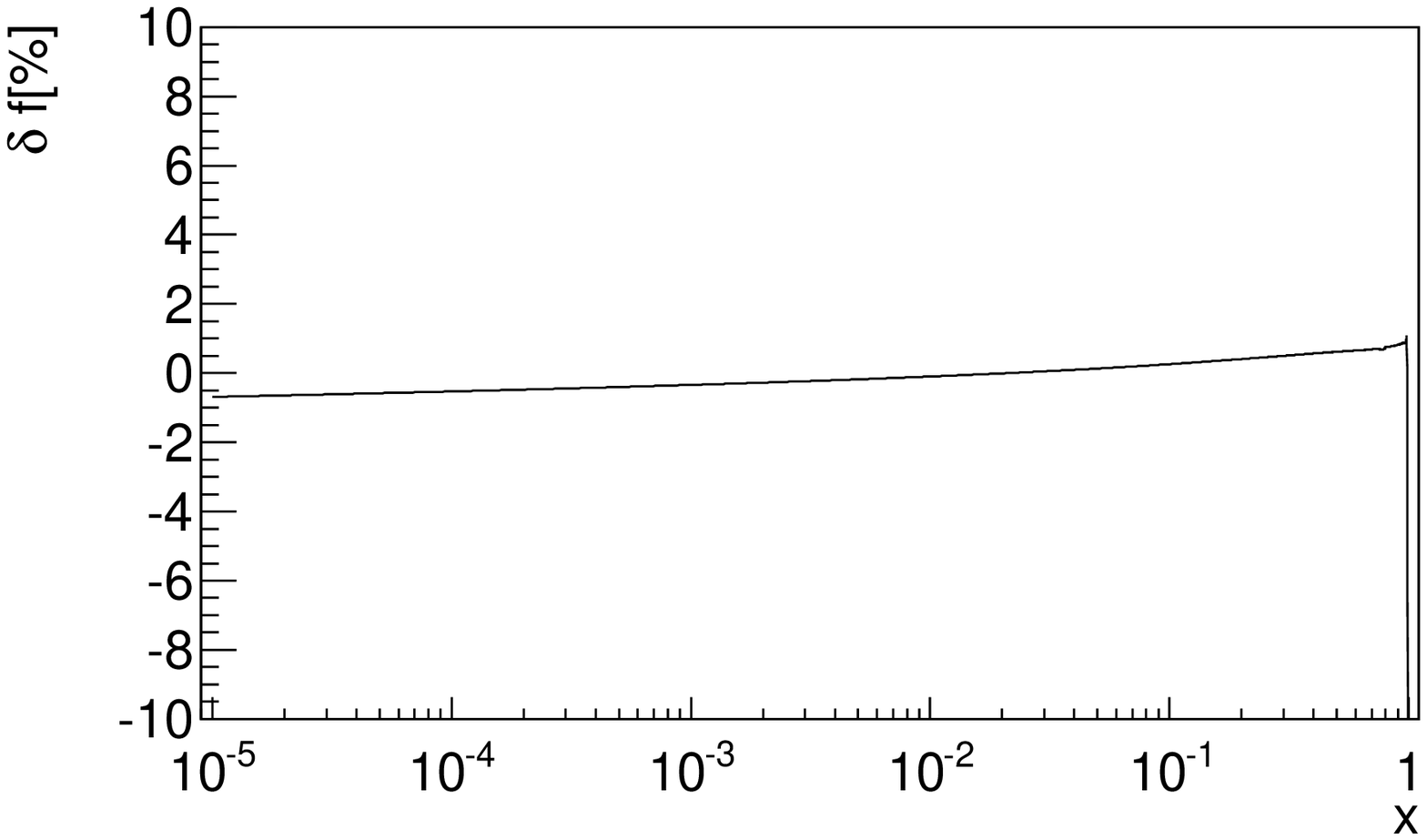}
\end{center}
\caption {Tuned comparison between {\tt QCDNUM+QED} and {\tt APFEL}.
Left plot shows the momentum distribution of $ g $ at $ \mu^2 = 10^4 \text{ GeV}^2 $.
The corresponding $ \delta f $ is shown on the right plot.}
\label{fig20}
\end{figure}

\begin{figure}
\begin{center}
\includegraphics[width = 0.45\textwidth]{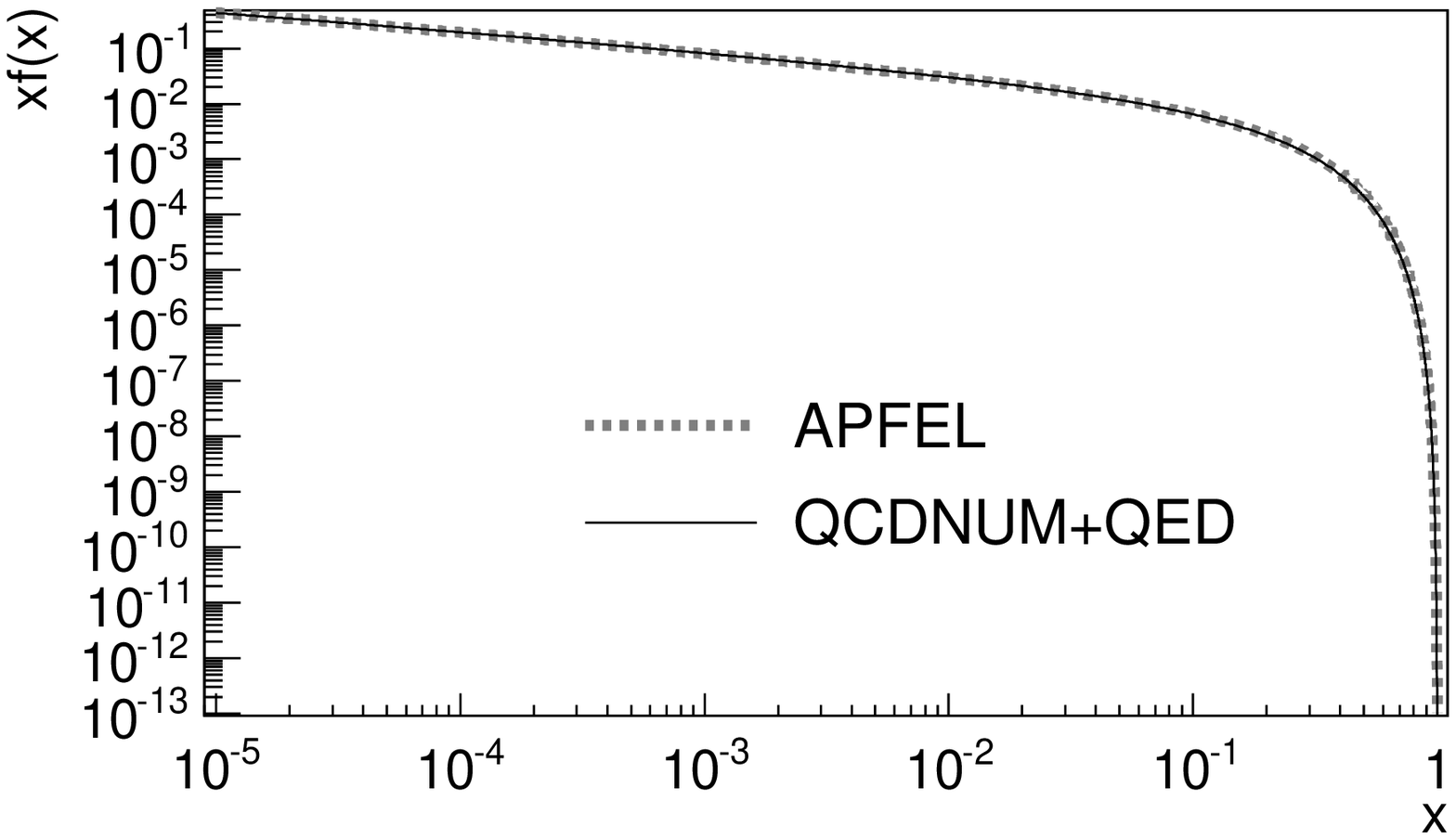}
\includegraphics[width = 0.45\textwidth]{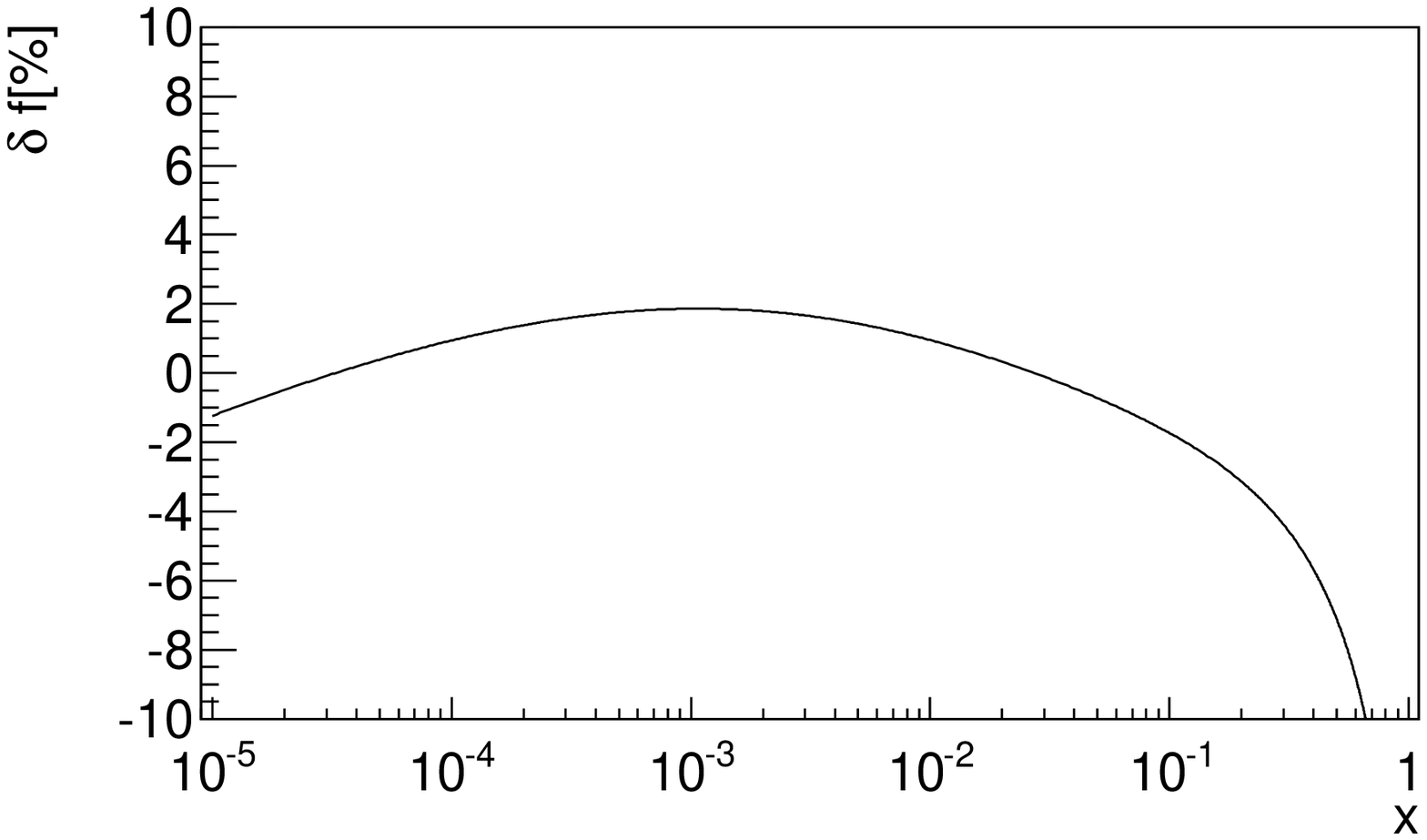}
\end{center}
\caption {Tuned comparison between {\tt QCDNUM+QED} and {\tt APFEL}.
Left plot shows the momentum distribution of $ \gamma $ at $ \mu^2 = 10^4 \text{ GeV}^2 $.
The corresponding $ \delta f $ is shown on the right plot.}
\label{fig21}
\end{figure}

\begin{figure}
\begin{center}
\includegraphics[width = 0.45\textwidth]{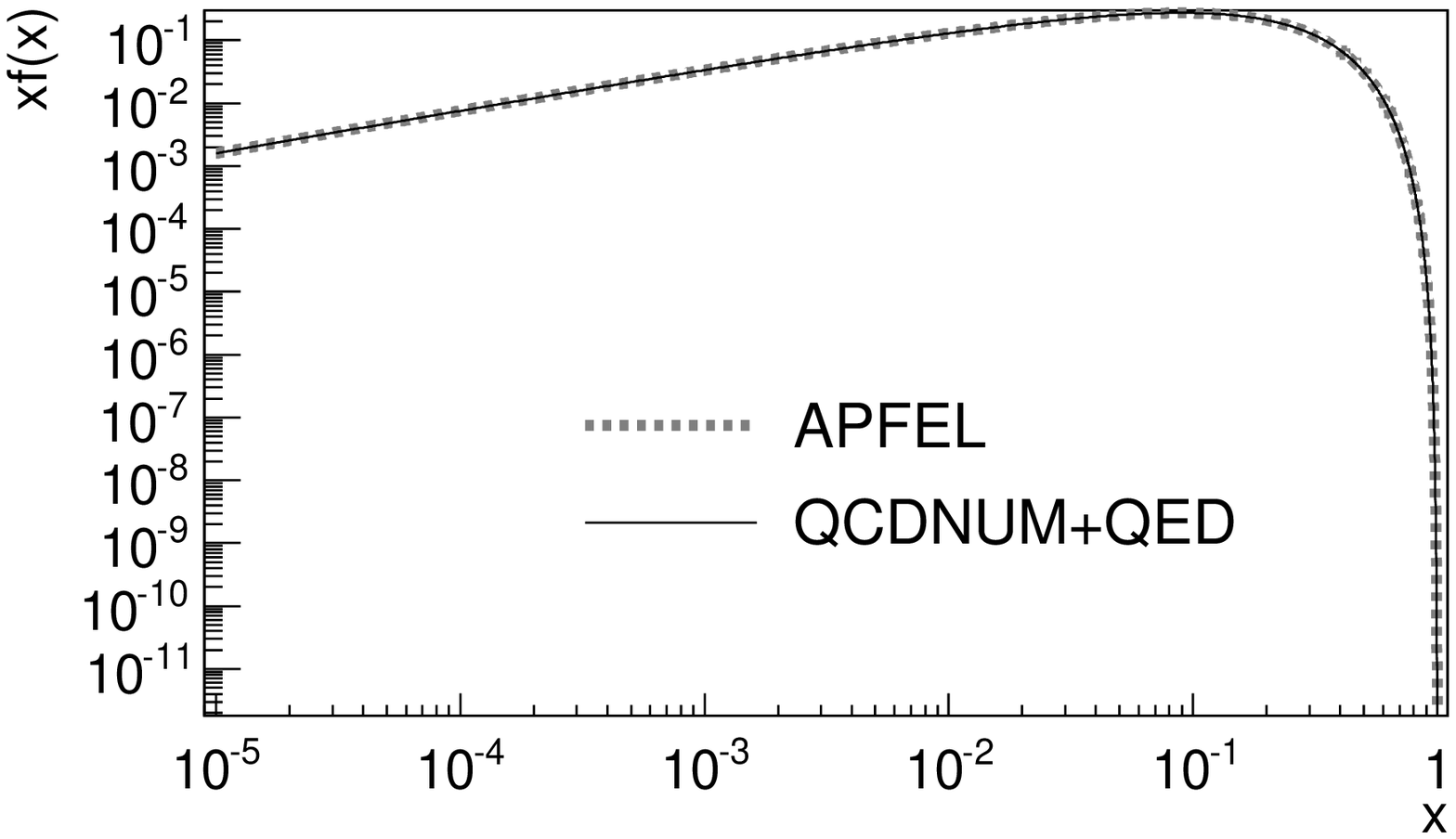}
\includegraphics[width = 0.45\textwidth]{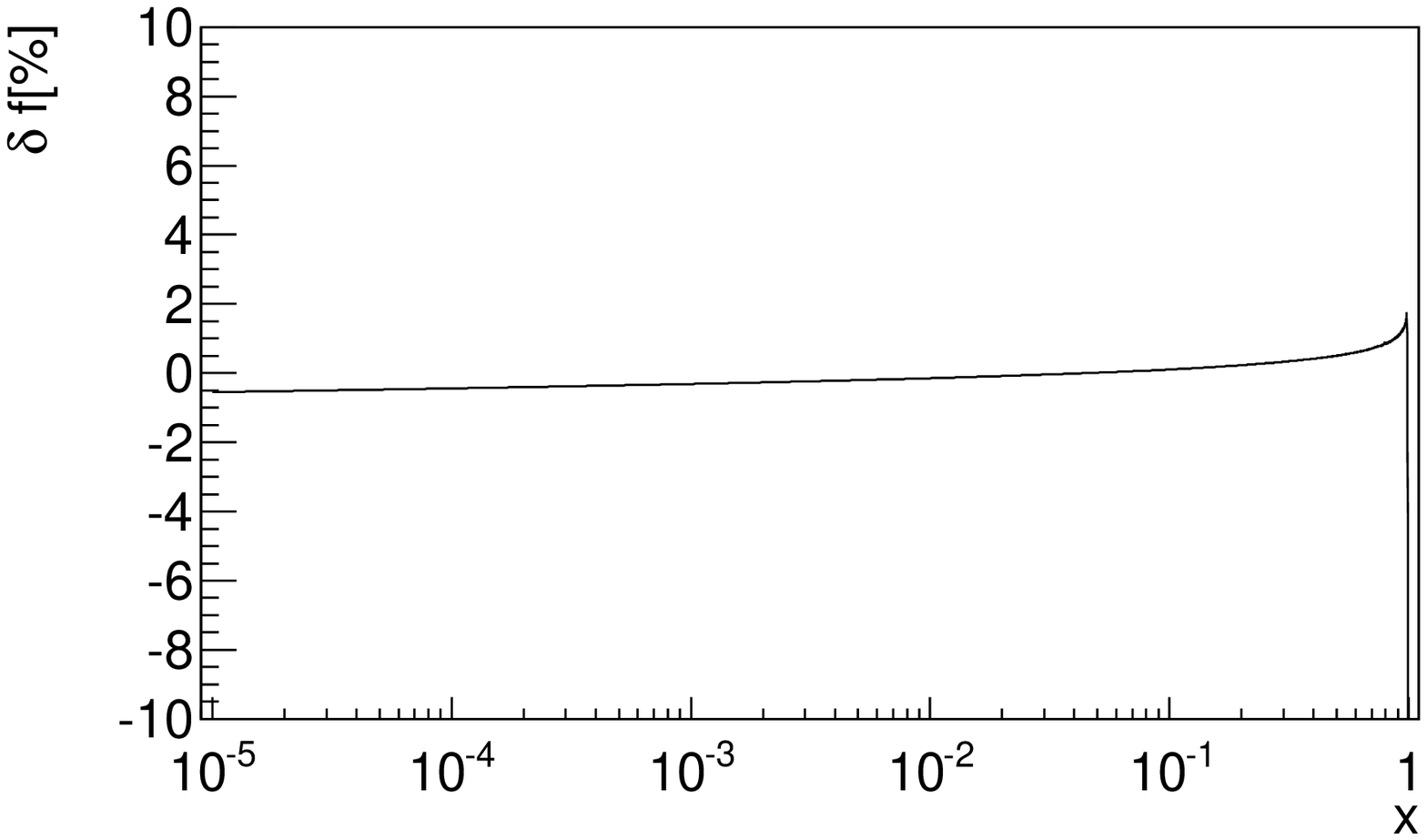}
\end{center}
\caption {Tuned comparison between {\tt QCDNUM+QED} and {\tt APFEL}.
Left plot shows the momentum distribution of $ d_v $ at $ \mu^2 = 10^4 \text{ GeV}^2 $.
The corresponding $ \delta f $ is shown on the right plot.}
\label{fig22}
\end{figure}

\begin{figure}
\begin{center}
\includegraphics[width = 0.45\textwidth]{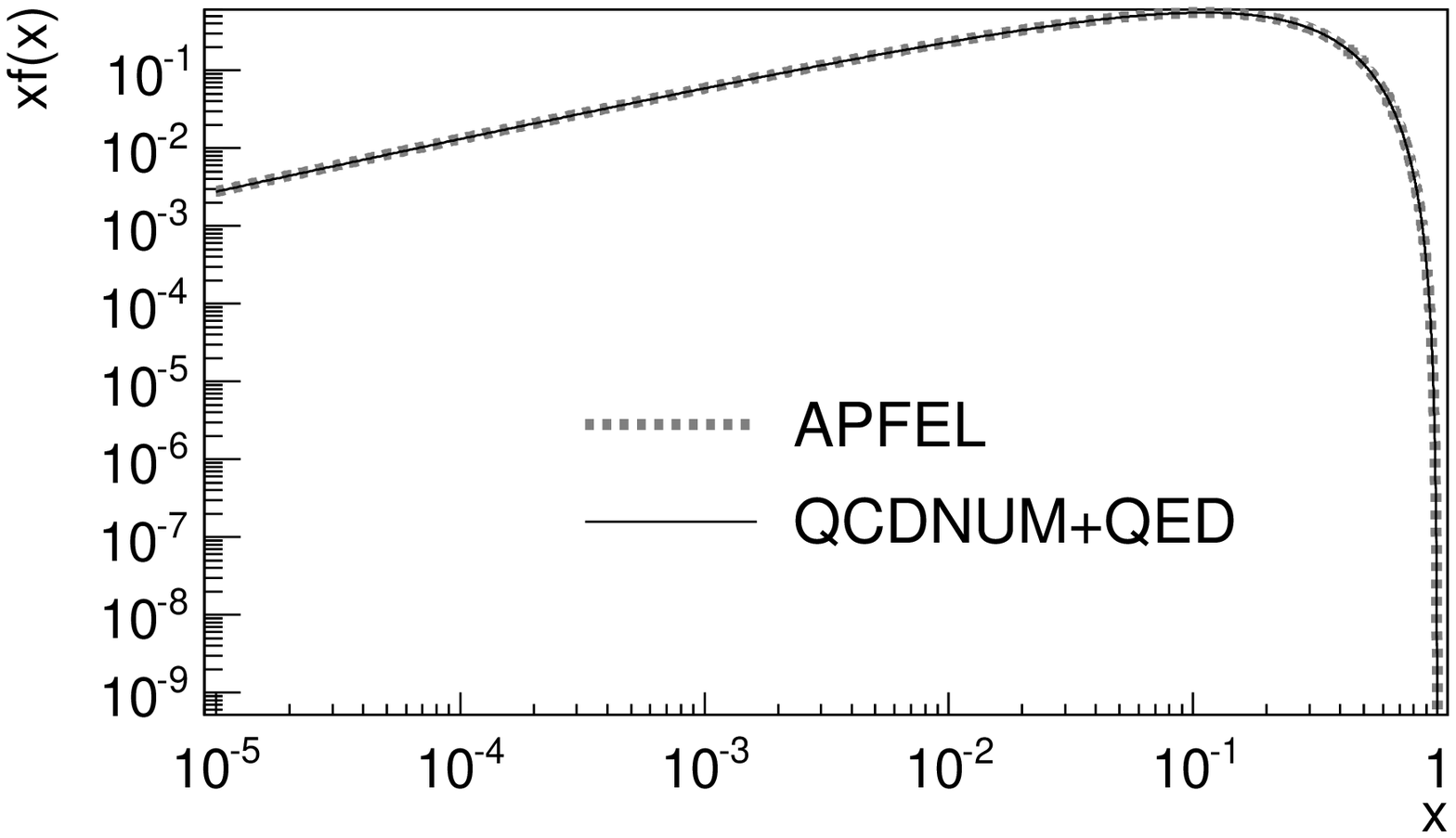}
\includegraphics[width = 0.45\textwidth]{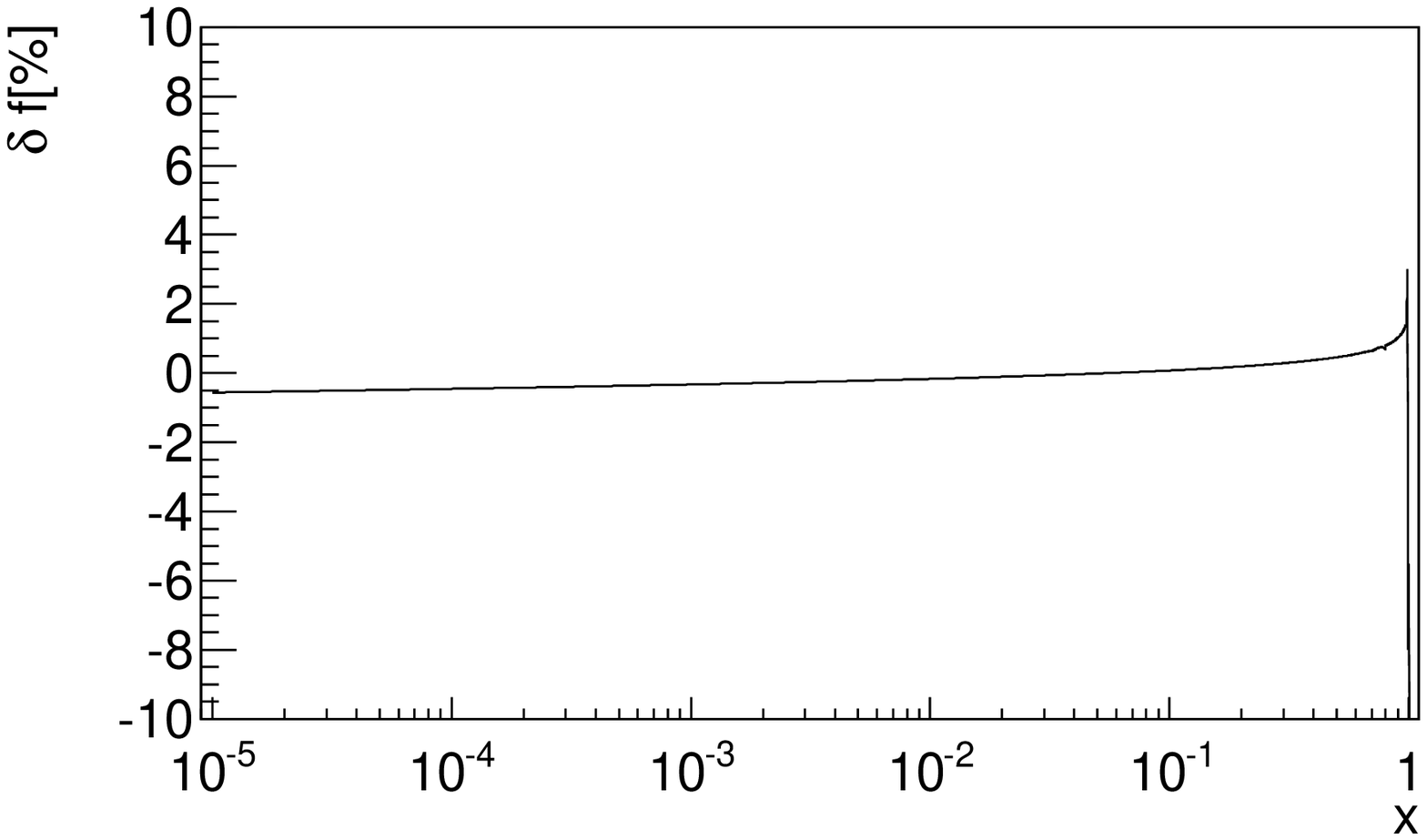}
\end{center}
\caption {Tuned comparison between {\tt QCDNUM+QED} and {\tt APFEL}.
Left plot shows the momentum distribution of $ u_v $ at $ \mu^2 = 10^4 \text{ GeV}^2 $.
The corresponding $ \delta f $ is shown on the right plot.}
\label{fig23}
\end{figure}

\begin{figure}
\begin{center}
\includegraphics[width = 0.45\textwidth]{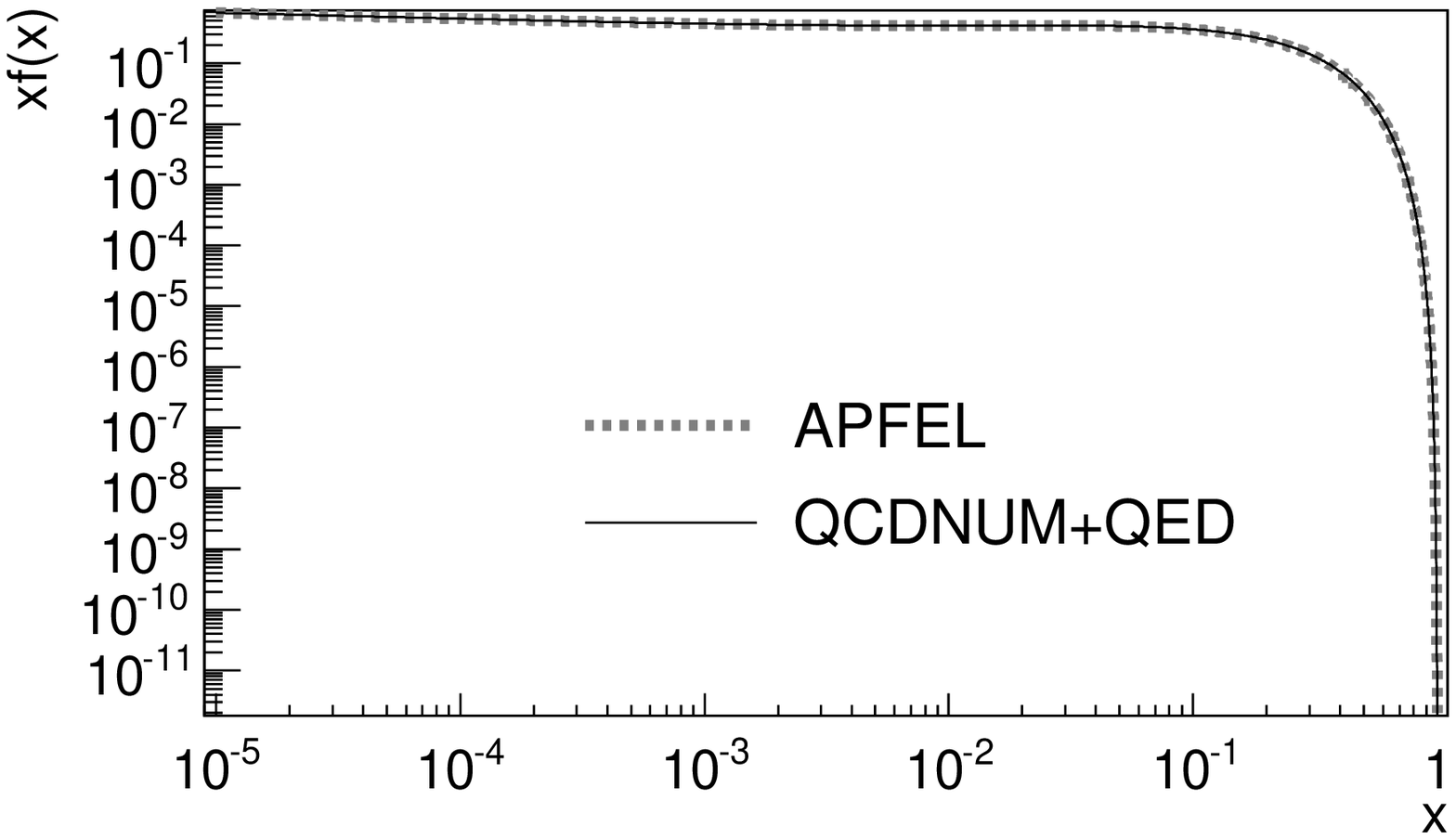}
\includegraphics[width = 0.45\textwidth]{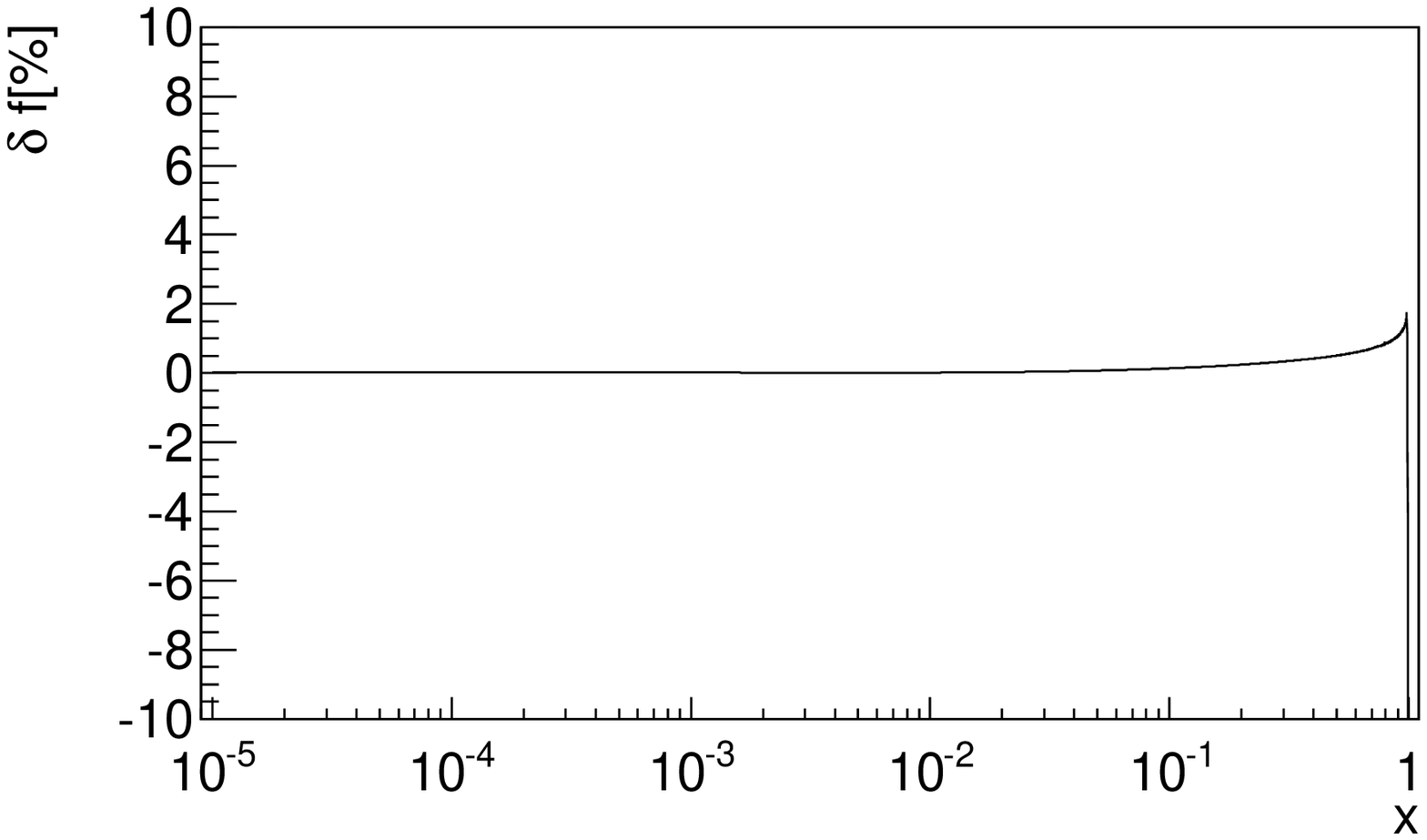}
\end{center}
\caption {Tuned comparison between {\tt QCDNUM+QED} and {\tt APFEL}.
Left plot shows the momentum distribution of $ \Delta_{ds} $ at $ \mu^2 = 10^4 \text{ GeV}^2 $.
The corresponding $ \delta f $ is shown on the right plot.}
\label{fig24}
\end{figure}

\begin{figure}
\begin{center}
\includegraphics[width = 0.45\textwidth]{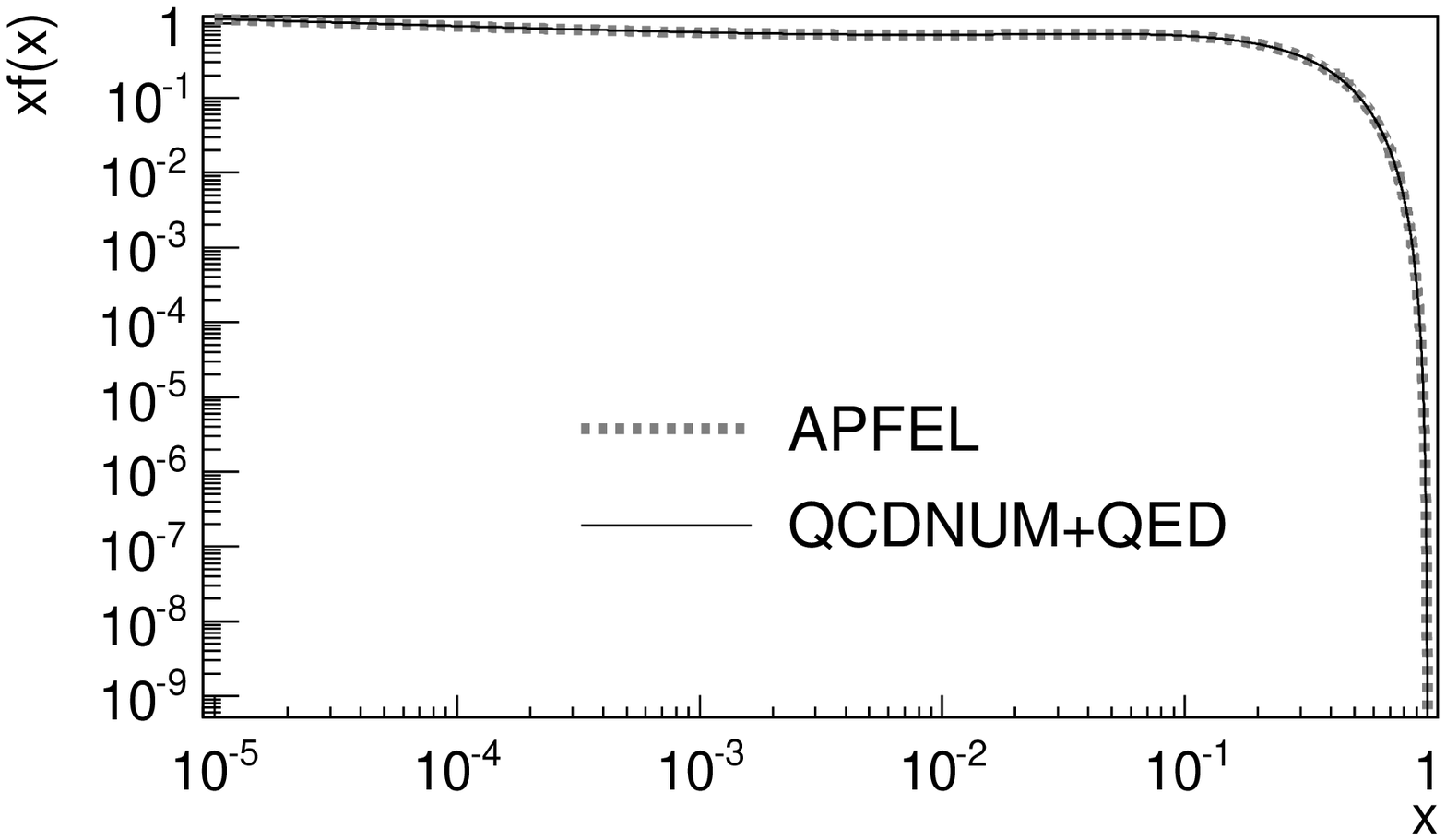}
\includegraphics[width = 0.45\textwidth]{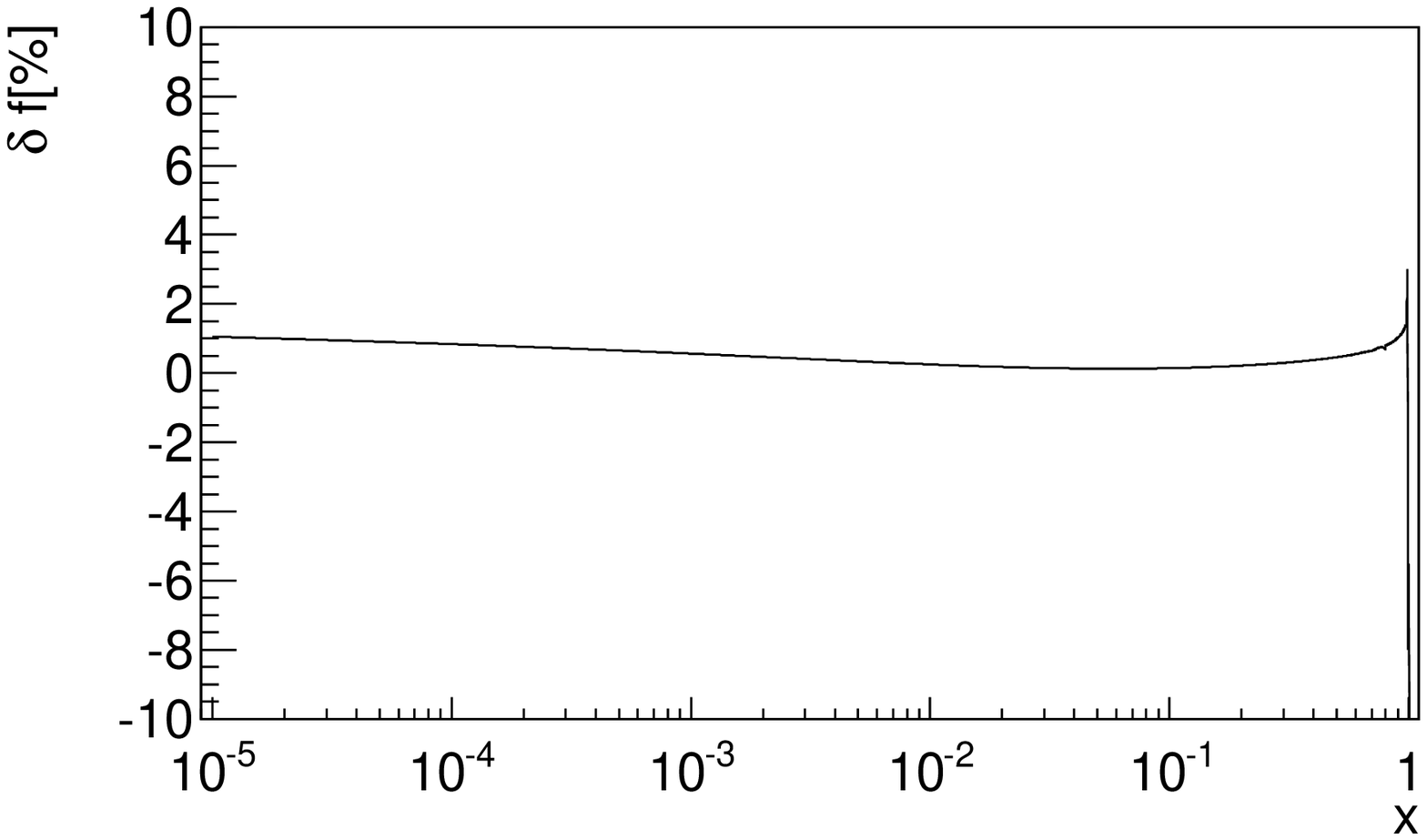}
\end{center}
\caption {Tuned comparison between {\tt QCDNUM+QED} and {\tt APFEL}.
Left plot shows the momentum distribution of $ \Delta_{uc} $ at $ \mu^2 = 10^4 \text{ GeV}^2 $.
The corresponding $ \delta f $ is shown on the right plot.}
\label{fig25}
\end{figure}

\begin{figure}
\begin{center}
\includegraphics[width = 0.45\textwidth]{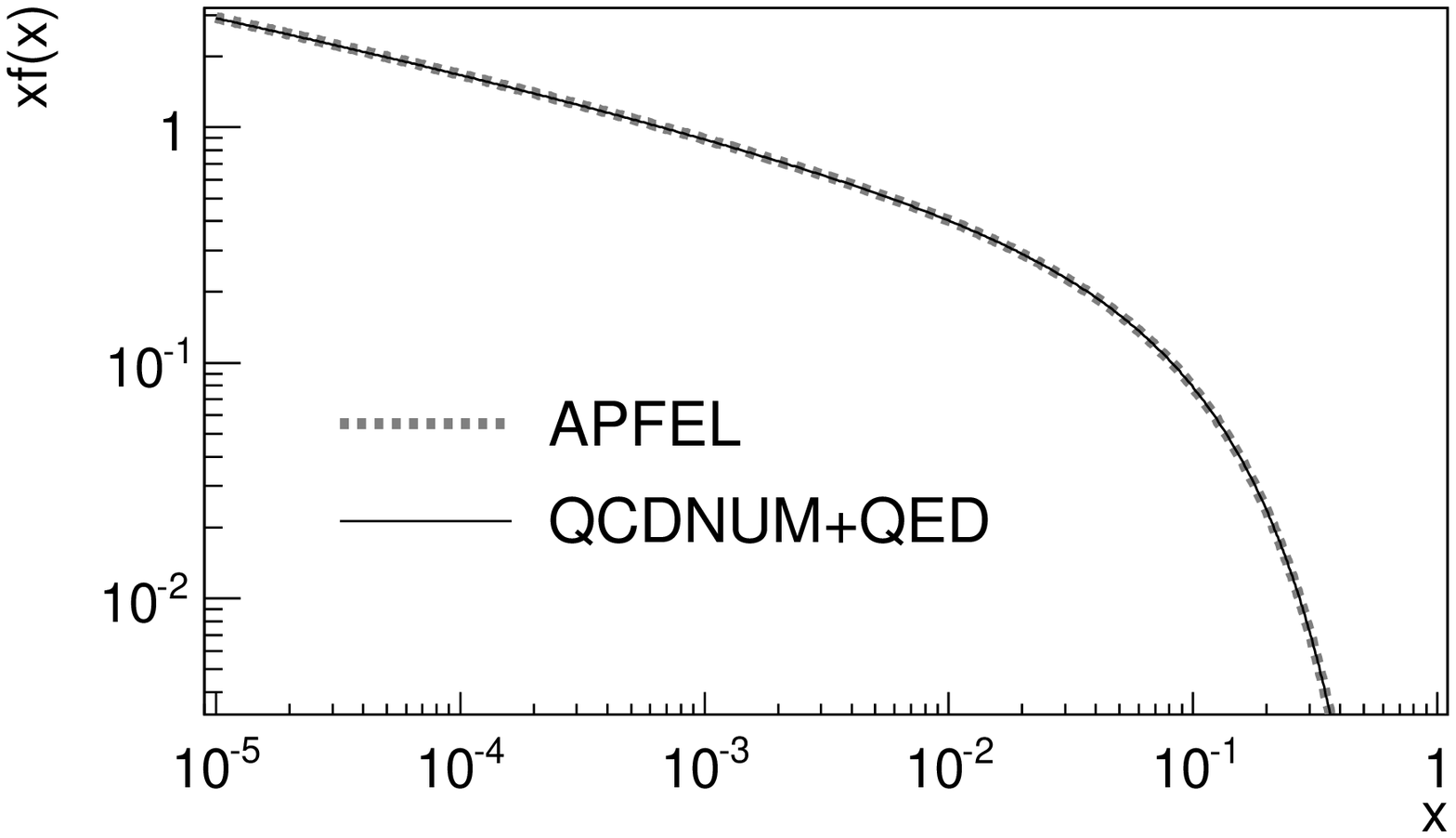}
\includegraphics[width = 0.45\textwidth]{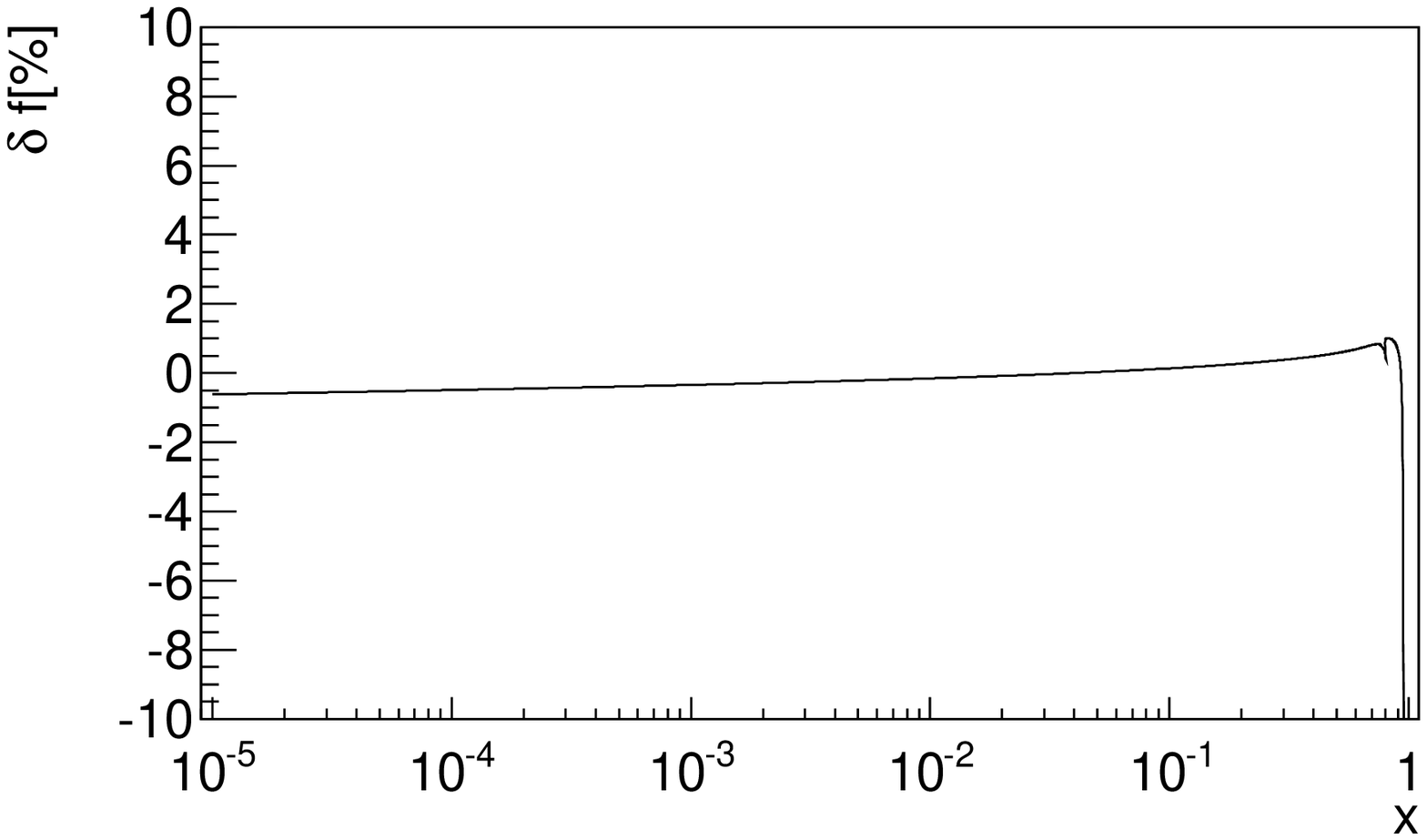}
\end{center}
\caption {Tuned comparison between {\tt QCDNUM+QED} and {\tt APFEL}.
Left plot shows the momentum distribution of $ \Delta_{sb} $ at $ \mu^2 = 10^4 \text{ GeV}^2 $.
The corresponding $ \delta f $ is shown on the right plot.}
\label{fig26}
\end{figure}

For the numerical comparison of {\tt QCDNUM+QED} and {\tt APFEL} in VFNS
the distributions were evolved from initial scale $ \mu_0 = 2 $ GeV
to $ \mu = 100 $ GeV on $ 1000 \times 1000 $ grid in $ x $ and $ \mu^2 $. The parametrization
for initial PDFs were taken from \cite{Giele:2002hx}. The results of comparison in VFNS for
momentum distributions $ xf $ and the corresponding relative differences
\begin{equation}
\delta f =
\frac{xf(\text{\tt QCDNUM+QED}) - xf(\text{\tt APFEL})}{xf(\text{\tt APFEL})}
\label{eq14}
\end{equation}
are presented in Figs. \ref{fig18}-\ref{fig26} where an agreement within $ \sim 1 \% $ is observed
excluding a region of large x. A larger discrepancy for photon distribution can be caused by
different treatment of $ \mathcal{O} (\alpha^2) $ effects.
\section{Conclusions}
QED-modified evolution equations are implemented into beta version of the next release of
{\tt QCDNUM} program and cross-checked with {\tt partonevolution} program in FFNS and with
{\tt MRST2004QED} PDF set and {\tt APFEL} program in VFNS. The {\tt APPLGRID} interface to
{\tt SANC} Monte Carlo generator was created for fast evaluation of LO photon-induced cross-section.
These developments are planned to implement into {\tt HERAFitter} package and can be used for QED
fits of LHC data.
\vskip 2 mm
{\bf \Large Acknowledgements}
\vskip 2 mm
I would like to thank the HERAFitter team and the ATLAS group in DESY for kind hospitality.
I am grateful to M. Botje for helpful discussions of the features of new {\tt QCDNUM} version
and to R. Placakyte and V. Radescu for their comments on the paper draft.
This work are supported in part by BMBF-JINR cooperation program and RFBR grant
{\verb"12-02-91526-ЦЕРН_а"}.
\clearpage

\clearpage

\clearpage

\clearpage

\clearpage

\clearpage

\clearpage

\clearpage

\clearpage

\providecommand{\href}[2]{#2}\end{document}